\def\thecol{2}
\if\thecol1
\documentclass[12pt, draftclsnofoot, onecolumn]{IEEEtran}
\else
\documentclass[journal,10pt]{IEEEtran}
\fi

\usepackage{enumerate}
\usepackage{amssymb}
\usepackage{color}
\usepackage{bm}

\usepackage{graphicx}
\usepackage{epsfig}
\usepackage{stfloats}
\ifCLASSOPTIONcompsoc
\usepackage[caption=false,font=normalsize,labelfon
t=sf,textfont=sf]{subfig}
\else
\usepackage[caption=false,font=footnotesize]{subfig}
\fi

\usepackage{amsfonts}
\usepackage{amsmath}
\usepackage{amsthm}
\usepackage{autobreak}
\usepackage{mathrsfs}
\usepackage{algorithm}
\usepackage{algorithmic}

\usepackage[adjust]{cite}
\usepackage[colorlinks,linkcolor=blue,citecolor=blue]{hyperref}
\usepackage{booktabs}
\usepackage{multirow}
\usepackage{amsthm}
\theoremstyle{definition}

\newcounter{eqlabel}[subsection]

\def\imagunit{\mathsf{j}}
\def\rq{\mathsf{r}}
\def\tq{\mathsf{t}}
\def\rrt{r_{\mathsf{rt}}}

\usepackage[mathscr]{eucal}
\allowdisplaybreaks[4]

\usepackage{xcolor}

\makeatletter
\def\ps@IEEEtitlepagestyle{
  \def\@oddfoot{\mycopyrightnotice}
  \def\@evenfoot{}
}
\def\mycopyrightnotice{
  {\scriptsize
  \begin{minipage}{\textwidth}
  \centering
  
\copyright~2023 IEEE. Personal use is permitted, but republication/redistribution requires IEEE permission.
See https://www.ieee.org/publications/rights/index.html for more information.
  \end{minipage}
  }
}

\begin{document}
\title{Cramér-Rao Bounds of Near-Field Positioning Based on Electromagnetic Propagation Model}
\author{Ang Chen,~Li Chen,~\IEEEmembership{Senior~Member, IEEE},~Yunfei Chen,~\IEEEmembership{Senior~Member, IEEE},\\~Changsheng You,~\IEEEmembership{Member, IEEE},~Guo Wei,~and F.~Richard~Yu,~\IEEEmembership{Fellow, IEEE}
\thanks{This research was supported by  
National Key R\&D Program of China (Grant No. 2021YFB2900302). (\emph{Corresponding author: Li Chen})}
\thanks{
		A.~Chen, L.~Chen, and G.~Wei are with the CAS Key Laboratory of Wireless Optical Communication, University of Science and Technology of China (USTC), Hefei 230027, China (e-mail: chenang1122@mail.ustc.edu.cn; \{chenli87, wei\}@ustc.edu.cn).
  
  Y.~Chen is with the Department of Engineering, University of Durham, Durham DH1 3LE, U.K. (e-mail: yunfei.chen@durham.ac.uk).

  C.~You is with the Department of Electronic and Electrical Engineering, Southern University of Science and Technology (SUSTech), Shenzhen, China (e-mail: youcs@sustech.edu.cn).

  F.~Richard~Yu is with the Department of Systems and Computer Engineering, Carleton University, Ottawa, ON, K1S 5B6, Canada (e-mail: richard.yu@carleton.ca).

  }}


\maketitle
\begin{abstract}
The adoption of large-scale antenna arrays at high-frequency bands is widely envisioned in the beyond 5G wireless networks. This leads to the near-field regime where the wavefront is no longer planar but spherical, bringing  new opportunities and challenges for communications and positioning. In this paper, we improve the near-field positioning technology from the  classical \textit{spherical wavefront model} (SWM) to the more accurate and true \textit{electromagnetic propagation model} (EPM). A generic near-field positioning model with different observation capabilities for three electric field types (\textit{vector}, \textit{scalar}, and \textit{overall scalar electric field}) is developed based on the complete EPM. For these three observed electric field types, the Cramér-Rao bound (CRB) is adopted to evaluate the achievable estimation accuracy. The expressions of the CRBs for different electric field observations are derived by combining electromagnetic propagation concepts with estimation theory. Closed-form expressions can be further obtained as the terminal is assumed to be on the central perpendicular line (CPL) of the receiving antenna surface. Moreover, the above discussions are extended to the system with multiple receiving antennas. In this case, the CRBs using various electric field types are  derived and the effect of different numbers of receiving antennas is deeply investigated. Numerical results are provided to quantify the CRBs and validate the analytical results. 
Also, the impact of different system parameters, including electric field type, wavelength,
size of the receiving antenna, and number of antennas, is evaluated.
\end{abstract}


\section{Introduction}
 The $\text{5}^\textrm{th}$ generation (5G) and beyond networks require real-time and high-accuracy positioning, since ubiquitous position information can be extracted from node-to-node communications in the networks\cite{del2017survey}. Traditional positioning technologies in the wireless networks typically consider the terminal located in the Fraunhofer (far-field) region, where the wavefront of an electromagnetic  wave can be approximated as planar.
 
 Envisioned as the key features of  the beyond 5G networks (B5G), the adoption of large-scale antenna arrays/surfaces\cite{bjornson2019massive,tang2020wireless}, and exploitation of high frequency bands\cite{akyildiz2018combating,rappaport2019wireless} will push the electromagnetic diffraction field from the far-field region towards the near-field region, 
where the propagated wavefront tends to be spherical and the uniform plane wave assumption will no longer hold\cite{zhang2022beam}. The near-field channel’s array manifold vectors contain more information on the terminal position, as both \textit{distance information} and direction of arrival (DoA) information can be inferred from the receiving array. Thus, wireless communication taking place in the near-field region provides both new opportunities and challenges for positioning. 

Since traditional positioning technologies are developed for far-field region, it is essential to develop new 
architectures and approaches to achieve high accuracy and resolution for near-field region.
Most works on near-field positioning have focused on three aspects: \textit{positioning model design}, \textit{signal processing algorithm}, and \textit{performance evaluation}. For the model design, \cite{1165222} proposed a model with an imperfectly calibrated array for near-field positioning and investigated a calibration method. To simplify the near-field model, many works applied the Fresnel approximation  to the antenna arrays with special geometries, e.g., uniform linear arrays (ULAs)\cite{grosicki2005weighted,chen2004new,deng2007closed}, and considered the model mismatch that was shown to reduce the estimation accuracy \cite{hsu2011mismatch} while analyzing the achievable precision.  

To improve the model accuracy further, the \textit{spherical wavefront model} ({SWM}) was  developed. An array was utilized to extract the distance and  DoA information based on the SWM. It was revealed that the spherical wavefront provided an underlying generic parametric model for near-field positioning\cite{haneda2006parametric}. 
In \cite{yin2017scatterer}, the SWM was extended to a practical scenario with large-scale antenna arrays. The result indicated that terminals in the near-field region could be identified by employing large-scale antenna arrays to estimate the wavefront curvature, i.e., curvature  arrival (CoA). In order
to reduce the complexity and implementation cost of large-scale antenna arrays, the authors in \cite{9335528} introduced the electromagnetic (EM) lens to the SWM.

Other works have studied signal processing algorithms for near-field positioning based on the SWM. For example, \cite{huang1991near,661337,4217610} developed a modified two-dimensional  MUSIC algorithm, high-order ESPRIT-like algorithm, and overlapping sub-arrays algorithm, respectively. In \cite{liang2009passive}, a two-stage MUSIC algorithm was proposed  to estimate the position of a mixed near-field and far-field terminal. The results demonstrated that the {curvature information} should be exploited when the terminal approaches the receiver.
A subspace-based algorithm without eigendecomposition was proposed in  \cite{8410024}, which could provide remarkable and satisfactory estimation performance compared with some existing near-field positioning algorithms. For the model using large-scale antenna
arrays equipped with EM-lens, a parameterized estimation algorithm was investigated  in \cite{9314267}, which directly reused receiving signals to extract position parameters. 

Based on the SWM, the performance (i.e., estimation accuracy) of near-field positioning could be evaluated. In practical scenarios, as electromagnetic waves encounter non-ideal phenomena such as noise, fading, and shadowing, the positioning performance is subject to uncertainty.
In the interest of system design and operation, it is crucial to obtain achievable accuracy in positioning operations to provide benchmarks for evaluating the performance of the actual systems. To this end, the Cramér-Rao bound (CRB) is the most commonly adopted tool, which describes the fundamental lower limits for estimation accuracy. In \cite{9314267,delmas2013crb,6705635,9145059,delmas2016crb,hu2018beyond2}, the SWM was used to derive the CRBs for the near-field estimator with ULA, uniform circular arrays, planar arrays, large-scale arrays, and large intelligent surfaces (LIS).

Most of the above mentioned works are based on the SWM. Although the SWM is widely utilized and relatively simple, it has been proven inaccurate in \cite{8736783}. Specifically, the SWM does not correspond to the equations governing the electromagnetic fields around an antenna or array while typically disregards the physical characteristics of the near-field source. This could profoundly impact the generated electromagnetic fields and the observations collected by the receiver. On the other hand, the \textit{electromagnetic propagation model} (EPM) is by far the most accurate electromagnetic theory-based model for investigating signals in the near-field region. Compared with SWM, it has the following three advantages: (i) EPM is a true and complete model on the basis of Maxwell's equations. It can intrinsically describe the dependence of the observed signal and physical characteristics (e.g., current distribution, type, and size) of the source. (ii) EPM contains more position, structure, and attitude information propagated outwards by electromagnetic radiation. (iii) Using EPM to model channels and signals is closer to the actual communication scenario and can explicitly consider the antenna's element design and radiation pattern. Thus, this work will develop a true EPM and use it for near-field positioning.

Utilizing EPM for near-field positioning may 
lead to higher estimation accuracy, but the EPM-based analysis is more challenging and complex. {The authors in \cite{de2021cramer,d2022cramertsp} investigated the EPM based on the \textit{radiation vector} \cite[Ch. 15]{orfanidis} and evaluated the near-field positioning performance by utilizing the EPM. In \cite{de2021cramer}, they computed the CRBs for the source dipole that is assumed to be located on the {central perpendicular line} (CPL) of the receiving antenna surface by measuring the \textit{vector electric field} (\textit{VEF}). In \cite{d2022cramertsp}, they further provided the expressions of the CRBs in two scenarios, in which the priori knowledge of the dipole orientation can be assumed known or
unknown to the receiver. However, it is more general for the terminal not to be located on the CPL since the CPL condition is not always satisfied in practical application. Although in \cite{d2022cramertsp}, the authors mentioned the case that the terminal is not on the CPL, they did not derive detailed expressions of the CRBs in this general case. Furthermore, in addition to \textit{vector electric field}, \textit{scalar electric field} and \textit{overall scalar electric field} observations are also possible due to the  different observation capabilities of various receiving antenna paradigms. Thus, compared to \cite{de2021cramer,d2022cramertsp}, a more comprehensive study of positioning the terminal  at an arbitrary position by measuring different electric fields is necessary. Consequently, it remains unclear how to evaluate the performance of near-field positioning in such a study using the \textit{electromagnetic propagation model} and estimation theory.}

{In this paper,
we will extend the work in 
\cite{de2021cramer} by developing a generic near-field positioning system based on the \textit{electromagnetic propagation model} for arbitrary terminal positions and three different  electric field types. Unlike \cite{de2021cramer}, the position of the terminal in front of the receiving antenna\footnote{The receiving antenna is a broad concept referring to antenna paradigms with different observation capabilities, such as a conventional surface antenna and intelligent surfaces with a large number of finely customizable antennas. Different observation capabilities refer to obtaining different electric fields.} is unrestricted such that it can be placed anywhere. {Unlike \cite{d2022cramertsp}, the detailed CRB expressions for arbitrary terminal positions are explicitly derived and the unknown effect of different observed electric field types on the positioning performance is investigated for the first time. Furthermore, the impact
of multiple distributed receiving antennas is extensively examined and this is a new analysis that cannot be found in prior near-field works.} The main contributions of this paper are summarized as follows.}

\begin{itemize}
	\item \textbf{Accurate near-field modeling based on EPM.} Unlike traditional near-field positioning technologies following the classical SWM, a complete EPM without any approximation is
 developed based on the \textit{electromagnetic theory}. This EPM can accurately model near-field channels and explicitly describe the functional dependence of the near-field signals on the physical characteristics of the source. In addition to the EPM in the near-field form, its Fresnel and plane wave approximation forms are also discussed. Moreover, the CRBs for estimating the terminal position are computed by combining the EPM with the estimation theory to provide fundamental limits for the performance of the actual near-field positioning system.
	\item \textbf{Generic CRB analysis and performance comparison.} A generic  near-field positioning model is developed  considering the \textit{variety} of observed electric fields and the \textit{universality} of the terminal position. In particular, three electric field observation types (\textit{vector}, \textit{scalar}, and \textit{overall scalar electric field}) are measured by receiving antennas with  different observation capabilities to derive the CRBs for a terminal located at an arbitrary position. To the best of the authors’ knowledge, such generic CRBs have never been studied, and they can generalize the existing results in \cite{de2021cramer}. {Also, we  first compare positioning performances using different observed electric fields through both theoretical analysis and simulation, which cannot be found in prior works, such as \cite{d2022cramertsp}}. We show that performance will decrease with the degradation of the receiving antenna's observation capability. 
 Moreover, the precise closed-form expressions or upper/lower bounds of the CRBs are given in the CPL case to obtain insights about the impact on the positioning performance for different system parameters. We show that the CRBs are proportional to the square of the wavelength. Also, we reveal that the estimation accuracy of some position coordinate components  approaches a fixed limit or improves infinitely when the ratio between the surface diagonal length  (size) of the
receiving antenna and its distance from the terminal increases. 
	\item \textbf{Extended discussion of SIMO system.} We have already
discussed the case of centralized deployment of receiving
antenna, in which the receiving antenna can be regarded
as the single antenna. To investigate the impact of multiple receiving antennas on the performance, the  generic positioning model is also extended to a system with multiple receiving antennas, i.e., the single-input multiple-output (SIMO) system. The expressions of CRBs are derived and the results reveal that multiple receiving antennas can significantly
improve the estimation accuracy of dimensions parallel to the
receiving surface. {This is a new result that cannot be found in prior works, e.g., \cite{d2022cramertsp}.}
\end{itemize}

The remainder of this paper is organized as follows. Section \ref{SEC:2} describes the generic system model, introduces three electric field observations and their physical implications, develops a complete EPM, and derives the specific CRB expressions. In Section \ref{sectionCPL}, the CPL case and two further simplified scenarios are studied. In Section \ref{sec:simo}, the generic
model is extended to the SIMO system. Numerical results and discussion are presented in Section \ref{sectionV}, and the conclusions are provided in Section \ref{sec:con}.

\textit{Notation:} Vectors and matrices are denoted in bold lowercase and uppercase, respectively, e.g., $\mathbf{a}$ and $\mathbf{A}$. We use $[\mathbf{A}]_{ij}$ to
denote the $(i,j)$th entry of $\mathbf{A}$ and $\mathbf{a}_{i}$ to denote the $i$th entry of $\mathbf{a}$. The superscripts $(\cdot)^{\dagger}$, $(\cdot)^{-1}$, and $(\cdot)^{\mathrm{T}}$ represent the matrix hermitian-transpose, inverse, and transpose, respectively. $(\cdot)^{*}$ and $\operatorname{Re}\{\cdot\}$ designate the complex conjugate and the real part of the input operations. The operator $\|\cdot\|$ means to compute $\mathcal{L}_{2}$-norm of the input and $|\cdot|$ stands for the modulo operator. The notations $\mathbb{C}$ and $\mathbb{R}$ represent the sets of complex numbers and of real numbers, respectively. The notation $\imagunit$ denotes the imaginary unit, and $\mathbf{I}_{N}$ indicates the $N\times N$ identity matrix. \textit{The suffix $\kappa = x, y, z$ represents the
$X$-, $Y$- and $Z$-dimension in the cartesian
coordinate system, respectively}.
\section{System Model and Performance Metric}
\label{SEC:2}
This section will first introduce a generic near-field positioning system aiming to estimate the terminal position based on the electric fields observed over the receiving antenna surface area. Since different paradigm selections and hardware settings of the receiving antenna have disparate {observation capabilities}, embodied in extracting various observations, i.e., \textit{vector}, \textit{scalar}, and \textit{overall scalar electric field}, we will consider all these  electric fields for the near-field positioning system. Then, a complete EPM without any approximation will be developed to accurately describe near-field signals. Finally, the CRB for the terminal position will be used as the performance metric and derived by combining the EPM with estimation theory. 
\subsection{Generic System Model of Near-Field Positioning}\label{subsection_SMNP}
\begin{figure}[!t]
\centering
\includegraphics[scale=0.59]{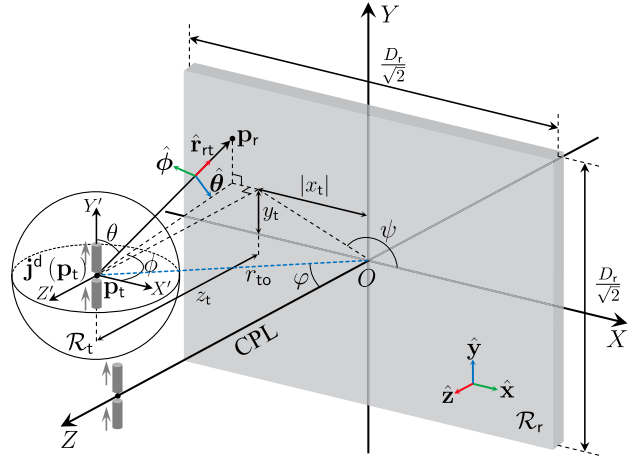}
\caption{ Illustration of the generic near-field positioning system.  The CPL case where the terminal is located in the central
perpendicular line (CPL) of the receiving
antenna surface, i.e., $\varphi=0$, $x_{\tq}=y_{\tq}=0$, is also illustrated. \emph{Note that} an electrically small source can be approximated as a single sizeless point source, i.e., we are not concerned with the geometric dimension and physical shape of the source. As will be seen later, we consider the terminal to be an \textit{elementary}
(i.e., \textit{Hertzian electric}, \textit{ideal} or \textit{short}) \textit{dipole}.}
\label{system1}
\end{figure}

Consider the near-field positioning system depicted in Fig. \ref{system1}. The terminal is an electrically small  
source equipped with
a monochromatic single-antenna located at an arbitrary point $\mathbf{p}_{\mathsf{t}}$\footnote{An arbitrary point $P$ in $\mathbb{R}^{3}$ Euclidean space can be represented by a spatial vector $\mathbf{p}$. Specifically, $P$ is the
endpoint of $\mathbf{p}$, whose starting point is fixed.} inside a three-dimensional source region $\mathcal{R}_{\mathsf{t}}$. The electric
current density ${\mathbf{j}^{\mathsf{d}}}(\mathbf{p}_{\tq},t)$ at the terminal generates an electric field ${\mathbf{e}^{\mathsf{f}}}(\mathbf{p}_{\mathsf{r}},t) \in \mathbb{C}^{3}$ at an arbitrary point $\mathbf{p}_{\mathsf{r}}$ on the surface $\mathcal{R}_{\mathsf{r}}$ of the receiving antenna through a homogeneous and isotropic medium with neither scatterers nor reflectors, and we consider time-harmonic fields and introduce phasor fields\footnote{In this case, Maxwell’s equations are considerably simplified and can be written only in terms of the current
and field phasors, ${\mathbf{j}^{\mathsf{d}}}\left(\mathbf{p}_{\tq}\right)$ and ${\mathbf{e}^{\mathsf{f}}}\left(\mathbf{p}_{\rq}\right)$.}: ${\mathbf{j}^{\mathsf{d}}}(\mathbf{p}_{\tq}, t)=\operatorname{Re}\left\{{\mathbf{j}^{\mathsf{d}}}\left(\mathbf{p}_{\tq}\right)\mathrm{e}^{-\imagunit \omega t}\right\}$ and ${\mathbf{e}^{\mathsf{f}}}(\mathbf{p}_{\rq}, t)=\operatorname{Re}\left\{{\mathbf{e}^{\mathsf{f}}}\left(\mathbf{p}_{\rq}\right)\mathrm{e}^{-\imagunit \omega t}\right\}$, where $\omega$ is the angular frequency in  radians/second.

We establish two cartesian coordinate systems, $OXYZ$ and $PX^{\prime}Y^{\prime}Z^{\prime}$, with 
a pure translational relationship. The center of $\mathcal{R}_{\mathsf{r}}$ ($O$) and $\mathbf{p}_{\mathsf{t}}$ are their origins, respectively. In the $OXYZ$ system, $\mathbf{p}_{\mathsf{t}}=\left(x_{\mathsf{t}},y_{\mathsf{t}},z_{\mathsf{t}}\right)^{\mathrm{T}}$, $\mathbf{p}_{\mathsf{r}}=(x_{\mathsf{r}},y_{\mathsf{r}},0)^{\mathrm{T}}$, and \textit{observation region} $\mathcal{R}_{\mathsf{r}}=\left \{(x_{\mathsf{r}},y_{\mathsf{r}},0):|x_{\mathsf{r}}|\leq D_{\mathsf{r}}/\sqrt{8},|y_{\mathsf{r}}|\leq D_{\mathsf{r}}/\sqrt{8} \right \}$, where $D_{\mathsf{r}}$ is the maximum geometric dimension of the receiving antenna, namely, the diagonal length of the square surface. Establish a spherical coordinate system $\left(\rrt,\theta,\phi\right)$ (with respect to $PX^{\prime}Y^{\prime}Z^{\prime}$) of point $\mathbf{p}_{\mathsf{t}}$ to facilitate the description of the radiation pattern. $\hat{\mathbf{x}}$, $\hat{\mathbf{y}}$, and $\hat{\mathbf{z}}$ are unit vectors along the $X$-, $Y$-, and $Z$-dimension in the $OXYZ$ system while $\hat{\bm{\theta}}$ and $\hat{\bm{\phi}}$ are unit vectors along the $\theta$ and $\phi$ coordinate curves. ${\hat{\mathbf{r}}_{\mathsf{rt}}}$ is a unit vector denoting the direction of ${\mathbf{r}_{\mathsf{rt}}}=\mathbf{p}_{\mathsf{r}}-\mathbf{p}_{\mathsf{t}}$, i.e., ${\hat{\mathbf{r}}_{\mathsf{rt}}}=\frac{{\mathbf{r}_{\mathsf{rt}}}} {\|{\mathbf{r}_{\mathsf{rt}}}\|}$. For the terminal,  let $r_{\mathsf{to}}$ denote its distance to the center of the receiving antenna, and $\varphi \in \left[0,\frac{\pi}{2}\right]$ and $\psi \in \left[0,2\pi\right)$ denote the zenith and azimuth
angles, respectively.

The near-field positioning system can estimate the terminal position by using the {electric field observations} obtained over the receiving antenna surface area (\textit{observation region} $\mathcal{R}_{\mathsf{r}}$). It is worth remarking that, depending on actual communication requirements, cost constraints or device technology limitations, the types and settings of receiving antennas may be different, leading to extraction of various types of observed electric fields and thus affecting the positioning performance. Next, we will discuss three different cases of the electric field observations.

\textbf{\textit{1) Vector Electric Field (VEF):}} The electric field generated by the source current distribution ${\mathbf{j}^{\mathsf{d}}}\left(\mathbf{p}_{\tq}\right)$ at the point $\mathbf{p}_{\tq}$ is a \textit{three-dimensional vector}, which has three \textit{components} along the directions of given orthonormal vectors in the reference system (cartesian or spherical).
The most ideal case is that the \textit{vector electric field} at \textit{each point} on the whole contiguous surface of the receiving antenna can be observed. In such a case, the receiving antenna should be modeled as a \textit{two-dimensional} metasurface, also known as  the emerging spatially-continuous electromagnetic (EM) surface, large
intelligent surface (LIS) \cite{hu2018beyond1}, and holographic MIMO surface \cite{huang2020holographic}, which is an electronically active surface consisting of arrays of reconfigurable elements of metamaterial. In fact, the metasurface concept can be seen as an extreme extension of earlier research in massive-MIMO or extremely large-scale MIMO concept \cite{marinello2020antenna}. From the technological
point of view, metamaterials represent appealing candidates for the creation of software-controlled metasurfaces since  
metasurfaces can be built from ultrathin two-dimensional metamaterials and \cite{tsilipakos2020toward} surveyed \textit{implementation} and practical application aspects of metasurfaces. In the cartesian $OXYZ$ system, the \textit{vector electric field} $\mathbf{e}^{\mathsf{v}}\left(\mathbf{p}_{\mathsf{r}} \right)$ can be written as
\begin{equation}
\small
\setlength\abovedisplayskip{3.5pt}
\setlength\belowdisplayskip{3.5pt}
\mathbf{e}^{\mathsf{v}}\left(\mathbf{p}_{\mathsf{r}} \right)= e_{x}^{\mathsf{v}}\left(\mathbf{p}_{\mathsf{r}} \right)\hat{\mathbf{x}}+e_{y}^{\mathsf{v}}\left(\mathbf{p}_{\mathsf{r}} \right)\hat{\mathbf{y}}+e_{z}^{\mathsf{v}}\left(\mathbf{p}_{\mathsf{r}} \right)\hat{\mathbf{z}}. \label{VEF11}
\end{equation}
The observation equation is $\tilde{\mathbf{e}}^{\mathsf{v}}\left(\mathbf{p}_{\mathsf{r}} \right)=\mathbf{e}^{\mathsf{v}}\left(\mathbf{p}_{\mathsf{r}} \right)+\mathbf{n}^{\mathsf{v}}\left( \mathbf{p}_{\mathsf{r}}\right)$, where $\tilde{\mathbf{e}}^{\mathsf{v}}\left(\mathbf{p}_{\mathsf{r}} \right)$ is the noisy \textit{VEF} and $\mathbf{n}^{\mathsf{v}}\left( \mathbf{p}_{\mathsf{r}}\right)\in \mathbb{C}^{3}$ is the random noise generated by electromagnetic sources outside $\mathcal{R}_{\tq}$.

\textbf{\textit{2) Scalar Electric Field (SEF):}} If a metasurface is selected as the receiving antenna, the electric field at \textit{each point} in the observation region $\mathcal{R}_{\rq}$ can be extracted spatially continuously. Sometimes all the three cartesian components of the measured electric field cannot be obtained accurately, but only one \textit{scalar electric field} can be acquired at \textit{each point}. We refer to this phenomenon as the \textit{observation capability degradation}\footnote{The reasons for the \textit{observation capability degradation} include: different selections or settings of the metamaterial elements, device hardware limitations, different requirements for communication and sensing, and so on.} of the receiving antenna. The simplest \textit{scalar electric field} can be defined as one of the three components of the \textit{vector electric field} $\mathbf{e}^{\mathsf{v}}\left(\mathbf{p}_{\mathsf{r}} \right)$, i.e., $e_{x}^{\mathsf{v}}\left(\mathbf{p}_{\mathsf{r}} \right)$, $e_{y}^{\mathsf{v}}\left(\mathbf{p}_{\mathsf{r}} \right)$, or $e_{z}^{\mathsf{v}}\left(\mathbf{p}_{\mathsf{r}} \right)$. In this paper, we consider the \textit{scalar electric field} defined from the power point of view. Specifically, we exploit the \textit{scalar electric field} that is a component of the \textit{Poynting vector} perpendicular to each point of the whole contiguous observation region $\mathcal{R}_{\mathsf{r}}$. This \textit{SEF} can be regarded as a scalar approximation to the \textit{VEF} in \eqref{VEF11}  and give an intermediate step to understand the {electric field model}. In the $OXYZ$ system, the \textit{SEF} $e^{\mathsf{s}}\left( \mathbf{p}_{\mathsf{r}}\right)$ is written as 
\begin{equation}\small \setlength\abovedisplayskip{3.5pt}
\setlength\belowdisplayskip{3.5pt}
e^{\mathsf{s}}\left(\mathbf{p}_{\mathsf{r}}\right)=\sqrt{\|\mathbf{e}^{\mathsf{v}}\left(\mathbf{p}_{\mathsf{r}} \right)\|^{2}\left(- {\hat{\mathbf{r}}_{\mathsf{rt}}}\cdot{\hat{\mathbf{z}}}\right)}\mathrm{e}^{-\imagunit k_{0}\rrt},
    \label{eq:Epr_definition}
\end{equation}
where $k_{0}=2\pi/\lambda$ is the wave number, $\lambda$ is the wavelength, $\cdot$ indicates inner product of vectors, and $\rrt=\|{\mathbf{r}_{\mathsf{rt}}}\|$. Then, the observation equation utilizing \textit{SEF} is
$\tilde{e}^{\mathsf{s}}\left(\mathbf{p}_{\mathsf{r}} \right)=e^{\mathsf{s}}\left(\mathbf{p}_{\mathsf{r}}\right)+n^{\mathsf{s}}\left( \mathbf{p}_{\mathsf{r}}\right)$, where $\tilde{e}^{\mathsf{s}}\left(\mathbf{p}_{\mathsf{r}} \right)$ is the observation of the \textit{SEF} with noise. 

\textbf{\textit{3) Overall Scalar Electric Field (OSEF):}} With a further decline in the
 \textit{observation capability} of the receiving antenna, we consider that only one \textit{overall scalar electric field} can be obtained through observation, which is defined as the double integral of the \textit{SEF} over the receiving antenna surface. In such a case, the receiving antenna degenerates from a metasurface to a conventional surface antenna\cite{bjornson2021primer}.  From \eqref{eq:Epr_definition}, the \textit{OSEF} $e^{\mathsf{o}}$ can be written as
\begin{equation}\setlength\abovedisplayskip{3.5pt}
\setlength\belowdisplayskip{3.5pt}\small
    e^{\mathsf{o}}=\sqrt{\frac{2}{D_{\mathsf{r}}^{2}}}\iint_{\mathcal{R}_{\mathsf{r}}}e^{\mathsf{s}}\left(\mathbf{p}_{\mathsf{r}} \right)d\mathbf{p}_{\mathsf{r}},
\label{eq:overall}
\end{equation}
where $D_{\mathsf{r}}^{2}/2$ is the area of the receiving surface antenna. Then, the observation equation using the \textit{OSEF} is $\tilde{e}^{\mathsf{o}}={e}^{\mathsf{o}}+n^{\mathsf{o}}$.

We aim to derive the CRBs for estimating the position of $\mathbf{p}_{\mathsf{t}}$ using the above three observation equations with noisy electric fields ($\tilde{\mathbf{e}}^{\mathsf{v}}\left(\mathbf{p}_{\mathsf{r}} \right)$, $\tilde{{e}}^{\mathsf{s}}\left(\mathbf{p}_{\mathsf{r}} \right)$, and $\tilde{{e}}^{\mathsf{o}}$) over the \textit{observation region} $\mathcal{R}_{\mathsf{r}}$. For this purpose, we provide the statistical model for the random noise fields $\mathbf{n}^{\mathsf{v}}\left( \mathbf{p}_{\mathsf{r}}\right)$, ${n}^{\mathsf{s}}\left( \mathbf{p}_{\mathsf{r}}\right)$, and $n^{\mathsf{o}}$ as follows.

\textbf{\textit{Random noise fields modeling:}} Following \cite{jensen2008capacity} and \cite{gruber2008new}, we model random noise fields as \textit{spatially uncorrelated}  circularly-symmetric zero-mean
complex-Gaussian processes with \textit{correlation functions}:
$\mathbb{E}\left\{\mathbf{n}^{\mathsf{v}}\left( \mathbf{p}_{\mathsf{r}}\right){\mathbf{n}^{\mathsf{v}}}^{\dagger}\left( \mathbf{p}^{\prime}_{\mathsf{r}}\right)\right\}=\sigma^{2}\mathbf{I}_{3}\delta\left({\mathbf{p}_{\mathsf{r}}-\mathbf{p}^{\prime}_{\mathsf{r}}}\right)$, $\mathbb{E}\left\{{n}^{\mathsf{s}}\left( \mathbf{p}_{\mathsf{r}}\right){{n}^{\mathsf{s}}}^{*}\left( \mathbf{p}^{\prime}_{\mathsf{r}}\right)\right\}=\sigma^{2}\delta\left({\mathbf{p}_{\mathsf{r}}-\mathbf{p}^{\prime}_{\mathsf{r}}}\right)$, $\mathbb{E}\left\{{n}^{\mathsf{o}}{{n}^{\mathsf{o}}}^{*}\right\}=\sigma^{2}$, where $\mathbb{E}\left\{\cdot\right\}$ denotes the expectation operator, $\delta(\cdot)$ is the Dirac's
delta function, $\mathbf{p}^{\prime}_{\mathsf{r}}$ is an arbitrary point different from $\mathbf{p}_{\mathsf{r}}$, and $\sigma^{2}$ is the variance measured in ${\mathrm{V}_{olt}^{2}}$ (${\mathrm{V}_{olt}}$ indicates volts). 

Based on the estimation theory of statistical signal processing, the computation of the CRBs is provided as follows.

\noindent \textbf{Proposition 1} (CRB using \textit{VEF})\textbf{.} Denote the real vector to be estimated as ${{\bm{\xi}}}\in \mathbb{R}^{3}=\left(x_{\mathsf{t}},y_{\mathsf{t}},z_{\mathsf{t}} \right)$, which collects the unknown cartesian coordinates of $\mathbf{p}_{\mathsf{t}}$. The Fisher's Information Matrix (FIM), denoted as $\mathbf{I}\left(\bm{\xi}\right)$, is a $3\times 3 $ Hermitian matrix, whose element on the $m$-th row and $n$-th column is given by:
    \begin{equation}\small\setlength\abovedisplayskip{3.5pt}
\setlength\belowdisplayskip{3.5pt}
\begin{aligned}
\left[\mathbf{I}\left(\bm{\xi}\right)\right]&_{mn}=\frac{2}{\sigma^{2}} \iint_{\mathcal{R}_{\mathsf{r}}} \operatorname{Re}\left\{\frac{\partial e_{x}^{\mathsf{v}}\left(\mathbf{p}_{\mathsf{r}}\right)}{\partial \bm{\xi}_{n}} \frac{\partial {e_{x}^{\mathsf{v}}}^{*}\left(\mathbf{p}_{\mathsf{r}}\right)}{\partial \bm{\xi}_{m}}+\right.\\
&\left.\frac{\partial e_{y}^{\mathsf{v}}\left(\mathbf{p}_{\mathsf{r}}\right)}{\partial \bm{\xi}_{n}} \frac{\partial {e_{y}^{\mathsf{v}}}^{*}\left(\mathbf{p}_{\mathsf{r}}\right)}{\partial \bm{\xi}_{m}}+\frac{\partial e_{z}^{\mathsf{v}}\left(\mathbf{p}_{\mathsf{r}}\right)}{\partial \bm{\xi}_{n}} \frac{\partial {e_{z}^{\mathsf{v}}}^{*}\left(\mathbf{p}_{\mathsf{r}}\right)}{\partial \bm{\xi}_{m}}\right\} d x_{\mathsf{r}} d y_{\mathsf{r}},
\label{eq:CRBvector}
\end{aligned}
\end{equation}
where $m,n=1,2,3$. The CRB for estimating the $i$th entry of ${\bm{\xi}}$ is
\begin{equation}\small
\setlength\abovedisplayskip{3.5pt}
\setlength\belowdisplayskip{3.5pt}
\mathrm{CRB}\left(\bm{\xi}_{i} \right)=\left[\mathbf{I}\left(\bm{\xi}\right)^{-1}\right]_{ii}.
\label{eq:CRB}
\end{equation}

\begin{IEEEproof}
	The result can be  derived from \cite[Appendix 15C]{kay1993fundamentals} by replacing the noisy observation and the estimated parameter with complex vector $\tilde{\mathbf{e}}^{\mathsf{v}}\left(\mathbf{p}_{\mathsf{r}} \right)$ and real
vector ${{\bm{\xi}}}$, respectively.
\end{IEEEproof}
From Proposition 1, the CRBs utilizing \textit{SEF} and \textit{OSEF} can be computed by Corollary 1 and Corollary 2.

\noindent \textbf{Corollary 1} (CRB using \textit{SEF})\textbf{.} Using the \textit{scalar electric field}, the elements of FIM can be computed as:
\begin{equation}\small
\setlength\abovedisplayskip{3.5pt}
\setlength\belowdisplayskip{3.5pt}
\left[\mathbf{I}\left(\bm{\xi}\right)\right]_{m n}=\frac{2}{\sigma^{2}}\iint_{\mathcal{R}_{\mathsf{r}}} \operatorname{Re}\left\{\frac{\partial e^{\mathsf{s}}\left(\mathbf{p}_{\mathsf{r}}\right)}{\partial \bm{\xi}_{n}} \frac{\partial {e^{\mathsf{s}}}^{*}\left(\mathbf{p}_{\mathsf{r}}\right)}{\partial \bm{\xi}_{m}}\right\}dx_{\mathsf{r}}dy_{\mathsf{r}}.
\label{eq:FIMscalar}
\end{equation}
By substituting \eqref{eq:FIMscalar} into \eqref{eq:CRB}, CRBs in this case can be derived.

\begin{IEEEproof}
	According to Proposition 1, FIM is additive since $e_{x}^{\mathsf{v}}\left(\mathbf{p}_{\mathsf{r}}\right)$, $e_{y}^{\mathsf{v}}\left(\mathbf{p}_{\mathsf{r}}\right)$, and $e_{z}^{\mathsf{v}}\left(\mathbf{p}_{\mathsf{r}}\right)$ are independent. Accordingly, if we only have one noisy observation $\tilde{e}^{\mathsf{s}}\left(\mathbf{p}_{\mathsf{r}} \right)$, \eqref{eq:FIMscalar} is derived.
\end{IEEEproof}

\noindent \textbf{Corollary 2} (CRB using \textit{OSEF})\textbf{.} Similar to Corollary 1, the elements of FIM can be derived as:
\begin{equation}\small
\setlength\abovedisplayskip{3.5pt}
\setlength\belowdisplayskip{3.5pt}
\left[\mathbf{I}\left(\bm{\xi}\right)\right]_{mn} =\frac{2}{\sigma^{2}}\operatorname{Re}\left\{\frac{\partial e^{\mathsf{o}}}{\partial \bm{\xi}_{n}} \frac{\partial {e^{\mathsf{o}}}^{*}}{\partial \bm{\xi}_{m}} \right\}.
\label{eq:FIMoverall}
\end{equation}
By substituting \eqref{eq:FIMoverall} into \eqref{eq:CRB}, CRBs in this case are computed.

\begin{IEEEproof}
	The only difference between \eqref{eq:FIMscalar} and \eqref{eq:FIMoverall} is that $e^{\mathsf{s}}\left(\mathbf{p}_{\mathsf{r}}\right)$ has already been integrated in \eqref{eq:overall}.
\end{IEEEproof}

\noindent {\textbf{Discussion 1} (Performance metrics)\textbf{.} In this paper, we derive the CRBs for lower bounding the mean square error (MSE) to evaluate the performance of near-field positioning estimators. Unfortunately, the CRB is a \textit{local bound} and only \textit{asymptotically tight} in small error estimation scenario (say, high signal-to-noise ratio (SNR)). To provide a \textit{global tight bound} on the MSE and show a threshold effect, the Ziv-Zakai bound (ZZB) is proposed, which relates the MSE to the probability of error in a binary hypothesis testing problem. Some excellent works \cite{ZZBB1,ZZBB2}
provide a comprehensive survey of ZZB for far-field compression time delay and DoA estimation. It is without any doubt an interesting extension of our  analysis to consider the ZZB in the near-field positioning, which is left for future work.}

\subsection{Electromagnetic Propagation Model (EPM)}\label{subsection_EFE} 
\subsubsection{EPM for near-field} From the fundamental electromagnetic (Maxwell's) equations, the \textit{vector electric field} $\mathbf{e}^{\mathsf{v}}\left(\mathbf{p}_{\mathsf{r}}\right)$ generated at point $\mathbf{p}_{\mathsf{r}}$ from the electrically small source at point $\mathbf{p}_{\mathsf{t}}$ is due to the electric
current density ${\mathbf{J}}\left({\mathbf{p}}_{\mathsf{t}}\right)$ and satisfies \cite{poon2005degrees}
\begin{equation}\setlength\abovedisplayskip{3.5pt}
\setlength\belowdisplayskip{3.5pt}\small
{\mathbf{e}}^{\mathsf{v}}\left({\mathbf{p}}_{\mathsf{r}}\right)=\iiint_{\mathcal{R}_{\mathsf{t}}}{\overline{\overline{{\mathbf{G}}}}_{t}}\left(\mathbf{p}_{\mathsf{r}}-\mathbf{p}_{\mathsf{t}}\right){\mathbf{J}}\left({\mathbf{p}}_{\mathsf{t}}\right)d\mathbf{p}_{\mathsf{t}},
        \label{eq:E}
    \end{equation}
where ${\mathbf{J}}\left({\mathbf{p}}_{\mathsf{t}}\right)$ is Fourier representation ${\mathbf{J}}\left({\mathbf{p}}_{\mathsf{t}},\omega \right)$ of the current ${\mathbf{j}^{\mathsf{d}}}\left(\mathbf{p}_{\mathsf{t}} \right)$ at $\mathbf{p}_{\mathsf{t}}$. ${\overline{\overline{{\mathbf{G}}}}_{t}}({\mathbf{r}_{\mathsf{rt}}}) \in \mathbb{C}^{3\times 3}$ is referred to as the \textit{tensor Green function} in electromagnetic theory and can be expressed as
\begin{equation}\setlength\abovedisplayskip{3.5pt}
\setlength\belowdisplayskip{3.5pt}\small
\frac{{\overline{\overline{{\mathbf{G}}}}_{t}}({\mathbf{r}_{\mathsf{rt}}})}{{G_{s}}(\rrt)}=\left[\left(1+\frac{\imagunit}{k_{0}\rrt}-\frac{1}{k_{0}^{2}{r_{\mathsf{rt}}^{2}}}\right)\mathbf{I}-\left(1+\frac{\imagunit 3}{k_{0}\rrt}-\frac{3}{k_{0}^{2}{r_{\mathsf{rt}}^{2}}}\right){\mathbf{\Xi}} \right],
\label{eq:green} 
\end{equation}where ${G_{s}}(\rrt)=-\frac{\imagunit \eta}{2 \lambda \rrt} \mathrm{e}^{-\imagunit  k_{0}\rrt}$ is the \textit{scalar Green function},  $\eta$ is the intrinsic impedance of the medium, and ${\mathbf{\Xi} \triangleq \hat{\mathbf{r}}_{\mathsf{rt}}{\hat{\mathbf{r}}_{\mathsf{rt}}}^{\dagger}}$.  It is evident from \eqref{eq:green} that when $\rrt \geq \lambda$\footnote{If $\rrt=\lambda$, $\Big|1+\frac{\imagunit}{k_{0}\rrt}-\frac{1}{k_{0}^{2}{r_{\mathsf{rt}}^{2}}}\Big|^{2}\approx 0.975$, $\Big|1+\frac{\imagunit 3}{k_{0}\rrt}-\frac{3}{k_{0}^{2}{r_{\mathsf{rt}}^{2}}}\Big|^{2}\approx 1.082$.}, the second and third terms\footnote{They decay
rapidly with $\rrt$ and thus are only influential in the
“\textit{reactive near-field}”, which is very close to the source and ends at $d_{\mathrm{f}}=0.5\sqrt{D_{\rq}^{3}/\lambda}$.} in two parentheses in \eqref{eq:green} can be neglected, and hence 
\begin{equation}\setlength\abovedisplayskip{3.5pt}
\setlength\belowdisplayskip{3.5pt}\small
{\overline{\overline{{\mathbf{G}}}}_{t}}({\mathbf{r}_{\mathsf{rt}}}) \simeq-\frac{\imagunit k_{0} \eta \mathrm{e}^{-\imagunit k_{0}\rrt} }{4 \pi \rrt}\left(\mathbf{I}-{\hat{\mathbf{r}}_{\mathsf{rt}}{\hat{\mathbf{r}}_{\mathsf{rt}}^{\dagger}}}\right).
\label{eq:Gr}
\end{equation}
Since $\rrt \geq \lambda$ always holds when the terminal is in the near-field\footnote{
 In the far-field region, the transceiver distance is larger than the \textit{Fraunhofer distance} $d_{\mathrm{F}}=2D_{\mathsf{r}}^{2}/\lambda$\cite{sherman1962properties}. In this paper, the term “near-field” refers to the “radiative near-field”, where the transceiver distance is smaller than $d_{\mathrm{F}}$, but larger than the \textit{Fresnel distance} $d_{\mathrm{f}}=0.5\sqrt{D_{\rq}^{3}/\lambda}$\cite{selvan2017fraunhofer}.} (between the reactive near-field and the far-field) of the receiving antenna,  \eqref{eq:Gr} is adopted in subsequent computation. Substituting \eqref{eq:Gr} into \eqref{eq:E} and following the definition of vector cross product, we give the \textit{electromagnetic propagation model}:
\begin{equation}\setlength\abovedisplayskip{3.5pt}
\setlength\belowdisplayskip{3.5pt}\small
{\mathbf{e}}^{\mathsf{v}}\left({\mathbf{p}}_{\mathsf{r}}\right)=-\frac{\imagunit k_{0} \eta \mathrm{e}^{-\imagunit k_{0}\rrt} }{4 \pi \rrt} \left({\hat{\mathbf{r}}_{\mathsf{rt}}} \times \iiint_{\mathcal{R}_{\mathsf{t}}}{\mathbf{J}}\left({\mathbf{p}}_{\mathsf{t}}\right)d\mathbf{p}_{\mathsf{t}}\right)\times {\hat{\mathbf{r}}_{\mathsf{rt}}}.
        \label{eq:E1}
    \end{equation}
Write $\iiint_{\mathcal{R}_{\mathsf{t}}}{\mathbf{J}}\left({\mathbf{p}}_{\mathsf{t}}\right)d\mathbf{p}_{\mathsf{t}}$ as ${\mathbf{J}}_{\mathcal{R}_{\mathsf{t}}}={{J}}_{\mathcal{R}_{\mathsf{t}}}^{r}{\hat{\mathbf{r}}_{\mathsf{rt}}}+{{J}}_{\mathcal{R}_{\mathsf{t}}}^{\theta}\hat{\bm{\theta}}+{{J}}_{\mathcal{R}_{\mathsf{t}}}^{\phi}\hat{\bm{\phi}}$, where ${{J}}_{\mathcal{R}_{\mathsf{t}}}^{r}$, ${{J}}_{\mathcal{R}_{\mathsf{t}}}^{\theta}$, and ${{J}}_{\mathcal{R}_{\mathsf{t}}}^{\phi}$ are three components of the \textit{source current integral vector} ${\mathbf{J}}_{\mathcal{R}_{\mathsf{t}}}$ along ${\hat{\mathbf{r}}_{\mathsf{rt}}}$, $\hat{\bm{\theta}}$, and $\hat{\bm{\phi}}$ directions. Since ${\hat{\mathbf{r}}_{\mathsf{rt}}}\times {\hat{\mathbf{r}}_{\mathsf{rt}}}=\mathbf{0}$, $({\hat{\mathbf{r}}_{\mathsf{rt}}}\times \hat{\bm{\theta}})\times {\hat{\mathbf{r}}_{\mathsf{rt}}}=\hat{\bm{\theta}}$, and $({\hat{\mathbf{r}}_{\mathsf{rt}}}\times \hat{\bm{\phi}})\times {\hat{\mathbf{r}}_{\mathsf{rt}}}=\hat{\bm{\phi}}$, we have:
\begin{equation}\setlength\abovedisplayskip{3.5pt}
\setlength\belowdisplayskip{3.5pt}\small
{\mathbf{e}}^{\mathsf{v}}\left({\mathbf{p}}_{\mathsf{r}}\right)={G_{s}}(\rrt)\left({{J}}_{\mathcal{R}_{\mathsf{t}}}^{\theta}\hat{\bm{\theta}}+{{J}}_{\mathcal{R}_{\mathsf{t}}}^{\phi}\hat{\bm{\phi}}\right)\triangleq {G_{s}}(\rrt){\mathbf{J}}_{\mathcal{R}_{\mathsf{t}}}^{\perp}.\label{eq:12}
\end{equation}
${\mathbf{e}}^{\mathsf{v}}\left({\mathbf{p}}_{\mathsf{r}}\right)$ in the near-field of the source is the product of ${G_{s}}(\rrt)$ and ${\mathbf{J}}_{\mathcal{R}_{\mathsf{t}}}^{\perp}$. In particular, ${G_{s}}({\rrt})$  represents the scalar spherical wave, which 
accounts for the distance $\rrt$ between the source
and $\mathbf{p}_{\rq}$. The transverse component ${\mathbf{J}}_{\mathcal{R}_{\mathsf{t}}}^{\perp}$ intrinsically captures the \textit{physical dependence} of ${\mathbf{e}}^{\mathsf{v}}\left({\mathbf{p}}_{\mathsf{r}}\right)$ on the current inside $\mathcal{R}_{\mathsf{t}}$ while this dependence is  typically ignored in the SWM\cite{8736783}. 

\noindent {\textbf{Discussion 2} (Typical approximations of ${G_{s}}(\rrt)$)\textbf{.} There are two typical approximations of the scalar spherical wave term ${G_{s}}(\rrt)$ contained in the near-field EPM, which we will discuss below.} The position can be written as $\mathbf{p}_{\mathsf{t}}=r_{\mathsf{to}}\left(\Psi,\Omega,\Phi\right)^{\mathrm{T}}$ with $\Psi \triangleq \sin{\varphi}\cos{\psi}$, $\Omega \triangleq \sin{\varphi}\sin{\psi}$, and $\Phi \triangleq \cos{\varphi}$. Thus, $\rrt$ can be written as
\begin{equation}\setlength\abovedisplayskip{3pt}
\setlength\belowdisplayskip{3pt}\small
    \rrt=\|\mathbf{p}_{\rq}-\mathbf{p}_{\tq} \|=r_{\mathsf{to}}\sqrt{1-\frac{2\left(x_{\rq}\Psi+y_{\rq}\Omega\right)}{r_{\mathsf{to}}}+\frac{x_{\rq}^{2}+y_{\rq}^{2}}{r_{\mathsf{to}}^{2}}}.
\end{equation}When $r_{\mathsf{to}} \sim \mathcal{O}(D_{\rq})$, $\rrt$ in the denominator of ${G_{s}}(\rrt)$ can be replaced by $r_{\mathsf{to}}$. In the Fresnel region, i.e., $r_{\mathsf{to}}>0.5\sqrt{D_{\rq}^{3}/\lambda}$, $\rrt$ in the exponent $\mathrm{e}^{-\imagunit  k_{0}\rrt}$ can be approximated as
\begin{equation}\setlength\abovedisplayskip{3.5pt}
\setlength\belowdisplayskip{3.5pt}\small
    \rrt\approx r_{\mathsf{to}}-\left(x_{\rq}\Psi+y_{\rq}\Omega\right)+\frac{x_{\rq}^{2}+y_{\rq}^{2}-\left(x_{\rq}\Psi+y_{\rq}\Omega\right)^{2}}{2r_{\mathsf{to}}},
\end{equation}
and ${G_{s}}(\rrt)$ thus becomes
\begin{equation}\setlength\abovedisplayskip{3.5pt}
\setlength\belowdisplayskip{3.5pt}\small
    {G_{s}}(\rrt)\approx -\frac{\imagunit k_{0}\eta }{4\pi r_{\mathsf{to}}}\mathrm{e}^{-\imagunit k_{0} \left[ r_{\mathsf{to}}-\left(x_{\rq}\Psi+y_{\rq}\Omega\right)+\frac{x_{\rq}^{2}+y_{\rq}^{2}-\left(x_{\rq}\Psi+y_{\rq}\Omega\right)^{2}}{2r_{\mathsf{to}}}\right]}. \label{eq:fresnel}
\end{equation}
\eqref{eq:fresnel} is called the \textit{Fresnel approximation} \cite{selvan2017fraunhofer}, which ignores the amplitude variations over the receiver aperture while the phase term is series expanded around $r_{\mathsf{to}}\to \infty$. In the case $r_{\mathsf{to}}>2D_{\mathsf{r}}^{2}/\lambda$, the second-order term in the exponent of \eqref{eq:fresnel} is  negligible and we obtain the well-known \textit{uniform plane wave approximation}\cite{lu2021communicating} of the  spherical wave ${G_{s}}(\rrt)$ as follows,
\begin{equation}\setlength\abovedisplayskip{3.5pt}
\setlength\belowdisplayskip{3.5pt}\small
   {G_{s}}(\rrt)\approx -\frac{\imagunit k_{0}\eta \mathrm{e}^{-\imagunit k_{0}r_{\mathsf{to}}} }{4\pi r_{\mathsf{to}}}\mathrm{e}^{\imagunit k_{0} \left(x_{\rq}\Psi+y_{\rq}\Omega\right)}. \label{eq:far}
\end{equation}
{It is worth noting that unlike many works \cite{bjornson2021primer,deshpande2022wideband,cui2022channel} using the \textit{Fresnel approximation}, this paper considers exact ${G_{s}}(\rrt)$.}

\noindent {\textbf{Discussion 3} (Prior adopted signal models)\textbf{.} The vast majority of previously adopted signal models is usually based on simple SWM and the received scalar field can be written as $e^{\mathsf{s}}\left(\mathbf{p}_{\mathsf{r}}\right)=\varepsilon {G_{s}}(\rrt)$ \cite{9335528,delmas2016crb,guerra2021near}, where $\varepsilon$ is a channel power scaling parameter. A more accurate SWM is adopted in \cite{hu2018beyond2,alegria2019cramer} and the received signal is provided as $e^{\mathsf{s}}\left(\mathbf{p}_{\mathsf{r}}\right)=\varepsilon {G_{s}}(\rrt)\sqrt{{z_{\tq}}/{\rrt}}$, where $\sqrt{{z_{\tq}}/{\rrt}}$  represents the angle-of-arrival of the transmitted signal. \textit{Note:} Compared with \eqref{eq:Epr_definition}, \eqref{eq:E1}, and \eqref{eq:12}, the above signal models overlook the functional dependence on physical properties of the source, although it may significantly affect the expression of the received electric field, as revealed in \cite{8736783}.}
\subsubsection{Electric field expressions} We consider that the source is a \textit{Hertzian  dipole} of length $l_{\tq}$ pointing in the direction of $Y$-axis. Hence, the electric current density ${\mathbf{J}}\left({\mathbf{p}}_{\mathsf{t}}\right)$ is written as
\begin{equation}\setlength\abovedisplayskip{3.5pt}
\setlength\belowdisplayskip{3.5pt}\small
{\mathbf{J}}\left({\mathbf{p}}_{\mathsf{t}}\right)=I_{\mathsf{in}}l_{\tq}\delta\left(\mathbf{p}_{\mathsf{t}}\right)\hat{\mathbf{y}}, \label{eq:dipole}
\end{equation}where $I_{\mathsf{in}}$ is the uniform current level in the dipole. Thus, we have ${\mathbf{J}}_{\mathcal{R}_{\mathsf{t}}}=I_{\mathsf{in}}l_{\tq}\hat{\mathbf{y}}=I_{\mathsf{in}}l_{\tq}\cos{\theta}{\hat{\mathbf{r}}_{\mathsf{rt}}}-I_{\mathsf{in}}l_{\tq}\sin{\theta}\hat{\bm{\theta}}$, which means that ${{J}}_{\mathcal{R}_{\mathsf{t}}}^{r}=I_{\mathsf{in}}l_{\tq}\cos{\theta}$, ${{J}}_{\mathcal{R}_{\mathsf{t}}}^{\theta}=-I_{\mathsf{in}}l_{\tq}\sin{\theta}$, and ${{J}}_{\mathcal{R}_{\mathsf{t}}}^{\phi}=0$. Based on \eqref{eq:12}, the  expressions of the three electric fields \textit{VEF}, \textit{SEF}, and \textit{OSEF} in the \textit{near-field} can be obtained as follows.

\noindent \textbf{Proposition 2} (\textit{Vector electric field})\textbf{.} In the coordinate system $OXYZ$, the three components of the \textit{VEF} can be derived as
{\begin{small}\setlength\abovedisplayskip{3.5pt}
\setlength\belowdisplayskip{3.5pt}
\begin{align}
&e_{x}^{\mathsf{v}}\left(\mathbf{p}_{\mathsf{r}}\right)=\imagunit E_{\mathsf{in}}\frac{\left(x_{\mathsf{r}}-x_{\mathsf{t}} \right)\left(y_{\mathsf{r}}-y_{\mathsf{t}} \right)}{{r_{\mathsf{rt}}^{3}}}\mathrm{e}^{-\imagunit k_{0}\rrt},\label{eq:exnear}\\
&e_{y}^{\mathsf{v}}\left(\mathbf{p}_{\mathsf{r}}\right)=-\imagunit E_{\mathsf{in}}\left[\dfrac{1}{\rrt}-\frac{\left(y_{\mathsf{r}}-y_{\mathsf{t}}\right)^{2}}{{r_{\mathsf{rt}}^{3}}}\right]\mathrm{e}^{-\imagunit k_{0}\rrt},\\
&e_{z}^{\mathsf{v}}\left(\mathbf{p}_{\mathsf{r}}\right)=-\imagunit E_{\mathsf{in}}\frac{z_{\mathsf{t}}\left(y_{\mathsf{r}}-y_{\mathsf{t}}\right)}{{r_{\mathsf{rt}}^{3}}}\mathrm{e}^{-\imagunit k_{0}\rrt},\label{eq:eznear}
\end{align}
\end{small}}where $E_{\mathsf{in}}=\frac{\eta I_{\mathsf{in}}l_{\tq}}{2 \lambda}$ is initial electric intensity measured in ${\mathrm{V}_{olt}}$. $\rrt=\sqrt{x_{\rq,\tq}^{2}+y_{\rq,\tq}^{2}+z_{\tq}^{2}}$, $x_{\rq,\tq}\triangleq x_{\rq}-x_{\tq}$, and $y_{\rq,\tq}\triangleq y_{\rq}-y_{\tq}$.

\begin{IEEEproof}
	Please see Appendix \ref{proof:VEF}.
\end{IEEEproof}

\noindent \textbf{Corollary 3} (\textit{Scalar electric field})\textbf{.} In the coordinate system $OXYZ$, the \textit{SEF} can be derived as
\begin{equation}\setlength\abovedisplayskip{3.5pt}
\setlength\belowdisplayskip{3.5pt}\small
e^{\mathsf{s}}\left(\mathbf{p}_{\rq}\right)=E_{\mathsf{in}}\frac{\sqrt{z_{\tq}\left(x_{\rq}-x_{\tq} \right)^{2}+z_{\tq}^{3}}}{{r_{\mathsf{rt}}^{5/2}}}\mathrm{e}^{-\imagunit k_{0}\rrt}.\label{eq:enfsca}
\end{equation}
\begin{IEEEproof}
 Since ${\hat{\mathbf{r}}_{\mathsf{rt}}} =\sin{\theta}\cos{\phi}\hat{\mathbf{x}}+\cos{\theta}\hat{\mathbf{y}}-\sin{\theta}\sin{\phi}\hat{\mathbf{z}}$, \eqref{eq:Epr_definition} can  be rewritten as
\begin{equation} \setlength\abovedisplayskip{3.5pt}
\setlength\belowdisplayskip{3.5pt}\small
\begin{aligned}
e^{\mathsf{s}}\left(\mathbf{p}_{\mathsf{r}}\right)=\sqrt{\|\mathbf{e}^{\mathsf{v}}\left(\mathbf{p}_{\mathsf{r}} \right)\|^{2}\sin{\theta}\sin{\phi}}\mathrm{e}^{-\imagunit k_{0}\rrt}.
    \label{eq:Epr_GJsinsine}
\end{aligned}
\end{equation}
Substituting \eqref{eq:Ex_simply} -- \eqref{eq:Ez_simply} in Appendix \ref{proof:VEF} into \eqref{eq:Epr_GJsinsine}, the \textit{scalar electric field} with respect to $\left(\rrt,\theta,\phi \right)$ can be expressed as 
\begin{equation}\setlength\abovedisplayskip{3.5pt}
\setlength\belowdisplayskip{3.5pt}\small
e^{\mathsf{s}}\left(\mathbf{p}_{\mathsf{r}}\right)=\left|{G_{s}}(\rrt)\right|I_{\mathsf{in}}l_{\tq}\sqrt{\sin^{3}\theta\sin{\phi}}\mathrm{e}^{-\imagunit k_{0}\rrt}.
    \label{eq:Epr_GJsinsine1}
\end{equation}
Substituting $\sin \theta=\frac{\sqrt{x_{\rq,\tq}^{2}+z_{\tq}^{2}}}{\rrt}$ and $\sin \phi=\frac{z_{\tq}}{\sqrt{x_{\rq,\tq}^{2}+z_{\tq}^{2}}}$ into \eqref{eq:Epr_GJsinsine1} yields \eqref{eq:enfsca}.
\end{IEEEproof}

\noindent \textbf{Corollary 4} (\textit{Overall scalar electric field})\textbf{.} In the coordinate system $OXYZ$, the \textit{OSEF} can be computed as
\begin{equation}\setlength\abovedisplayskip{3.5pt}
\setlength\belowdisplayskip{3.5pt}\small
    e^{\mathsf{o}}=E_{\mathsf{in}}\sqrt{\frac{2}{D_{\mathsf{r}}^{2}}}\iint_{\mathcal{R}_{\mathsf{r}}}\frac{\sqrt{z_{\mathsf{t}}\left(x_{\mathsf{r}}-x_{\mathsf{t}} \right)^{2}+z_{\mathsf{t}}^{3}}}{{r_{\mathsf{rt}}^{5/2}}}\mathrm{e}^{-\imagunit k_{0}\rrt}dx_{\mathsf{r}}dy_{\mathsf{r}}.
\end{equation}

\begin{IEEEproof}
Based on \eqref{eq:overall} and \eqref{eq:Epr_GJsinsine1}, Corollary 4 is proven.
\end{IEEEproof}

\subsection{Performance Metric Computation and Analysis} \label{CRB_NCPL}
Utilizing results in Sec. \ref{subsection_SMNP} and \ref{subsection_EFE}, the expressions of the CRBs for estimating the position of $\mathbf{p}_{\mathsf{t}}$ in Fig. \ref{system1} are provided.

\noindent \textbf{Proposition 3} (CRB expressions, $\mathbf{e}^{\mathsf{v}}\left(\mathbf{p}_{\mathsf{r}} \right)$)\textbf{.} Using the observed \textit{vector electric field}, the CRBs can be computed as
{\begin{small}\setlength\abovedisplayskip{3.5pt}
\setlength\belowdisplayskip{3.5pt}
\begin{align}
&\mathrm{CRB}_{1}\left(x_{\mathsf{t}} \right)=\mathrm{SNR}^{-1}\cdot\frac{-\mathscr{I}_{23}^{2}+\mathscr{I}_{22}\mathscr{I}_{33}}{2\mathscr{I}_{\mathrm{sum}}},\label{eq:CRB1x}\\
&\mathrm{CRB}_{1}\left(y_{\mathsf{t}} \right)=\mathrm{SNR}^{-1}\cdot\frac{-\mathscr{I}_{13}^{2}+\mathscr{I}_{11}\mathscr{I}_{33}}{2\mathscr{I}_{\mathrm{sum}}},\\
      &\mathrm{CRB}_{1}\left(z_{\mathsf{t}} \right)=\mathrm{SNR}^{-1}\cdot\frac{-\mathscr{I}_{12}^{2}+\mathscr{I}_{11}\mathscr{I}_{22}}{2\mathscr{I}_{\mathrm{sum}}}\label{eq:CRB1z},
\end{align}\end{small}}where ${\mathrm{SNR}}=\frac{{|E_{\mathsf{in}}|^{2}}}{\sigma^{2}}$ is the signal-to-noise ratio (SNR), $\mathscr{I}_{mn}=\rho_{11}^{mn}+\rho_{12}^{mn}$, $\rho^{mn}_{11}$ and $\rho^{mn}_{12}$ are computed in \eqref{eq:rho_11^11} -- \eqref{eq:rho_12^23}, and
\begin{equation}\setlength\abovedisplayskip{3.5pt}
\setlength\belowdisplayskip{3.5pt}\small
\mathscr{I}_{\mathrm{sum}}=2\mathscr{I}_{12}\mathscr{I}_{13}\mathscr{I}_{23}+\mathscr{I}_{11}\mathscr{I}_{22}\mathscr{I}_{33}-\mathscr{I}_{13}^{2}\mathscr{I}_{22}-\mathscr{I}_{11}\mathscr{I}_{23}^{2}-\mathscr{I}_{12}^{2}\mathscr{I}_{33}.
\end{equation}

\begin{IEEEproof}
According to Proposition 1 and 2, the first-order derivatives ${\partial h_{x}^{\mathsf{v}}\left(\mathbf{p}_{\rq}\right)}/{\partial x_{\tq}}$, $\cdots$, ${\partial h_{z}^{\mathsf{v}}\left(\mathbf{p}_{\rq}\right)}/{\partial z_{\tq}}$ in FIM, where $h_{\kappa}^{\mathsf{v}}\left(\mathbf{p}_{\rq}\right)\triangleq e_{\kappa}^{\mathsf{v}}\left(\mathbf{p}_{\rq}\right)/E_{\mathsf{in}}$, should be first computed. For their specific expressions, please see \eqref{eq:hx_xt} -- \eqref{eq:hz_zt} in Appendix \ref{experssion}. Then by substituting these expressions into \eqref{eq:CRBvector}, we can derive the elements of FIM as $\left[\mathbf{I}\left(\bm{\xi}\right)\right]_{mn}
=2{\mathrm{SNR}}\left(\rho_{11}^{mn}+\rho_{12}^{mn}\right)$, $m\leq n$. Since FIM is a symmetric matrix, we have $\left[\mathbf{I}\left(\bm{\xi}\right)\right]_{mn,m\neq n}=\left[\mathbf{I}\left(\bm{\xi}\right)\right]_{nm,m\neq n}$. By applying the matrix inversion lemma, we obtain the inverse of $\mathbf{I}\left(\bm{\xi}\right)$, denoted as $\mathbf{I}\left(\bm{\xi}\right)^{-1}$, whose diagonal elements are the CRBs for estimating $x_{\tq}$, $y_{\tq}$, and $z_{\tq}$.
\end{IEEEproof}

From the above expressions of the CRBs for \textit{VEF}, the CRBs for \textit{SEF} and \textit{OSEF} are provided in the following corollaries. 

\noindent \textbf{Corollary 5} (CRB expressions, $e^{\mathsf{s}}\left(\mathbf{p}_{\rq}\right)$)\textbf{.} If utilizing the \textit{scalar electric field} observation, the specific expressions of the CRBs can also be computed by \eqref{eq:CRB1x} -- \eqref{eq:CRB1z}, and we represent them as $\mathrm{CRB}_{2}\left(\kappa_{\tq} \right)$. The only difference from Proposition 3 is the computation of $\mathscr{I}_{mn}$, where $\mathscr{I}_{mn}=\rho_{21}^{mn}+\rho_{22}^{mn}$.  $\rho_{21}^{mn}$ and $\rho_{22}^{mn}$ are given in \eqref{eq:rho_21^11} -- \eqref{eq:rho_22^23} in Appendix \ref{experssion}.

\begin{IEEEproof}
Based on Corollary 3, the first-order derivatives ${\partial h^{\mathsf{s}}\left(\mathbf{p}_{\tq}\right)}/{\partial \kappa_{\tq}}$ involved in $\mathbf{I}\left(\bm{\xi}\right)$, where $h^{\mathsf{s}}\left(\mathbf{p}_{\tq}\right)\triangleq e^{\mathsf{s}}\left(\mathbf{p}_{\tq}\right)/E_{\mathsf{in}}$, are computed in \eqref{eq:h_xt} -- \eqref{eq:h_zt} in Appendix \ref{experssion}. According to Corollary 1, $\mathrm{CRB}_{2}\left(\kappa_{\tq} \right)$ can be derived.
\end{IEEEproof}

\noindent \textbf{Corollary 6} (CRB expressions, $e^{\mathsf{o}}$)\textbf{.} If we can only capture the \textit{overall scalar electric field} observation. The CRBs, denoted as $\mathrm{CRB}_{3}\left(\kappa_{\tq} \right)$, can also be computed by \eqref{eq:CRB1x} -- \eqref{eq:CRB1z}, but the expression of $\mathscr{I}_{mn}$ is different. Specifically, $\mathscr{I}_{mn}=\rho_{3}^{mn}$, where
\begin{equation}\setlength\abovedisplayskip{3.5pt}
\setlength\belowdisplayskip{3.5pt}\small
    \rho_{3}^{mn}=\frac{2}{D_{\rq}^{2}}\operatorname{Re}\left\{\frac{\partial {h^{\mathsf{o}}}}{\partial \bm{\xi}_{n}}\frac{\partial {h^{\mathsf{o}}}^{*}}{\partial \bm{\xi}_{m}}\right\},
    \label{eq:rho3mn}
\end{equation}
and $h^{\mathsf{o}}\triangleq \frac{D_{\rq}e^{\mathsf{o}}}{\sqrt{2}E_{\mathsf{in}}}$.

\begin{IEEEproof}
The result is derived from Corollary 2 and 4.
\end{IEEEproof}

Notice that it is hard to compute the value of $\rho_{3}^{mn}$ due to the double integral in the molecule of partial derivative ${\partial h^{\mathsf{o}}}/{\partial \kappa_{\tq}}$ in \eqref{eq:rho3mn}. By using the \textit{Riemann integral
method}, we approximate the integral as a summation, and thus a simpler expression of  $\rho_{3}^{mn}$ is acquired. Specifically, we divide the receiving surface $\mathcal{R}_{\rq}$ into $\alpha$ parts, where $\sqrt{\alpha}$  is assumed to be a positive integer and an odd number for simplicity. We denote the coordinate of each small part as $(x_{i},y_{j})$, in which $\{x_{1},x_{2},\ldots,x_{\sqrt{\alpha}}\}$ is the arithmetic sequence, the common difference is \begin{small}${D_{\rq}}/{\sqrt{2\alpha}}$\end{small}, and the first item is $x_{1}=\frac{D_{\rq}}{2\sqrt{2\alpha}}-\frac{D_{\rq}}{2\sqrt{2}}$. Similarly, the arithmetic sequence $\{y_{1},y_{2},\ldots,y_{\sqrt{\alpha}}\}$ has the same common difference and the first item as $\{x_{i}\}$. Thus, ${h^{\mathsf{o}}}$ is approximated as ${h}^{\mathsf{o}}_{d}$,
\begin{equation}\setlength\abovedisplayskip{3.5pt}
\setlength\belowdisplayskip{3.5pt}\small
 {h}^{\mathsf{o}}_{d}=\frac{D_{\rq}^{2}}{2\alpha}\sum_{i=1}^{\sqrt{\alpha}} \sum_{j=1}^{\sqrt{\alpha}} \frac{\sqrt{z_{\tq}\left(x_{i}-x_{\tq}\right)^{2}+z_{\tq}^{3}}}{{r_{\mathsf{rt};i,j}^{5/2}}}{\mathrm{e}}^{-\imagunit k_{0}{r_{\mathsf{rt};i,j}}},
 \label{eq:hd}
\end{equation}in which ${r_{\mathsf{rt};i,j}}\triangleq\sqrt{\left(x_{i}-x_{\tq}\right)^{2}+\left(y_{j}-y_{\tq}\right)^{2}+z_{\tq}^{2}}$. Therefore, $\rho_{3}^{mn}$ can be computed by replacing $h^{\mathsf{o}}$ in \eqref{eq:rho3mn} with ${h}^{\mathsf{o}}_{d}$. The expressions of $\rho_{3}^{mn}$ are given in \eqref{eq:rho_3^11} -- \eqref{eq:rho_3^23} in Appendix \ref{experssion}.

\section{Performance for a Terminal on the CPL}\label{sectionCPL}
To validate the results derived in Sec. \ref{CRB_NCPL} and gain further insights into the performance, a simplified case of the general system is considered, where the terminal is located on the CPL of the receiving surface. This is also for comparison with \cite{de2021cramer}. In particular, the CPL is the boresight line perpendicular to the receiving surface $\mathcal{R}_{\rq}$ passing through the center point $O$ while the three-dimensional source region $\mathcal{R}_{\tq}$ degenerates into one-dimensional region, as shown in Fig. \ref{system1}.

\subsection{Performance Analysis for CPL Case}\label{subsection:CRB_CPL}
In CPL case, we have $x_{\tq}=y_{\tq}=0$ (but they are unknown), and hence $\rrt=\sqrt{x_{\rq}^{2}+y_{\rq}^{2}+z_{\tq}^{2}}$. Since $\rrt$ is an even function with respect to $x_{\rq}$ and $y_{\rq}$, and the integration domain $\mathcal{R}_{\rq}$ is symmetric, the cross-terms of different dimensions in the FIM $\mathbf{I}\left(\bm{\xi}\right)$ are zero, meaning that the FIM $\mathbf{I}\left(\bm{\xi}\right)$ is a diagonal matrix. Utilizing  the properties of the diagonal matrix inversion, the process of computing CRBs will be  considerably simplified.

We denote a useful parameter $\tau\triangleq{D_{\rq}}/{z_{\tq}}$, which measures the maximum geometric dimension $D_{\rq}$ of the receiving surface normalized by the distance from the terminal position to the receiver. For a terminal in the near-field region, the value of $\tau$ is large, and for a terminal far away from the receiving antenna, $\tau$ becomes small. We define a new integration domain $\mathcal{R}_{\tau}=\left \{(u,v):|u|\leq \tau/\sqrt{8},|v|\leq \tau/\sqrt{8} \right \}$. Based on Proposition 3, Corollary 5, and 6, the following results are obtained.

\noindent \textbf{Corollary 7} (CRB, \textit{VEF}, CPL)\textbf{.} If the terminal is on the CPL, the CRBs for the estimation of $x_{\tq}$, $y_{\tq}$, and $z_{\tq}$ using the \textit{vector electric field}, denoted as $\mathrm{CRB}_{1}^{\mathsf{C}}\left(\kappa_{\tq} \right)$, are computed as
\begin{equation}\small\setlength\abovedisplayskip{3.5pt}
\setlength\belowdisplayskip{3.5pt}
\mathrm{CRB}_{1}^{\mathsf{C}}\left(\kappa_{\tq} \right)=\frac{{\mathrm{SNR}}^{-1}}{2\left(k_{0}^{2}\rho_{11\kappa}+z_{\tq}^{-2}\rho_{12\kappa}\right)},
\end{equation}where{\begin{small}\setlength\abovedisplayskip{3pt}
\setlength\belowdisplayskip{3pt}\begin{align}
&\rho_{11x}\triangleq \iint_{\mathcal{R}_{\tau}} \frac{u^{2}(u^{2}+1)}{(u^{2}+v^{2}+1)^{3}}dudv\label{eq:11x},\\
&\rho_{12x}\triangleq \iint_{\mathcal{R}_{\tau}}\frac{u^{4}+v^{4}+u^{2}+v^{2}-u^{2}v^{2}}{(u^{2}+v^{2}+1)^{4}}dudv\label{eq:12x},\\
&\rho_{11y}\triangleq \iint_{\mathcal{R}_{\tau}} \frac{v^{2}(u^{2}+1)}{(u^{2}+v^{2}+1)^{3}}dudv\label{eq:11y},\\
&\rho_{12y}\triangleq \iint_{\mathcal{R}_{\tau}}\frac{(u^{2}+1)(u^{2}+4v^{2}+1)}{(u^{2}+v^{2}+1)^{4}}dudv\label{eq:12y},\\
&\rho_{11z}\triangleq \iint_{\mathcal{R}_{\tau}} \frac{u^{2}+1}{(u^{2}+v^{2}+1)^{3}}dudv\label{eq:11z},\\
&\rho_{12z}\triangleq \iint_{\mathcal{R}_{\tau}}\frac{v^{4}+u^{2}v^{2}+1}{(u^{2}+v^{2}+1)^{4}}dudv\label{eq:12z}.
\end{align}\end{small}}

\begin{IEEEproof}
Since $\mathbf{I}\left(\bm{\xi}\right)$ is 
 diagonal, \eqref{eq:CRB} can be rewritten as
\begin{equation}\setlength\abovedisplayskip{3.5pt}
\setlength\belowdisplayskip{3.5pt}\small
\mathrm{CRB}\left(\bm{\xi}_{i} \right)=\left[\mathbf{I}(\bm{\xi})\right]_{ii}^{-1}=\mathcal{I}_{ii}^{-1},
\label{eq:CRB_CPL}
\end{equation}
where $\mathcal{I}_{ii}=2\mathrm{SNR}(\rho_{11}^{ii}+\rho_{12}^{ii})$, $\rho_{11}^{ii}$ and $\rho_{12}^{ii}$ can be computed by replacing $x_{\rq,\tq}$ and $y_{\rq,\tq}$ in \eqref{eq:rho_11^11} -- \eqref{eq:rho_12^33} with $x_{\rq}$ and $y_{\rq}$.
\end{IEEEproof}

\noindent {\textbf{Remark 1} (The  generalizability of proposition 3)\textbf{.}} Proposition 3 can be simplified to Corollary 7  by utilizing diagonal matrix inversion and simplification of $\rho_{11}^{11}$ -- $\rho_{12}^{33}$ when the terminal is on the CPL. {\textit{Additionally, the expressions of $\mathrm{CRB}_{1}^{\mathsf{C}}\left(\kappa_{\tq} \right)$ are consistent with the results in \cite[Eqs. (28)--(36)]{de2021cramer}}. The only difference is that we have replaced the integration variables $x_{\tq}$ and $y_{\tq}$ with $u$ and $v$ for a more intuitive analysis of the effect of $\lambda$ and $z_{\tq}$ on the CRBs.} {Consequently, the CRBs (using the \textit{vector electric field}) derived in proposition 3 are more general than \cite{de2021cramer}. In fact, compared with the CPL case \cite{de2021cramer,d2022cramertsp}, Sec. \ref{subsection_SMNP} provides a generic near-field positioning model since the terminal does not have to be located on the CPL.} 

\noindent \textbf{Remark 2} (Closed-form expressions of $\mathrm{CRB}_{1}^{\mathsf{C}}\left(\kappa_{\tq} \right)$)\textbf{.} Different from \cite[Eqs. (39)--(46)]{de2021cramer}, the more precise closed-form expressions for $\rho_{12x}$, $\rho_{12y}$, $\rho_{11z}$, and $\rho_{12z}$ are provided in \eqref{eq:12x_close} -- \eqref{eq:12z_close} in Appendix \ref{ap:remark2}. Since the closed-form expressions of $\rho_{11x}$ and $\rho_{11y}$ are hard to obtain, their closed-form upper and lower bounds are provided in \eqref{eq:11x+} -- \eqref{eq:11y-} in Appendix \ref{ap:remark2}.

\noindent \textbf{Corollary 8} (CRB, $\textit{SEF}$, CPL)\textbf{.} For the CPL case, the CRBs for estimating $x_{\tq}$, $y_{\tq}$, and $z_{\tq}$ utilizing the \textit{scalar electric field}, denoted as $\mathrm{CRB}_{2}^{\mathsf{C}}\left(\kappa_{\tq} \right)$, are given by
\begin{equation}\small\setlength\abovedisplayskip{3.5pt}
\setlength\belowdisplayskip{3.5pt}
\mathrm{CRB}_{2}^{\mathsf{C}}\left(\kappa_{\tq} \right)=\frac{{\rm{SNR}}^{-1}}{2\left(k_{0}^{2}\rho_{21\kappa}+z_{\tq}^{-2}\rho_{22\kappa}\right)},
\end{equation}where{\begin{small}\setlength\abovedisplayskip{3pt}
\setlength\belowdisplayskip{3pt}\begin{align}
&\rho_{21x}\triangleq\iint_{\mathcal{R}_{\tau}} \frac{u^{2}(u^{2}+1)}{(u^{2}+v^{2}+1)^{7/2}}dudv\label{eq:21x},\\
&\rho_{22x}\triangleq\iint_{\mathcal{R}_{\tau}}\frac{u^{2}(3u^{2}-2v^{2}+3)^{2}}{4(u^{2}+1)(u^{2}+v^{2}+1)^{9/2}}dudv,\\
&\rho_{21y}\triangleq\iint_{\mathcal{R}_{\tau}} \frac{v^{2}(u^{2}+1)}{(u^{2}+v^{2}+1)^{7/2}}dudv,\\
&\rho_{22y}\triangleq\iint_{\mathcal{R}_{\tau}}\frac{25v^{2}(u^{2}+1)}{4(u^{2}+v^{2}+1)^{9/2}}dudv,\\
&\rho_{21z}\triangleq\iint_{\mathcal{R}_{\tau}} \frac{u^{2}+1}{(u^{2}+v^{2}+1)^{7/2}}dudv,\\
&\rho_{22z}\triangleq\iint_{\mathcal{R}_{\tau}}\frac{(u^{4}+u^{2}v^{2}+3v^{2}-u^{2}-2)^{2}}{4(u^{2}+1)(u^{2}+v^{2}+1)^{9/2}}dudv\label{eq:22z}.
\end{align}\end{small}}

\begin{IEEEproof}
The diagonal elements of $\mathbf{I}(\bm{\xi})$ in \eqref{eq:CRB_CPL} are written as $\mathcal{I}_{ii}=2\mathrm{SNR}(\rho_{21}^{ii}+\rho_{22}^{ii})$, $\rho_{21}^{ii}$ and $\rho_{22}^{ii}$ can be computed by replacing $x_{\rq,\tq}$ and $y_{\rq,\tq}$ in \eqref{eq:rho_21^11} -- \eqref{eq:rho_22^33} with $x_{\rq}$ and $y_{\rq}$.
\end{IEEEproof}

The closed-form expressions of $\rho_{21\kappa}$ and $\rho_{22\kappa}$ are complicated and lengthy, so we provided their closed-form upper and lower bounds in \eqref{eq:21xu} -- \eqref{eq:22zl} in Appendix \ref{ap:remark2}.

Corollary 7 and Corollary 8 clearly demonstrate the effects of the wavelength $\lambda = 2\pi/k_{0}$ and the propagation distance $r_{\mathsf{to}}=z_{\tq}$ on the CRB for fixed values of $\tau$ and $\mathrm{SNR}$ in the near-field positioning
system (using the \textit{vector} or \textit{scalar electric field}). In particular, the CRBs for all dimensions decrease as $\lambda$ or $z_{\tq}$ decreases. In other words, the estimation accuracy of the positioning
system increases as the carrier frequency becomes higher or as the propagation distance becomes smaller.

\noindent \textbf{Corollary 9} (CRB, $\textit{OSEF}$, CPL)\textbf{.} When we employ the \textit{overall scalar electric field}, the CRBs for the CPL case, denoted as $\mathrm{CRB}_{3}^{\mathsf{C}}\left(\kappa_{\tq} \right)$, can be computed as follows.
\begin{equation}\small
\setlength\abovedisplayskip{3.5pt}
\setlength\belowdisplayskip{3.5pt}
\mathrm{CRB}_{3}^{\mathsf{C}}\left(\kappa_{\tq} \right)=\frac{{\mathrm{SNR}}^{-1}}{\frac{4}{D_{\rq}^{2}}\left|\rho_{3\kappa} \right|^{2}},
\end{equation}
where $\rho_{3\kappa}\triangleq\frac{\partial {h}^{\mathsf{o}}}{\partial \kappa_{\tq}}$. By utilizing ${h}^{\mathsf{o}}_{d}$ to discretize ${h}^{\mathsf{o}}$, we have
\begin{equation} \small \setlength\abovedisplayskip{3.5pt}
\setlength\belowdisplayskip{3.5pt}
\mathrm{CRB}_{3}^{\mathsf{C}}\left(\kappa_{\tq} \right)\approx\frac{\frac{\alpha^{2}}{D_{\rq}^{2}}{\rm{SNR}}^{-1}}{\left| \sum_{i=1}^{\sqrt{\alpha}} \sum_{j=1}^{\sqrt{\alpha}}\sqrt{z_{\tq}\left(x_{i}^{2}+z_{\tq}^{2}\right)}\rho_{3\kappa}^{i,j}\mathrm{e}^{-\imagunit k_{0} {r_{\mathsf{rt};i,j}}}\right|^{2}},\label{eq:CRB3c}
\end{equation}
where 
{\begin{small}
\setlength\abovedisplayskip{3pt}
\setlength\belowdisplayskip{3pt}
\begin{align}
    &\rho_{3x}^{i,j}\triangleq {\imagunit k_{0}}x_{i}{r_{\mathsf{rt};i,j}^{-\frac{7}{2}}}+\frac{5}{2}x_{i}{r_{\mathsf{rt};i,j}^{-\frac{9}{2}}}-\frac{x_{i}}{z_{\tq}^{2}+x_{i}^{2}} {r_{\mathsf{rt};i,j}^{-\frac{5}{2}}},\label{eq:rho3xij} \\
    &\rho_{3y}^{i,j}\triangleq {\imagunit k_{0}}y_{j} {r_{\mathsf{rt};i,j}^{-\frac{7}{2}}}+\frac{5}{2}y_{j} {r_{\mathsf{rt};i,j}^{-\frac{9}{2}}},\\
    &\rho_{3z}^{i,j}\triangleq-\imagunit k_{0} z_{\tq} {r_{\mathsf{rt};i,j}^{-\frac{7}{2}}}-\frac{5}{2}z_{\tq} {r_{\mathsf{rt};i,j}^{-\frac{9}{2}}}+\frac{3 z_{\tq}^{2}+x_{i}^{2}}{2 z_{\tq}\left(x_{i}^{2}+z_{\tq}^{2}\right)}{r_{\mathsf{rt};i,j}^{-\frac{5}{2}}}\label{eq:rho3zij}.
\end{align}\end{small}}

\begin{IEEEproof}
It follows from  Corollary 6 and \eqref{eq:hd} by using the property of the inverse of a diagonal matrix $\mathbf{I}(\bm{\xi})$.
\end{IEEEproof}

\noindent \textbf{Remark 3} ($\mathrm{CRB}_{2}^{\mathsf{C}}\left(\kappa_{\tq} \right)<\mathrm{CRB}_{3}^{\mathsf{C}}\left(\kappa_{\tq} \right)$)\textbf{.} We can either compute \eqref{eq:CRB3c} numerically or apply the Cauchy-Schwarz inequality:
\begin{equation*}\small 
\setlength\abovedisplayskip{3pt}
\setlength\belowdisplayskip{3pt}
\begin{aligned}
&\mathrm{CRB}_{3}^{\mathsf{C}}\left(\kappa_{\tq} \right)> \frac{\frac{\alpha^{2}}{D_{\rq}^{2}}{\mathrm{SNR}}^{-1}}{ \sum_{i=1}^{\sqrt{\alpha}} \sum_{j=1}^{\sqrt{\alpha}}\alpha\left|\sqrt{z_{\tq}\left(x_{i}^{2}+z_{\tq}^{2}\right)}\rho_{3\kappa}^{i,j}\mathrm{e}^{-\imagunit k_{0} {r_{\mathsf{rt};i,j}}}\right|^{2}}\\
    &=\frac{{\mathrm{SNR}}^{-1}}{\frac{D_{\rq}^{2}}{\alpha} \sum_{i=1}^{\sqrt{\alpha}} \sum_{j=1}^{\sqrt{\alpha}}\left(k_{0}^{2}\rho_{21\kappa}^{i,j}+z_{\tq}^{-2}\rho_{22\kappa}^{i,j}\right)}=\mathrm{CRB}_{2}^{\mathsf{C}}\left(\kappa_{\tq} \right),
\end{aligned}
\end{equation*}
where $\rho_{21\kappa}^{i,j}$ and $\rho_{22\kappa}^{i,j}$ are defined as the discretized sampling of the integrand functions in \eqref{eq:21x} -- \eqref{eq:22z}. It shows that, under the same condition, the CRBs using \textit{SEF} are the lower bounds of the CRBs using \textit{OSEF}. We get a conclusion that is in line with intuition: using \textit{OSEF} can significantly reduce the complexity of the system, but at the cost of reducing estimation accuracy.

\subsection{Two  Further Simplified Scenarios}\label{CRB further simplified}
\subsubsection{Performance analysis for \texorpdfstring{$z_{\tq}\gg \lambda$}{zt}} \label{CRB analysis for} Consider the scenario in which the distance from the terminal located on the CPL to the receiver is much larger than the wavelength, i.e., $z_{\tq}\gg \lambda$\footnote{Since $z_{\tq}\ll 2D_{\rq}^{2}/\lambda$ is always satisfied when $z_{\tq}\gg \lambda$ and $D_{\rq}$ is not very small, we know that $z_{\tq}\gg \lambda$ corresponds to the near-field region when the size of the receiving antenna is on the order of meters.}. It generally holds in the wireless communication systems with carrier frequencies in the range of GHz  or above. Expressions of the CRBs in Corollary 7 and 8 can be  simplified as follows.

\noindent \textbf{Corollary 10} (CRB, CPL, $z_{\tq}\gg \lambda$)\textbf{.} If $z_{\tq}\gg \lambda$, the CRBs for the CPL case can be further simplified as

a) {Using the \textit{vector electric field}, $\mathrm{CRB}_{1}^{\mathsf{C}}\left(\kappa_{\tq} \right)$ reduces to
\begin{equation}\small \setlength\abovedisplayskip{3pt}
\setlength\belowdisplayskip{3pt}
\mathrm{CRB}_{1}^{\mathsf{C}}\left(\kappa_{\tq} \right)\approx \frac{{\mathrm{SNR}}^{-1}}{2k_{0}^{2}\rho_{11\kappa}}.
\end{equation}}

b) {Using the \textit{scalar electric field}, $\mathrm{CRB}_{2}^{\mathsf{C}}\left(\kappa_{\tq} \right)$ reduces to
\begin{equation}\small
\setlength\abovedisplayskip{3pt}
\setlength\belowdisplayskip{3pt}
\mathrm{CRB}_{2}^{\mathsf{C}}\left(\kappa_{\tq} \right)\approx \frac{{\mathrm{SNR}}^{-1}}{2k_{0}^{2}\rho_{21\kappa}}.
\end{equation}}

\begin{IEEEproof}
Please refer to Appendix \ref{ap:prop4}. 
\end{IEEEproof}

Corollary 10 clearly shows that the estimation accuracy for all dimensions is
completely determined by the values of $\lambda$ and
$\tau$ when $z_{\tq}\gg\lambda$. In particular, when we keep $\tau$ and $\mathrm{SNR}$ fixed, $\mathrm{CRB}_{1}^{\mathsf{C}}\left(\kappa_{\tq} \right)$ and $\mathrm{CRB}_{2}^{\mathsf{C}}\left(\kappa_{\tq} \right)$ will be proportional to the square of $\lambda$. Additionally, for a fixed value of
$\tau$, if $z_{\tq}$ increases
by a factor $\epsilon$, the surface diagonal length $D_{\rq}$ needs to be  scaled by the same
factor $\epsilon$ (the surface area of the receiving antenna increases by the factor $\epsilon^{2}$) to keep the CRBs unchanged.

\noindent \textbf{Remark 4} (Comparison of estimation accuracy)\textbf{.} From Corollary 7 and Corollary 8, we find that $\rho_{11\kappa}>\rho_{12\kappa}$. Accordingly, based on Corollary 10 and Remark 3, we can derive that
\begin{equation}\small \setlength\abovedisplayskip{3.5pt}
\setlength\belowdisplayskip{3.5pt}
\mathrm{CRB}_{1}^{\mathsf{C}}\left(\kappa_{\tq} \right)<\mathrm{CRB}_{2}^{\mathsf{C}}\left(\kappa_{\tq} \right)<\mathrm{CRB}_{3}^{\mathsf{C}}\left(\kappa_{\tq} \right).\label{eq:CRB_vec_sca_overall}
\end{equation}
Inequality \eqref{eq:CRB_vec_sca_overall} shows that using the \textit{vector electric field} at each point on the contiguous receiving surface  renders lower CRBs, i.e., higher estimation accuracy. Using the \textit{scalar electric field} will reduce the complexity of the observed electric fields, but the CRBs will increase accordingly.
If the conventional surface antenna is employed as the receiver, the system can only obtain the \textit{overall scalar electric field}, which will further reduce the complexity of the system but the accuracy decreases too.

\subsubsection{Asymptotic performance  analysis for \texorpdfstring{$\tau \to \infty$}{zt}}\label{Asymptotic CRB analysis for}
Based on the above analysis, it is interesting to analyze the behavior of the asymptotic CRBs if the surface diagonal length  $D_{\rq}$ is much larger than the distance $z_{\tq}$ from the terminal to the receiver. Corollary 11 gives the CRBs in the asymptotic regime  $\tau \to \infty$.

\noindent \textbf{Corollary 11} (CRB, CPL, $\tau \to \infty$)\textbf{.} For the CPL case and $z_{\tq}\gg \lambda$, in the asymptotic regime $\tau \to \infty$, the CRBs for the estimation of $x_{\tq}$, $y_{\tq}$, and $z_{\tq}$ are given by

a) Using the \textit{vector electric field}, we have
{\begin{small}\setlength\abovedisplayskip{3pt}
\setlength\belowdisplayskip{3pt}
\begin{align}
&\lim_{\tau\to \infty}\mathrm{CRB}_{1}^{\mathsf{C}}\left(x_{\tq} \right)=\frac{{\mathrm{SNR}}^{-1}}{6\pi^{3}}\frac{\lambda^{2}}{\ln{\tau}}\label{eq:CRB1xttau},\\
&\lim_{\tau\to \infty}\mathrm{CRB}_{1}^{\mathsf{C}}\left(y_{\tq} \right)=\frac{{\mathrm{SNR}}^{-1}}{2\pi^{3}}\frac{\lambda^{2}}{\ln{\tau}}\label{eq:CRB1yttau},\\
&\lim_{\tau\to \infty}\mathrm{CRB}_{1}^{\mathsf{C}}\left(z_{\tq} \right)=\frac{{\mathrm{SNR}}^{-1}}{6\pi^{3}}{\lambda^{2}}.\label{eq:CRB1zttau}
\end{align}\end{small}}

b) Using the \textit{scalar electric field}, we have
{\begin{small}\setlength\abovedisplayskip{3pt}
\setlength\belowdisplayskip{3pt}
\begin{align}
&\lim_{\tau\to \infty}\mathrm{CRB}_{2}^{\mathsf{C}}\left(x_{\tq} \right)=\frac{15}{64}\frac{{\mathrm{SNR}}^{-1}}{\pi^{3}}{\lambda^{2}}\label{eq:CRB2xttau},\\
&\lim_{\tau\to \infty}\mathrm{CRB}_{2}^{\mathsf{C}}\left(y_{\tq} \right)=\frac{15}{32}\frac{{\mathrm{SNR}}^{-1}}{\pi^{3}}\lambda^{2}\label{eq:CRB2yttau},\\
&\lim_{\tau\to \infty}\mathrm{CRB}_{2}^{\mathsf{C}}\left(z_{\tq} \right)=\lim_{\tau\to \infty}\mathrm{CRB}_{2}^{\mathsf{C}}\left(x_{\tq} \right)\label{eq:CRB2zttau}.
\end{align}\end{small}}

\begin{IEEEproof}
 We have provided the closed-form expressions or upper and lower bounds in Appendix \ref{ap:remark2}, making it possible to compute and analyze the asymptotic CRBs. By computing the limit values of \eqref{eq:11x+} and \eqref{eq:11x-}, we derive that $\rho_{11x}\sim\frac{3\pi}{4}\ln{\tau}$ for $\tau\to\infty$. Then, according to \eqref{eq:11y+} and \eqref{eq:11y-}, we derive  that $\rho_{11y}\sim\frac{\pi}{4}\ln{\tau}$ for $\tau\to\infty$. Similarly, based on \eqref{eq:11z_close} and \eqref{eq:21xu} -- \eqref{eq:21zl}, we have $\lim \rho_{11z}=\frac{3\pi}{4}$, $\lim\rho_{21x}=\frac{8\pi}{15}$, $\lim\rho_{21y}=\frac{4\pi}{15}$ and $\lim\rho_{21z}=\frac{8\pi}{15}$, where we use $\lim$ to represent $\lim_{\tau\to \infty}$. Thus, Corollary 11 holds.
\end{IEEEproof}
\begin{figure}[!t]
\centering
\includegraphics[scale=0.59]{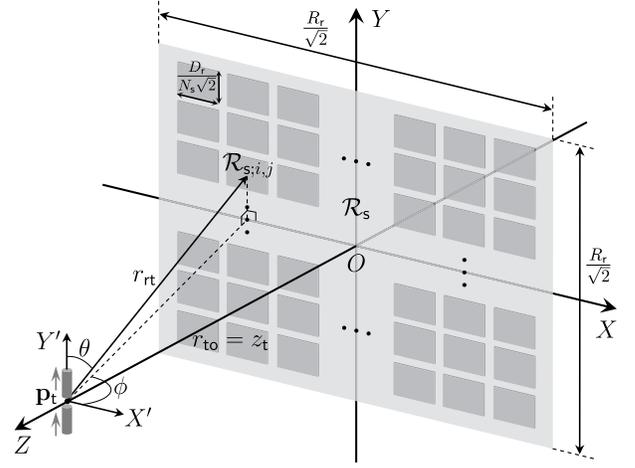}
\caption{Illustration of the SIMO near-field positioning system.}
\label{fig3}
\end{figure}
From Corollary 11, the following observations can be made. Firstly, if we use the observed \textit{vector electric field}, the CRBs for estimating $x_{\tq}$ and $y_{\tq}$ will decrease in the form of $\ln^{-1} \tau$ and go to zero as $\tau$ increases infinitely. But $\mathrm{CRB}_{1}^{\mathsf{C}}\left(z_{\tq} \right)$ tends to a fixed value which depends uniquely on the $\lambda$ and $\mathrm{SNR}$, and does not change with $\tau$. In the CPL case, $z_{\tq}$ represents the propagation distance, so equation \eqref{eq:CRB1zttau} provides a fundamental lower limit to the near-field ranging precision. Secondly, when we utilize the \textit{scalar electric field}, the CRBs for the estimation of $x_{\tq}$ and $z_{\tq}$
are identical and these three CRBs are completely  determined by $\lambda$ and $\mathrm{SNR}$ as $\tau$ increases. Finally, in order to get more insights on the difference of fundamental limit of the estimation accuracy between \textit{VEF} and \textit{SEF} as $\tau$ increases, we represent their difference as $\Delta C_{\kappa}=p_{\kappa} \mathrm{SNR}^{-1}\lambda^{2}$ with $p_{x}=15/(64\pi^{3})\approx7.56\times10^{-3}$, $p_{y}=15/(32\pi^{3})\approx1.512\times10^{-2}$, and $p_{z}=13/(192\pi^{3})\approx6.77\times10^{-5}$. This indicates that using \textit{SEF} has a smaller performance penalty for the estimation of $z_{\tq}$ than $x_{\tq}$ and $y_{\tq}$ compared to utilizing \textit{VEF}.
\section{Performance  of the SIMO Positioning System}
\label{sec:simo}
The receiving antenna adopted in the previous sections is a single antenna or intelligent surface\footnote{The single intelligent surface denotes a \textit{centralized-deployment} LIS \cite{hu2018beyond2} or holographic MIMO surface \cite{huang2020holographic}, which can observe $\textit{VEF}$ and $\textit{SEF}$. Besides, the single antenna represents a conventional surface antenna\cite{bjornson2021primer} and it can only obtain $\textit{OSEF}$. For simplicity, we define both of them as  “single-output”.}, in which the positioning system can be defined as the single-input single-output (SISO) system. In this section, a new system with multiple distributed receiving antennas will be investigated extensively, referred to as the single-input multiple-output (SIMO) system depicted in Fig. \ref{fig3}. This SIMO system is specifically interpreted as follows.
\begin{itemize}
	\item \textbf{Space constraints:} Each of the small receiving antenna is an  intelligent surface or  a conventional surface antenna as previously described and they are distributed on a large rectangular surface $\mathcal{R}_{\mathsf{s}}$ with size $\frac{R_{\rq}}{\sqrt{2}}\times\frac{R_{\rq}}{\sqrt{2}}$, in which $R_{\rq}$ is usually a fixed value (a few meters to tens of meters) due to space constraints\footnote{The receiving antenna, such as LIS, can be easily embedded in daily life objects with limited size such as buildings, walls, cars, etc.} of the  actual positioning system.
	\item \textbf{Total surface area:} The total surface area is assumed to be the same for different numbers of the small receiving antennas and each of them has the same surface area and property. In particular, we consider that the total surface area is $\frac{D_{\rq}^{2}}{2}$ and the number of the antennas is $N_{\mathsf{s}}^{2}$. Thus, the size of each receiving antenna is $\frac{D_{\rq}}{N_{\mathsf{s}}\sqrt{2}}\times\frac{D_{\rq}}{N_{\mathsf{s}}\sqrt{2}}$. 
\item \textbf{Terminal position:} For simplicity, the terminal is located on the CPL with coordinates $(0, 0, z_{\tq})$, which makes the FIM matrix diagonalize as will be shown in Lemma 1.
\end{itemize}

Note
that if $N_{\mathsf{s}}=1$, the SIMO
system degenerates into the SISO system, where the CRBs for all three dimensions using the three electric fields have been computed and analyzed in Sec. \ref{subsection_EFE} and Sec. \ref{sectionCPL}. In this section, we assume $N_{\mathsf{s}}\geq2$. To derive the CRBs of the SIMO system, Lemma 1 is given.

\noindent \textbf{Lemma 1} (Properties of the Fisher's information)\textbf{.} The FIM of the SIMO system becomes a diagonal matrix, and the Fisher's information is identical for every four small receiving antennas rotationally symmetric about the origin (rotation angle is $90^{\circ}$).

\begin{IEEEproof}
Since $\rho_{11}^{12}$ -- $\rho_{12}^{23}$ in \eqref{eq:rho_11^12} -- \eqref{eq:rho_12^23} and $\rho_{21}^{12}$ -- $\rho_{22}^{23}$ in \eqref{eq:rho_21^12} -- \eqref{eq:rho_22^23} (items in FIM  off-diagonal elements) contain at least an odd power term of either $x_{\rq}$ or $y_{\rq}$, and $\rrt$ is an even function with respect to $x_{\rq}$ and $y_{\rq}$, we can demonstrate that  even though the integral domains of $\rho_{11}^{12}$ -- $\rho_{12}^{23}$ and $\rho_{21}^{12}$ -- $\rho_{22}^{23}$ are no longer symmetric about the origin, due to the additivity of the Fisher's information, there can be a symmetric integral of each integral whose sum equals zero. Consequently, the off-diagonal elements of the FIM matrix are canceled. Similarly, $\rho_{11}^{11}$ -- $\rho_{12}^{33}$ in \eqref{eq:rho_11^11} -- \eqref{eq:rho_12^33} and $\rho_{21}^{11}$ -- $\rho_{22}^{33}$ in \eqref{eq:rho_21^11} -- \eqref{eq:rho_22^33} (items in FIM diagonal elements) contain even power terms of $x_{\rq}$ and/or $y_{\rq}$, 
so the diagonal elements are non-zero, and the values of $\rho_{11}^{11}$ -- $\rho_{12}^{33}$ and $\rho_{21}^{11}$ -- $\rho_{22}^{33}$ remain unchanged if $x_{\rq}$ becomes $-x_{\rq}$ and/or $y_{\rq}$ becomes $-y_{\rq}$. Therefore, Lemma 1 holds.
\end{IEEEproof}

 Based on Lemma 1, we divide the large
rectangular surface into four equal parts using the $X$- and $Y$- axes as their boundaries. Then, we only need to study one of the four parts, which contains $\frac{N_{\mathsf{s}}^{2}}{4}$ small receiving antennas with index $({i},{j}),i,j=1,\cdots,\frac{N_{\mathsf{s}}}{2}$. The integral domain of the small receiving antenna with index $({i},{j})$ is denoted as ${\mathcal{R}_{\mathsf{s};i,j}}=\left(x_{\rq},y_{\rq},0\right)$, where $x_{\rq}\in \big[\frac{\left(2i-1\right)R_{\rq}-D_{\rq}}{2\sqrt{2}N_{\mathsf{s}}},\frac{\left(2i-1\right)R_{\rq}+D_{\rq}}{2\sqrt{2}N_{\mathsf{s}}}\big]$, $y_{\rq}\in\big[\frac{\left(2j-1\right)R_{\rq}-D_{\rq}}{2\sqrt{2}N_{\mathsf{s}}},\frac{\left(2j-1\right)R_{\rq}+D_{\rq}}{2\sqrt{2}N_{\mathsf{s}}}\big]$. Additionally, ${\mathcal{R}_{\mathsf{s};i,j}}$ can be written as ${\mathcal{R}_{\mathsf{s};i,j}^{\tau}}=\left(u,v,0\right)$, where $u\in \frac{1}{2\sqrt{2}N_{\mathsf{s}}}\big[\frac{\left(2i-1\right)R_{\rq}}{z_{\tq}}-\tau,\frac{\left(2i-1\right)R_{\rq}}{z_{\tq}}+\tau\big]$, $v\in \frac{1}{2\sqrt{2}N_{\mathsf{s}}}\big[\frac{\left(2j-1\right)R_{\rq}}{z_{\tq}}-\tau,\frac{\left(2j-1\right)R_{\rq}}{z_{\tq}}+\tau\big]$. As will be seen later, unlike the SISO system, the integral operation of the Fisher’s information is carried out in each small integral domain ${\mathcal{R}_{\mathsf{s};i,j}^{\tau}}$ and accumulated at the end. The CRBs of the SIMO positioning system utilizing \textit{VEF}, \textit{SEF}, and \textit{OSEF} are computed in the following proposition.

\noindent \textbf{Proposition 4} (CRB, SIMO)\textbf{.} For the defined SIMO positioning system depicted in Fig. \ref{fig3}, we have that:

a) {Using the \textit{vector electric field}, the CRBs can be given by
\begin{equation}\small\setlength\abovedisplayskip{3pt}
\setlength\belowdisplayskip{3pt}
\mathrm{CRB}_{1}^{\mathsf{M}}\left(\kappa_{\tq}\right)=\frac{{\mathrm{SNR}}^{-1}}{8\sum_{j=1}^{\frac{N_{\mathsf{s}}}{2}}\sum_{i=1}^{\frac{N_{\mathsf{s}}}{2}}\left(k_{0}^{2}{\rho_{11\kappa}^{\mathsf{s};i,j}}+z_{\tq}^{-2}{\rho_{12\kappa}^{\mathsf{s};i,j}}\right)},
\label{eq:CRB_SIMO_1}
\end{equation}
where ${\rho_{11\kappa}^{\mathsf{s};i,j}}$, ${\rho_{12\kappa}^{\mathsf{s};i,j}}$ have the same integrands as $\rho_{11\kappa}$, $\rho_{12\kappa}$ in \eqref{eq:11x} -- \eqref{eq:12z}, but their integral domain is ${\mathcal{R}_{\mathsf{s};i,j}^{\tau}}$. }

b) {Using the \textit{scalar electric field}, the CRBs are derived by
\begin{equation}\small\setlength\abovedisplayskip{3pt}
\setlength\belowdisplayskip{3pt}
\mathrm{CRB}_{2}^{\mathsf{M}}\left(\kappa_{\tq}\right)=\frac{{\mathrm{SNR}}^{-1}}{8\sum_{j=1}^{\frac{N_{\mathsf{s}}}{2}}\sum_{i=1}^{\frac{N_{\mathsf{s}}}{2}}\left(k_{0}^{2}{\rho_{21\kappa}^{\mathsf{s};i,j}}+z_{\tq}^{-2}{\rho_{22\kappa}^{\mathsf{s};i,j}}\right)},
\label{eq:CRB_SIMO_2}
\end{equation}
where ${\rho_{21\kappa}^{\mathsf{s};i,j}}$, ${\rho_{22\kappa}^{\mathsf{s};i,j}}$ have the same integrands as $\rho_{21\kappa}$, $\rho_{22\kappa}$ in \eqref{eq:21x} -- \eqref{eq:22z}, but their integral domain is ${\mathcal{R}_{\mathsf{s};i,j}^{\tau}}$.}

c) {Using the \textit{overall scalar electric field}, the CRBs are
\begin{equation}\small\setlength\abovedisplayskip{3pt}
\setlength\belowdisplayskip{3pt}
\mathrm{CRB}_{3}^{\mathsf{M}}\left(\kappa_{\tq}\right)=\frac{{\mathrm{SNR}}^{-1}}{\frac{16}{D_{\rq}^{2}}\sum_{j=1}^{\frac{N_{\mathsf{s}}}{2}}\sum_{i=1}^{\frac{N_{\mathsf{s}}}{2}}\big|{\rho_{3\kappa}^{\mathsf{s};i,j}} \big|^{2}},\label{eq:CRB3M}
\end{equation}
where ${\rho_{3\kappa}^{\mathsf{s};i,j}}=\frac{\partial {{h}^{\mathsf{o}}_{\mathsf{s};i,j}}}{\partial \kappa_{\tq}}$, and ${{h}^{\mathsf{o}}_{\mathsf{s};i,j}}$ contains the same integrand as $h^{\mathsf{o}}$ in Corollary 6 while its integral domain is ${\mathcal{R}_{\mathsf{s};i,j}}$. On the basis of  \eqref{eq:CRB3c}, we provide the more feasible discretized form of \eqref{eq:CRB3M}. Similarly, we divide each small receiving surface region ${\mathcal{R}_{\mathsf{s};i,j}}$ into $\alpha$ parts, then denote that ${x_{\mathsf{s};m;i}}=\frac{\left(2i-1\right)R_{\rq}-D_{\rq}}{2\sqrt{2}N_{\mathsf{s}}}+\frac{(2m-1)D_{\rq}}{2\sqrt{2\alpha}N_{\mathsf{s}}}$, ${y_{\mathsf{s};n;j}}=\frac{\left(2j-1\right)R_{\rq}-D_{\rq}}{2\sqrt{2}N_{\mathsf{s}}}+\frac{(2n-1)D_{\rq}}{2\sqrt{2\alpha}N_{\mathsf{s}}}$,  and ${r_{\mathsf{rt};mn;ij}}=\sqrt{{{x_{\mathsf{s};m;i}^{2}}+{y_{\mathsf{s};n;j}^{2}}+z_{\tq}^{2}}}$, thus $\mathrm{CRB}_{3}^{\mathsf{M}}\left(\kappa_{\tq}\right)$ is further written as 
\begin{equation}\small\setlength\abovedisplayskip{2.5pt}
\setlength\belowdisplayskip{2.5pt}
\mathrm{CRB}_{3}^{\mathsf{M}}\left(\kappa_{\tq}\right)\approx\frac{\frac{\alpha^{2}}{4D_{\rq}^{2}}{\mathrm{SNR}}^{-1}}{\sum_{j=1}^{\frac{N_{\mathsf{s}}}{2}}\sum_{i=1}^{\frac{N_{\mathsf{s}}}{2}}\left| \sum_{m=1}^{\sqrt{\alpha}} \sum_{n=1}^{\sqrt{\alpha}}{\varrho_{3\kappa}^{\mathsf{s};mn;ij}}\right|^{2}},
\end{equation}
in which ${\varrho_{3\kappa}^{\mathsf{s};mn;ij}}=\sqrt{z_{\tq}({x_{\mathsf{s};m;i}^{2}}+z_{\tq}^{2})}{\rho_{3\kappa}^{\mathsf{s};mn;ij}}\mathrm{e}^{-\imagunit k_{0} {r_{\mathsf{rt};mn;ij}}}$ and ${\rho_{3\kappa}^{\mathsf{s};mn;ij}}$ is given in \eqref{eq:rho3xij} -- \eqref{eq:rho3zij}, but $x_{i}$, $y_{j}$,  and ${r_{\mathsf{rt};i,j}}$ need to be modified to ${x_{\mathsf{s};m;i}}$, ${y_{\mathsf{s};n;j}}$, and ${r_{\mathsf{rt};mn;ij}}$, respectively.}
\begin{IEEEproof}
Corollary 7, 8, and 9
have computed the CRBs for all three
dimensions utilizing the three observed electric field types in the SISO system and the crux of the computation is to derive the values of double integrals $\rho_{11\kappa}$, $\rho_{12\kappa}$, $\rho_{21\kappa}$, $\rho_{22\kappa}$, and $\rho_{3\kappa}$, whose integral domains are $\mathcal{R}_{\tau}$ or $\mathcal{R}_{\rq}$. In the SIMO system, the domain of each small receiving antenna is distinct and spatially discontinuous, thus  we modify  the domains from $\mathcal{R}_{\tau}$/$\mathcal{R}_{\rq}$ to ${\mathcal{R}_{\mathsf{s};i,j}^{\tau}}$/${\mathcal{R}_{\mathsf{s};i,j}}$. Besides, the electric fields observed in each small receiving antenna are independent, so the Fisher's
information is additive. Hence, Proposition 4 holds.
\end{IEEEproof}
From \eqref{eq:CRB_SIMO_1} and \eqref{eq:CRB_SIMO_2}, we see that $\mathrm{CRB}_{1}^{\mathsf{M}}\left(\kappa_{\tq}\right)$ and $\mathrm{CRB}_{2}^{\mathsf{M}}\left(\kappa_{\tq}\right)$ decrease as $\lambda$ or $z_{\tq}$ decreases for fixed values of $N_{\mathsf{s}}$ and $\tau$ or, equivalently,
of the functions ${\rho_{ab\kappa}^{\mathsf{s};i,j},a,b=1,2}$. The impact of the number $N_{\mathsf{s}}^{2}$ of small receiving antennas on the CRBs will be investigated in Sec. \ref{subsection:CRB_simo_numer}. Similar to Remark 3, $\mathrm{CRB}_{2}^{\mathsf{M}}\left(\kappa_{\tq}\right)$ can be  verified as the lower bounds of the $\mathrm{CRB}_{3}^{\mathsf{M}}\left(\kappa_{\tq}\right)$ by using the Cauchy-Schwarz inequality. 

Next, we analyze
the behavior of the CRBs in the SIMO system if $z_{\tq}\gg\lambda$ and $\tau\to\infty$. The main results are as follows.

\noindent \textbf{Corollary 12} (SIMO, $z_{\tq}\gg\lambda$)\textbf{.} If $z_{\tq}\gg \lambda$, the CRBs of the
SIMO positioning system can be simplified as
{\begin{small}\setlength\abovedisplayskip{3pt}
\setlength\belowdisplayskip{3pt}
\begin{align}
&\mathrm{CRB}_{1}^{\mathsf{M}}\left(\kappa_{\tq}\right)\approx\frac{{\mathrm{SNR}}^{-1}}{8\sum_{j=1}^{\frac{N_{\mathsf{s}}}{2}}\sum_{i=1}^{\frac{N_{\mathsf{s}}}{2}}k_{0}^{2}{\rho_{11\kappa}^{\mathsf{s};i,j}}}\label{CRB1Mzt},\\
&\mathrm{CRB}_{2}^{\mathsf{M}}\left(\kappa_{\tq}\right)\approx\frac{{\mathrm{SNR}}^{-1}}{8\sum_{j=1}^{\frac{N_{\mathsf{s}}}{2}}\sum_{i=1}^{\frac{N_{\mathsf{s}}}{2}}k_{0}^{2}{\rho_{21\kappa}^{\mathsf{s};i,j}}}\label{CRB2Mzt}.
\end{align}\end{small}}

\begin{IEEEproof}
We observe that:  1) ${\rho_{11\kappa}^{\mathsf{s};i,j}}>{\rho_{12\kappa}^{\mathsf{s};i,j}}$ or ${\rho_{11\kappa}^{\mathsf{s};i,j}}$ has the same order of magnitude as ${\rho_{12\kappa}^{\mathsf{s};i,j}}$; 2) both of them are positive. Then, $k_{0}^{2}{\rho_{11\kappa}^{\mathsf{s};i,j}}\gg z_{\tq}^{-2}{\rho_{12\kappa}^{\mathsf{s};i,j}}$ for $z_{\tq}\gg \lambda$. Similarly, we can prove that $k_{0}^{2}{\rho_{21\kappa}^{\mathsf{s};i,j}}\gg z_{\tq}^{-2}{\rho_{22\kappa}^{\mathsf{s};i,j}}$ for $z_{\tq}\gg \lambda$. Hence, expression \eqref{eq:CRB_SIMO_1} and \eqref{eq:CRB_SIMO_2} can be simplified to \eqref{CRB1Mzt} and \eqref{CRB2Mzt}.
\end{IEEEproof}

Notice that $\mathrm{CRB}_{1}^{\mathsf{M}}\left(\kappa_{\tq} \right)<\mathrm{CRB}_{2}^{\mathsf{M}}\left(\kappa_{\tq} \right)<\mathrm{CRB}_{3}^{\mathsf{M}}\left(\kappa_{\tq} \right)$ can be derived based on Corollary 12, which is similar to inequality \eqref{eq:CRB_vec_sca_overall}. It clearly indicates that using multiple distributed receiving antennas does not affect the order of estimation accuracy of exploiting different electric field observations. 

\noindent \textbf{Corollary 13} (SIMO, $\tau \to \infty$)\textbf{.} 
{If $z_{\tq}\gg \lambda$ and $\tau \to \infty$, the CRBs of the SIMO positioning system can be given by}
{\begin{small}\setlength\abovedisplayskip{3pt}
\setlength\belowdisplayskip{3pt}
\begin{align}
&\lim_{\tau\to \infty}\mathrm{CRB}_{1}^{\mathsf{M}}\left(\kappa_{\tq}\right)=\lim_{\tau\to \infty}{\mathrm{CRB}_{1}^{\mathsf{C}}\left(\kappa_{\tq}\right)}/{N_{\mathsf{s}}^{2}},\\
&\lim_{\tau\to \infty}\mathrm{CRB}_{2}^{\mathsf{M}}\left(\kappa_{\tq}\right)=\lim_{\tau\to \infty}{\mathrm{CRB}_{2}^{\mathsf{C}}\left(\kappa_{\tq}\right)}/{N_{\mathsf{s}}^{2}}.
\end{align}\end{small}}

\begin{IEEEproof}
It follows from Corollary 11 and 12. Particularly, we have $\lim{\rho_{11\kappa}^{\mathsf{s};i,j}}=\lim\rho_{11\kappa}$ and $\lim{\rho_{21\kappa}^{\mathsf{s};i,j}}=\lim\rho_{21\kappa}$, where $\lim$ stands for $\lim_{\tau\to \infty}$. Thus, Corollary 13 holds.
\end{IEEEproof}
It can be seen from Corollary 13 that the CRBs of the SIMO positioning system will be one-$N_{\mathsf{s}}^{2}$th of the SISO system as $\tau$ increases unboundedly. Configured on the surface of fixed size, adjacent small receiving antennas will be stacked on top of each other with $\tau$ increasing, resulting in multiplexing benefits and lower CRBs. Besides, the total area of the small receiving antennas will be larger than  $\mathcal{R}_{\mathsf{s}}$ as $\tau\to \infty$, which ignores the space constraints. In fact, the more practical and meaningful case is $\tau \leq R_{\rq}/z_{\tq}$, which will be analyzed in Sec. \ref{subsection:CRB_simo_numer}.

\section{Numerical Results and Discussion}\label{sectionV}
In this section, we will provide numerical results to illustrate the propositions and corollaries derived in previous sections. 
We set the signal-to-noise ratio as
$\mathrm{SNR}=\frac{{|E_{\mathsf{in}}|^{2}}}{\sigma^{2}} = -10\textrm{dB}$ and the wavelength as $\lambda= 0.01\textrm{m}$ (corresponding to $f_{c} = 30 \textrm{GHz}$), unless otherwise specified. Other values can also be examined, as the results are very general.
 
\subsection{CRB for CPL Case}\label{subsection:CRB_CPL_numer}
\begin{figure}
\centering
\includegraphics[scale=0.45]{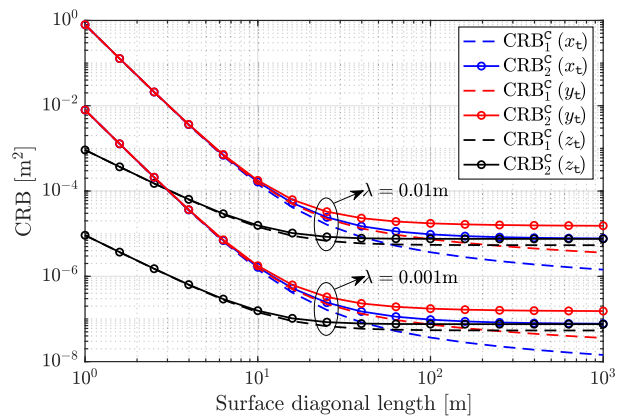}
\caption{CRBs versus surface diagonal length $D_{\rq}$, with $\lambda= 0.01\textrm{m}$ or $0.001\textrm{m}$, $z_{\tq}=6\textrm{m}$, when $\mathbf{p}_{\tq}$ is on the CPL and using $\textit{VEF}$ or $\textit{SEF}$.}
\label{fig:cpl111}
\end{figure}

\begin{figure}
\centering
\includegraphics[scale=0.45]{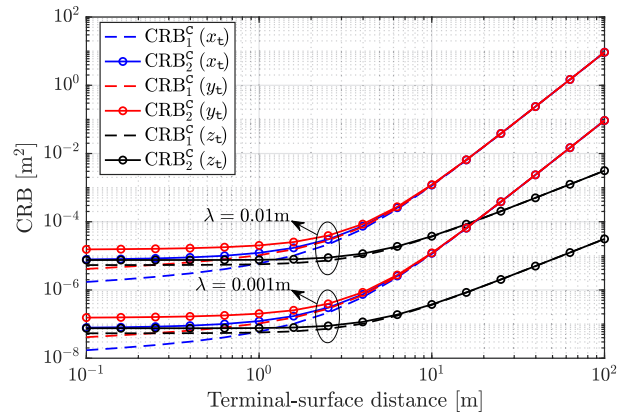}
\caption{CRBs versus terminal-surface distance $z_{\tq}$, with $D_{\rq}=9\textrm{m}$, $\lambda= 0.01\textrm{m}$ or $0.001\textrm{m}$, when $\mathbf{p}_{\tq}$ is on the CPL and using $\textit{VEF}$ or $\textit{SEF}$.}
\label{fig:cpl222}
\end{figure}

We first show the CRBs for a terminal on the CPL computed in Sec. \ref{sectionCPL}. To illustrate the impact of the wavelength on the CRBs, we consider two different values, i.e., $\lambda=0.01\textrm{m}$ and $\lambda=0.001\textrm{m}$ (corresponding to $f_{c} = 300 \textrm{GHz}$). Fig. \ref{fig:cpl111} and Fig. \ref{fig:cpl222} demonstrate the CRBs, measured in square meters $[\textrm{m}^{2}]$, versus the surface diagonal length $D_{\rq}$ or the distance 
from the terminal to the receiving antenna (terminal-surface distance) $r_{\mathsf{to}}=z_{\tq}$ when $z_{\tq}=6\textrm{m}$ or $D_{\rq}=9\textrm{m}$, respectively.

Fig. \ref{fig:cpl111} shows that all the CRBs decrease dramatically with the surface diagonal length $D_{\rq}$ in the range $1\textrm{m}\sim 10\textrm{m}$, which contains the values of $D_{\rq}$ commonly used in the actual system. In addition, the CRBs for $z_{\tq}$ are much lower than those for $x_{\tq}$ and $y_{\tq}$ in the above range. More interestingly, the CRBs utilizing \textit{SEF} are greater than CRBs using \textit{VEF} for all values of $D_{\rq}$, which agrees with Remark 4. The difference between $\mathrm{CRB}_{1}^{\mathsf{C}}\left(\kappa_{\tq}\right)$ and $\mathrm{CRB}_{2}^{\mathsf{C}}\left(\kappa_{\tq}\right)$ is  negligible 
 if $D_{\rq}$ is smaller than $10\textrm{m}$, but it increases progressively with the increase of $D_{\rq}$. As for the CRBs in the asymptotic regime, we observe that: (i) $\mathrm{CRB}_{1}^{\mathsf{C}}\left(x_{\tq}\right)$
and $\mathrm{CRB}_{1}^{\mathsf{C}}\left(y_{\tq}\right)$ decrease infinitely with the trend of the $\ln^{-1}$ function provided in \eqref{eq:CRB1xttau} and \eqref{eq:CRB1yttau}; (ii) $\mathrm{CRB}_{1}^{\mathsf{C}}\left(z_{\tq}\right)$ and $\mathrm{CRB}_{2}^{\mathsf{C}}\left(z_{\tq}\right)$ approach the asymptotic limit in \eqref{eq:CRB1zttau} and \eqref{eq:CRB2zttau} from $D_{\rq}\approx20\textrm{m}$; (iii) $\mathrm{CRB}_{2}^{\mathsf{C}}\left(x_{\tq}\right)$ and $\mathrm{CRB}_{2}^{\mathsf{C}}\left(y_{\tq}\right)$ converge to the asymptotic limit in \eqref{eq:CRB2xttau} and \eqref{eq:CRB2yttau} when $D_{\rq}>10^{3}\textrm{m}$. These phenomena are consistent with Corollary 11. Fig. \ref{fig:cpl222} shows that the CRBs for all three
dimensions increase very slowly with the terminal-surface distance $r_{\mathsf{to}}$ in the range $0.1\textrm{m}\leq z_{\tq}\leq 1\textrm{m}$, but they increase considerably ($\mathrm{CRB}_{1}^{\mathsf{C}}\left(z_{\tq}\right)$ and $\mathrm{CRB}^{\mathsf{C}}_{2}\left(z_{\tq}\right)$ are much lower than the CRBs for $x_{\tq}$ and $y_{\tq}$) when $z_{\tq}> 1\textrm{m}$.

\begin{figure}
\centering
\includegraphics[scale=0.45]{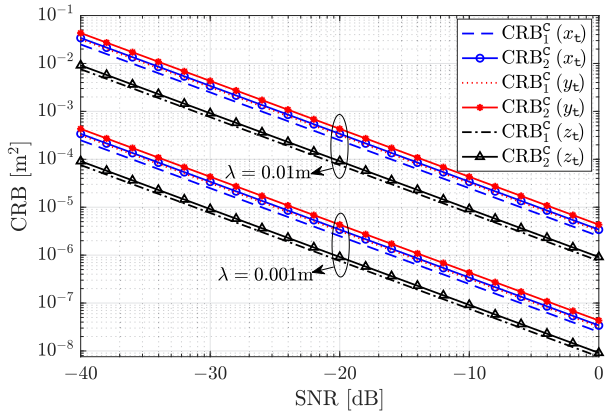}
\caption{{CRBs versus $\mathrm{SNR}$, with $D_{\rq}=10\textrm{m}$, $\lambda= 0.01\textrm{m}$ or $0.001\textrm{m}$, and $z_{\tq}=3\textrm{m}$, when $\mathbf{p}_{\tq}$ is on the CPL and using $\textit{VEF}$ or $\textit{SEF}$.}}
\label{fig:SNR}
\end{figure}

{In Fig. \ref{fig:SNR}, we perform a simulation to show the CRBs with respect to the $\mathrm{SNR}$ when $z_{\tq}=3\textrm{m}$ and $D_{\rq}=10\textrm{m}$. It can be observed that all the CRBs are \textit{inversely proportional} to the $\mathrm{SNR}$. In particular, the CRBs will decrease by a factor of $10$ if the $\mathrm{SNR}$ increases by $10\textrm{dB}$, which can be derived analytically by considering the results in Corollary 7 and Corollary 8. This holds true also for the general scenario and the SIMO system (simulations are no longer shown due to space limitations), as revealed by \eqref{eq:CRB1x} -- \eqref{eq:CRB1z} and \eqref{eq:CRB_SIMO_1} -- \eqref{eq:CRB3M}. It is worth mentioning that the CRB is an asymptotically tight lower bound for MSE only in \textit{high} $\mathrm{SNR}$ regions. Considering the global tight lower bound, e.g., Ziv-Zakai bound (ZZB)\cite{ZZBB1,ZZBB2}, in \textit{low} $\mathrm{SNR}$ regions, will be an attractive extension of our work. In Fig. \ref{fig:cpl111}, Fig. \ref{fig:cpl222}, and Fig. \ref{fig:SNR}, it is also observed that all the CRBs depend \textit{linearly} on the square of wavelength regardless of using \textit{VEF} or \textit{SEF}, as in Corollary 10. Indeed, reducing the wavelength by a factor of $\epsilon$ reduces the CRBs of the factor of $\epsilon^{2}$.}

Table. \ref{tab:performance} gives the square root of the CRBs (RCRB, denoted as ${\mathrm{R}_{crb}}\left(\kappa_{\tq}\right)$), measured
in centimeters (i.e., $[\textrm{cm}]$), for the three components $x_{\tq}$, $y_{\tq}$, and $z_{\tq}$, for terminals located on the CPL. ${D_{1}^{case}}$, ${D_{2}^{case}}$, ${D_{3}^{case}}$, and ${D_{4}^{case}}$ represent that the surface diagonal length $D_{\rq}$ is $0.5\textrm{m}$,  $1\textrm{m}$,  $2\textrm{m}$,  and $3\textrm{m}$ when $z_{\tq}=6\textrm{m}$, respectively. To evaluate the average positioning performance, we adopt the receiving antenna with $D_{\rq}=3\textrm{m}$ to compute the average RCRB of $1000$ terminals with coordinates of $z_{\tq}$ dimension uniformly distributed in $[1\textrm{m}, 20\textrm{m}]$, which is denoted as ${A_{ver}^{case}}$. It can be seen that using  \textit{VEF} or \textit{SEF} can guarantee
 a centimeter/$\textrm{cm}$-level accuracy (within a few centimeters) for estimating all three dimensions in the mmWave and  sub-THz band. Unfortunately, although accuracy on the order of tens of centimeters for $z_{\tq}$ can be achieved by utilizing \textit{OSEF}, we are unable to estimate $x_{\tq}$ and $y_{\tq}$ with acceptable accuracy. This reveals that the single conventional surface antenna possesses the near-field \textit{ranging} function, which can be considered a one-dimensional special case of near-field positioning.

\begin{table}[!t]
\center
\caption{Comparison of estimation accuracy between using \textit{VEF}, \textit{SEF}, and \textit{OSEF}, $\lambda=0.01{\textrm{m}}$ (- means that the value is too large).}
\label{tab:performance}
\begin{tabular}{c|c|c|c|c|c||c}
\toprule
\multicolumn{2}{c|}{} & \multicolumn{5}{c}{RCRB $[\textrm{cm}]$} \\
\cline{3-7}
\multicolumn{2}{c|}{} & ${D_{1}^{case}}$& ${D_{2}^{case}}$& ${D_{3}^{case}}$ & ${D_{4}^{case}}$ & ${A_{ver}^{case}}$ \\ 
\midrule
\multirow{3}{*}{\textit{VEF}}& ${\mathrm{R}_{crb}}\left(x_{\tq}\right)$&35.5 &8.91 &2.25 & 1.02& \textbf{3.88} \\
 &${\mathrm{R}_{crb}}\left(y_{\tq}\right)$& 35.5& 8.91& 2.26&1.02  & \textbf{3.88}\\
 &${\mathrm{R}_{crb}}\left(z_{\tq}\right)$& 0.604& 0.303& 0.153&0.103 & \textbf{0.179}\\
\midrule
\multirow{3}{*}{\textit{SEF}}& ${\mathrm{R}_{crb}}\left(x_{\tq}\right)$&35.5 &8.92 &2.26 &1.03 &  \textbf{3.89} \\
 &${\mathrm{R}_{crb}}\left(y_{\tq}\right)$& 35.6& 8.92& 2.26&1.03  & \textbf{3.89}\\
 &${\mathrm{R}_{crb}}\left(z_{\tq}\right)$& 0.605& 0.303& 0.153&0.104 & \textbf{0.179}\\
\midrule
\multirow{3}{*}{\textit{OSEF}}& ${\mathrm{R}_{crb}}\left(x_{\tq}\right)$&- &- &- &- & \textbf{-} \\
 &${\mathrm{R}_{crb}}\left(y_{\tq}\right)$& -& -& -&-  & \textbf{-}\\
 &${\mathrm{R}_{crb}}\left(z_{\tq}\right)$& 11.8& 21.1& 20.4&23.7 & \textbf{18.0}\\
\bottomrule
\end{tabular}
\end{table}

\subsection{CRB for the General Scenario}\label{subsection:CRB_general_numer}
\begin{figure}
\centering
\includegraphics[scale=0.45]{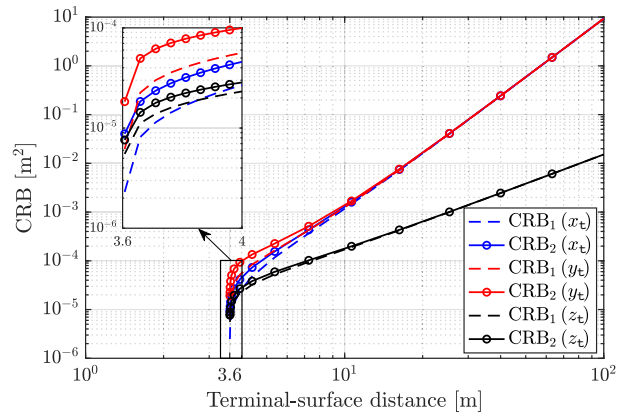}
\caption{CRBs as a function of the terminal-surface distance for a terminal at $\left(2,3,z_{\tq}\right)$ when using \textit{VEF} or \textit{SEF}, $D_{\rq}=9\textrm{m}$, and $\lambda=0.01\textrm{m}$.}
\label{fig:6}
\end{figure}

We will evaluate the positioning performance for a terminal not restricted to the CPL as discussed in Proposition 3, Corollary 5 and 6. Fig. \ref{fig:6} illustrates the CRBs as a function of the distance $r_{\mathsf{to}}=\sqrt{x_{\tq}^{2}+y_{\tq}^{2}+z_{\tq}^{2}}$ for a terminal at $(2,3,z_{\tq})$ when $D_{\rq}=9\textrm{m}$. It can be found that the estimation accuracy reduces as the terminal-surface distance increases, which is consistent with our intuition. Particularly, the CRBs for estimating $x_{\tq}$ and $y_{\tq}$ increase faster than those for $z_{\tq}$ regardless of \textit{VEF} or \textit{SEF}. Furthermore, all the CRBs increase rapidly when the terminal is close to the receiving antenna ($0<z_{\tq}\leq \sqrt{3}\textrm{m}$). This occurs since the estimation for all three dimensions is nearly perfect (CRBs are approaching $0$) when the terminal approaches the receiving antenna ($z_{\tq}\to0$, $|x_{\tq}|$ and $|y_{\tq}|$ are less than $\frac{D_{\rq}}{2\sqrt{2}}$), and as $z_{\tq}$ increases from zero, CRBs will rapidly increase to greater orders of magnitude. In addition, it can be seen that $\mathrm{CRB}_{2}\left(\kappa_{\tq}\right)$ is greater than $\mathrm{CRB}_{1}\left(\kappa_{\tq}\right)$ when the terminal-surface distance is less than $10\textrm{m}$, otherwise they are equal. This indicates that for a receiving antenna with fixed size, there is a considerable performance gap between utilizing \textit{VEF} and \textit{SEF}, only when the terminal is close to the receiving antenna. 

\begin{figure}[!t]
\centering
\includegraphics[scale=0.45]{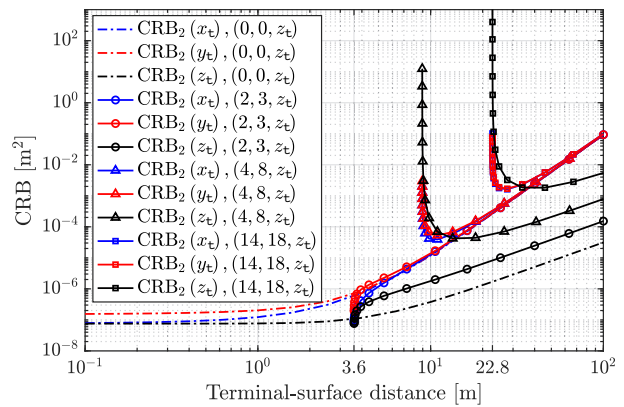}
\caption{CRBs as a function of the terminal-surface distance for terminals with different $x_{\tq}$ and $y_{\tq}$ when using \textit{SEF}, $D_{\rq}=9\textrm{m}$ and $\lambda=0.001\textrm{m}$. Since using \textit{VEF} or \textit{SEF} has the same rules, we take the use of \textit{SEF} as an example.}
\label{fig:7}
\end{figure}
Fig. \ref{fig:7} illustrates the CRBs for terminals with different $x_{\tq}$ and $y_{\tq}$ versus the terminal-surface distance $r_{\mathsf{to}}$ when  utilizing \textit{SEF}, $D_{\rq}=9\textrm{m}$ and $\lambda=0.001\textrm{m}$. It indicates that the CRBs possess different trends and the curve shapes vary from each other for different $x_{\tq}$ and $y_{\tq}$ when the terminal is close to the receiving antenna. For instance, if the terminal is on the CPL ($x_{\tq}=y_{\tq}=0$, $r_{\mathsf{to}}=z_{\tq}$),  the CRBs for all three dimensions are almost unchanged in the range $0.1\textrm{m}<r_{\mathsf{to}}<1\textrm{m}$. However, if $x_{\tq}$ or $y_{\tq}$ are greater than $\frac{D_{\rq}}{2\sqrt{2}}$ and $z_{\tq}$ is small, which means the vertical projection of the terminal along the $Z$-dimension is not on the receiving antenna surface, and the distance from the terminal to the CPL is much larger than $z_{\tq}$, the CRBs sharply decrease from infinity. We refer to this interesting phenomenon as the near-field positioning \textit{blocking zone} effect, which always exists for a fixed-size receiving antenna. Moreover, extensive numerical simulation as Fig. \ref{fig:cpl111} for terminals not on the CPL demonstrate that the result obtained in the analysis of the CPL case in Sec. \ref{sectionCPL} is also applicable to the sophisticated generic near-field system proposed in Sec. \ref{SEC:2}, which provides support for the generality of our insights and results.

\begin{figure}
	\centering
	\vspace{-1em}
	\subfloat[$\mathrm{CRB}_{1}^{\mathsf{N}}\left(x_{\tq} \right)$ in $\textrm{dB}$.]{
		\includegraphics[scale=0.164]{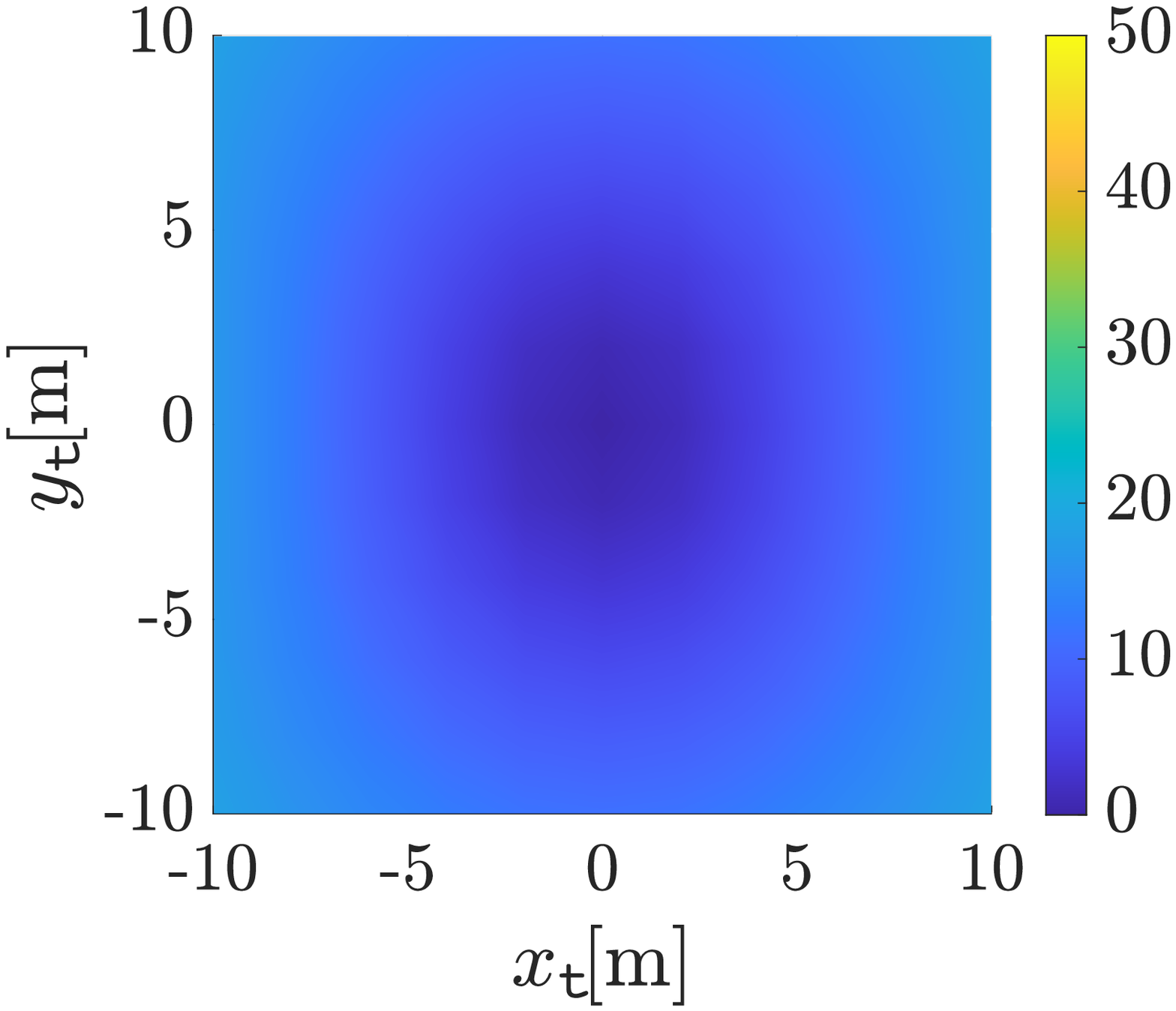}
		\label{fig:CRBxa}
	}
	\subfloat[$\mathrm{CRB}_{2}^{\mathsf{N}}\left(x_{\tq} \right)$ in $\textrm{dB}$.]{
		\includegraphics[scale=0.164]{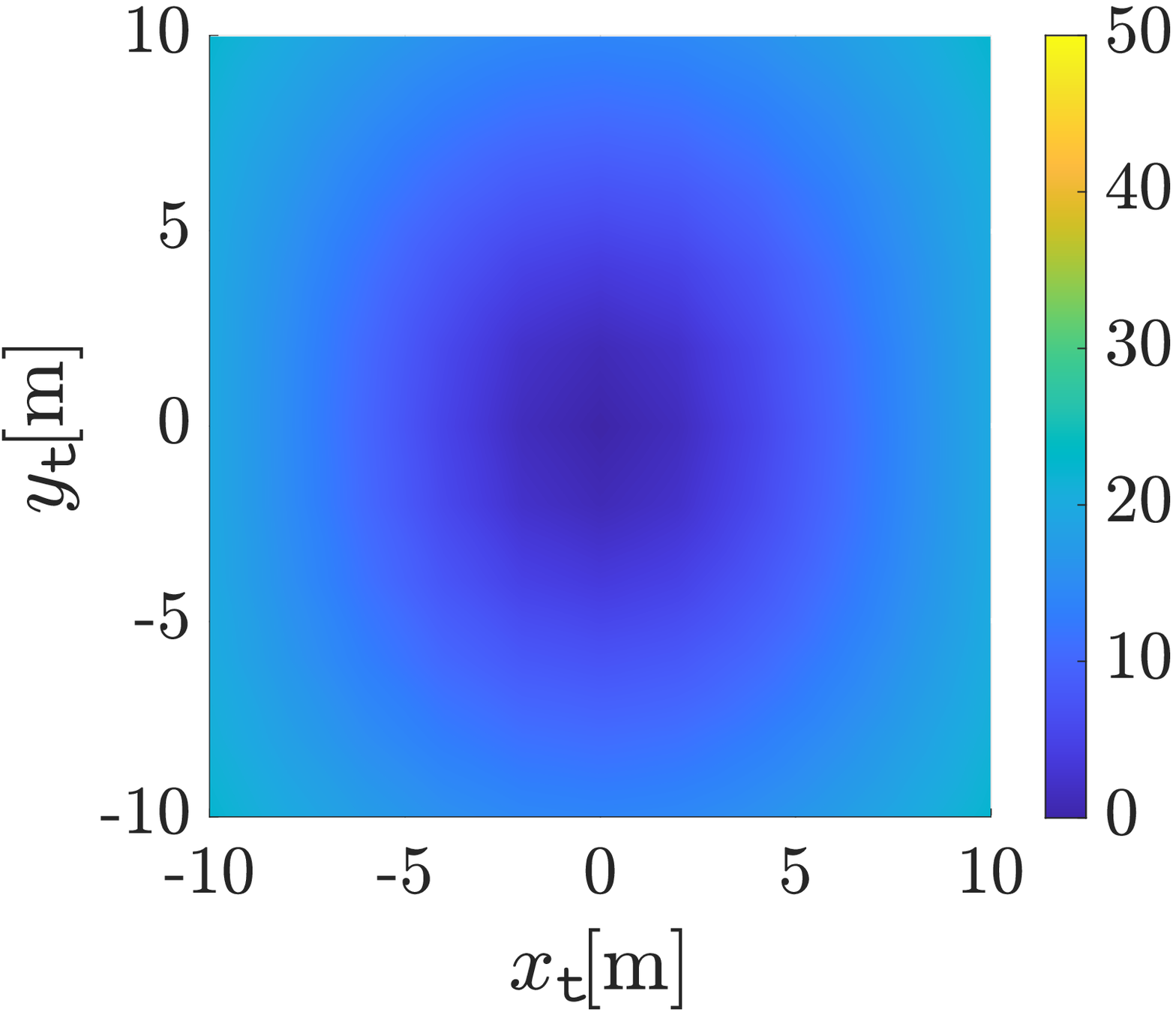}
		\label{fig:CRBxb}
	}
	\subfloat[$\mathrm{CRB}_{1}^{\mathsf{N}}\left(y_{\tq} \right)$ in $\textrm{dB}$.]{
		\includegraphics[scale=0.164]{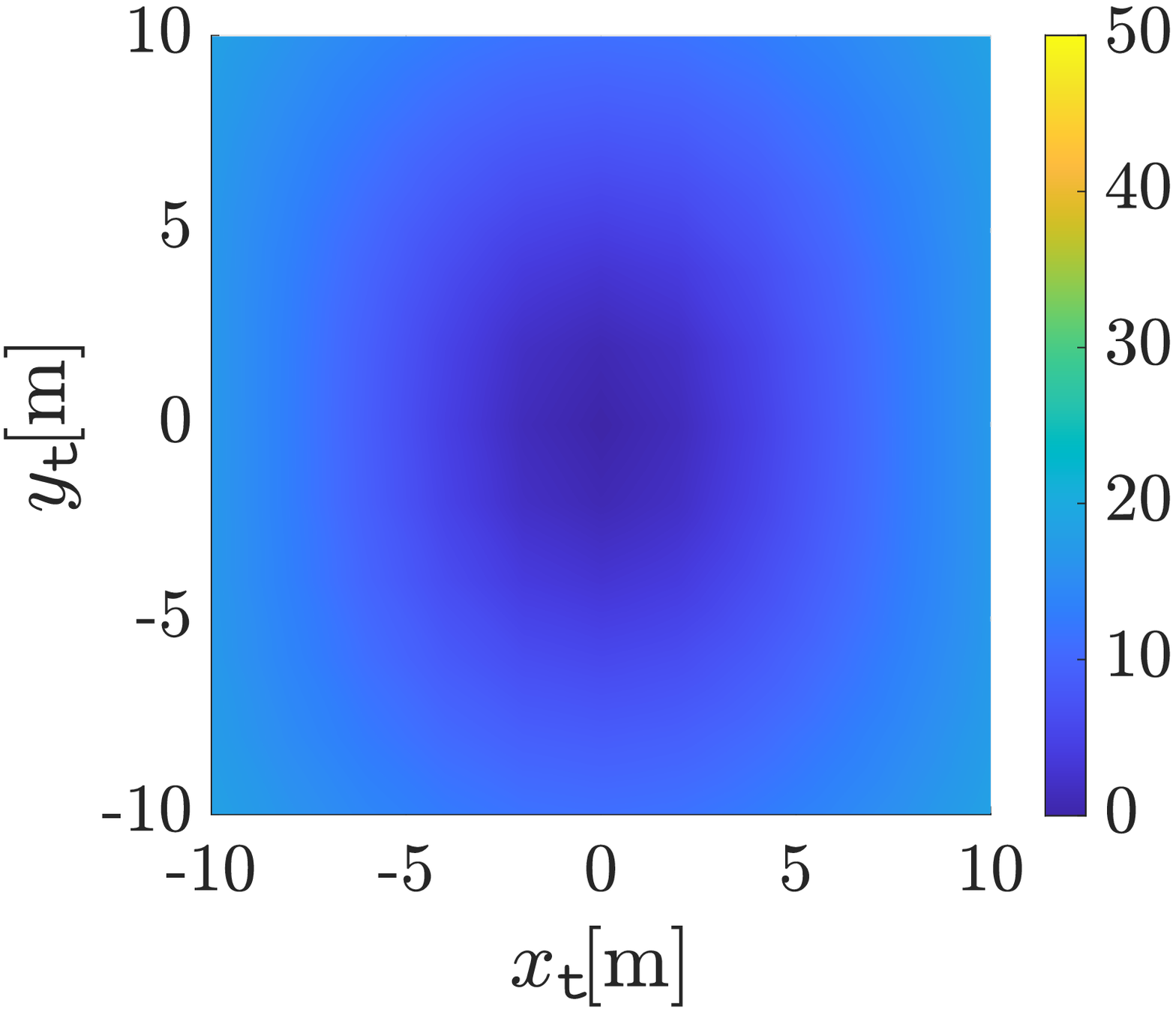}
		\label{fig:CRBya}
	}
\vfill

\subfloat[$\mathrm{CRB}_{2}^{\mathsf{N}}\left(y_{\tq} \right)$ in $\textrm{dB}$.]{
		\includegraphics[scale=0.164]{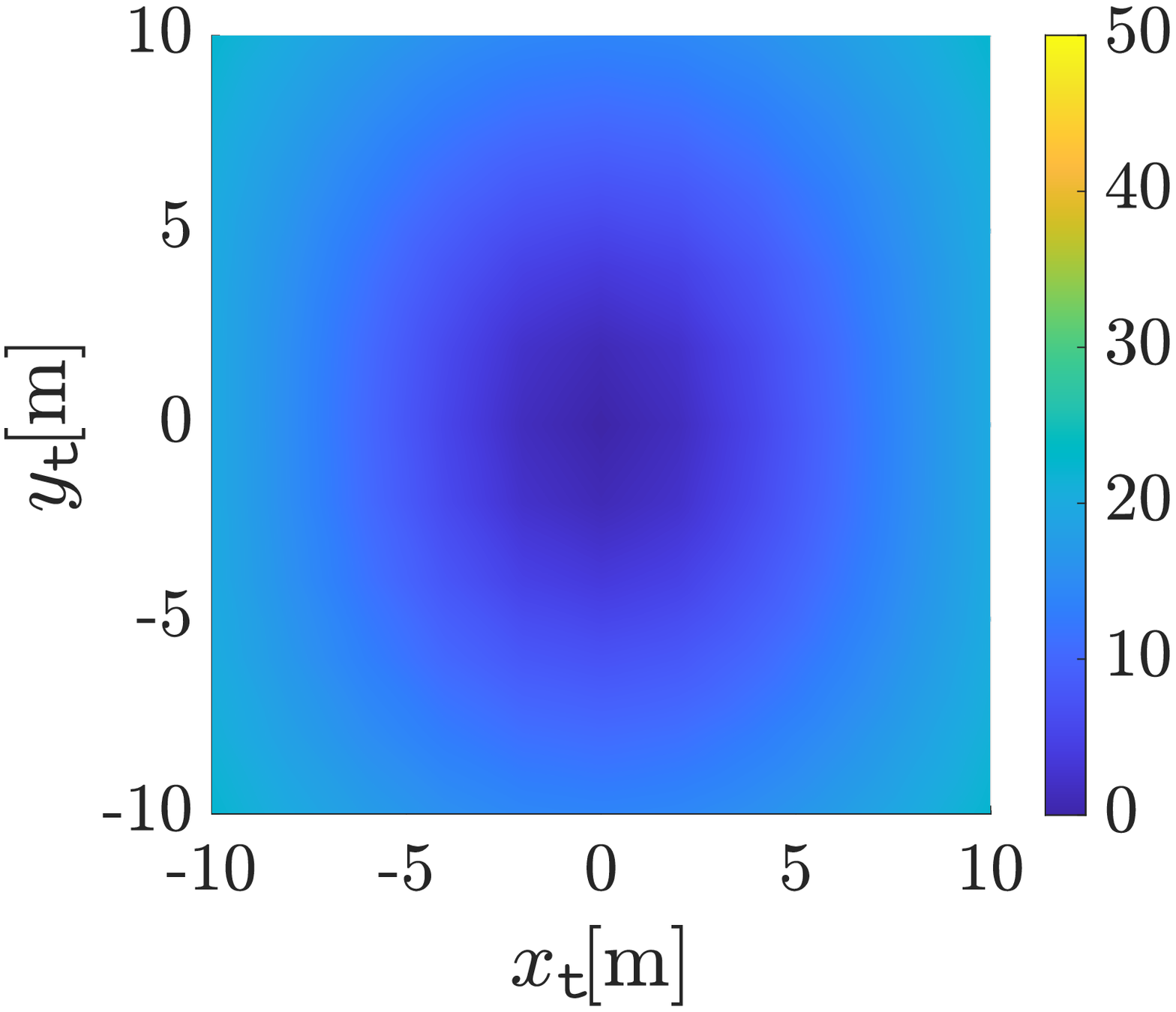}
		\label{fig:CRByb}
	}
	\subfloat[$\mathrm{CRB}_{1}^{\mathsf{N}}\left(z_{\tq} \right)$ in $\textrm{dB}$.]{
		\includegraphics[scale=0.164]{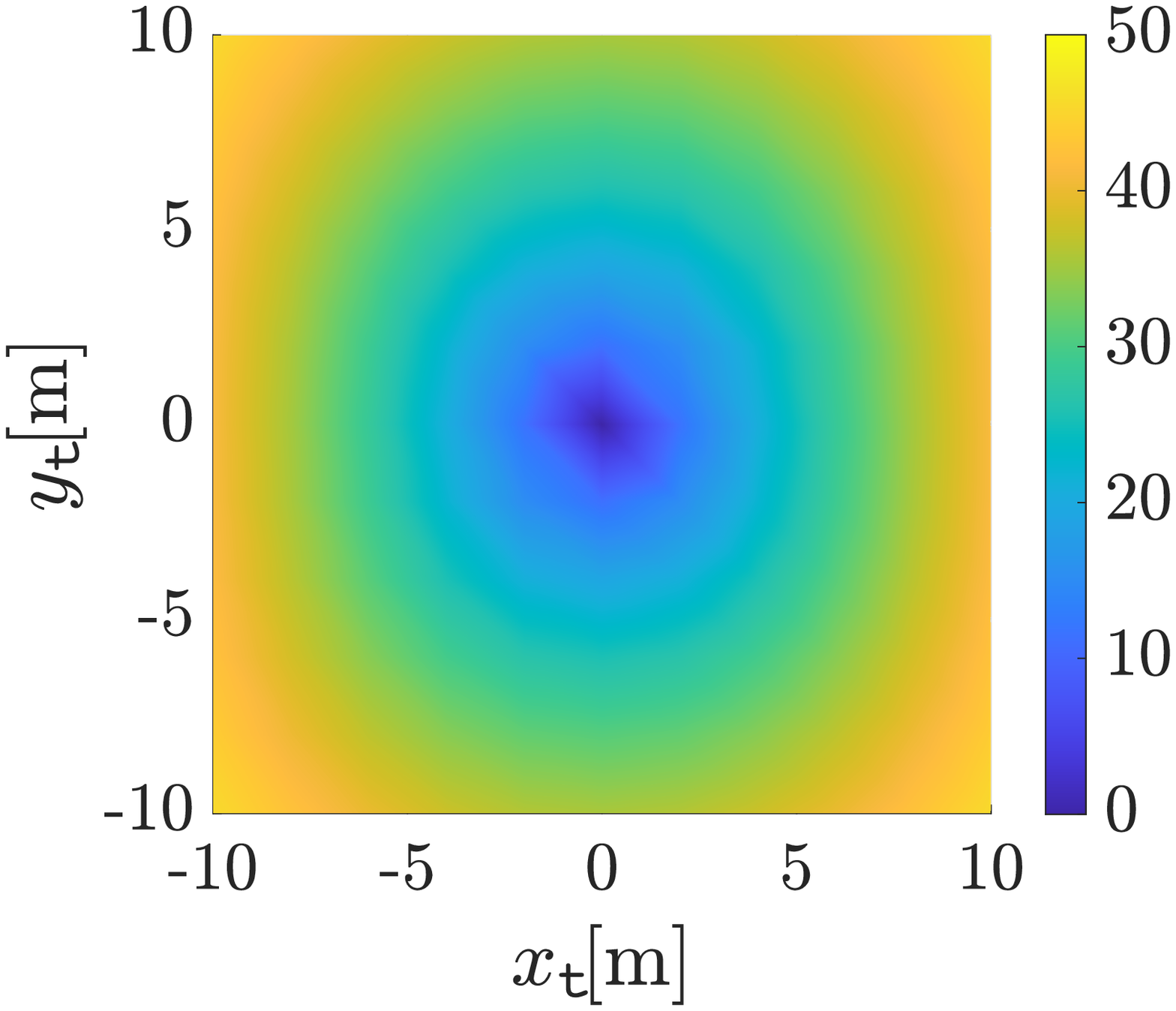}
		\label{fig:CRBza}
	}
	\subfloat[$\mathrm{CRB}_{2}^{\mathsf{N}}\left(z_{\tq} \right)$ in $\textrm{dB}$.]{
		\includegraphics[scale=0.164]{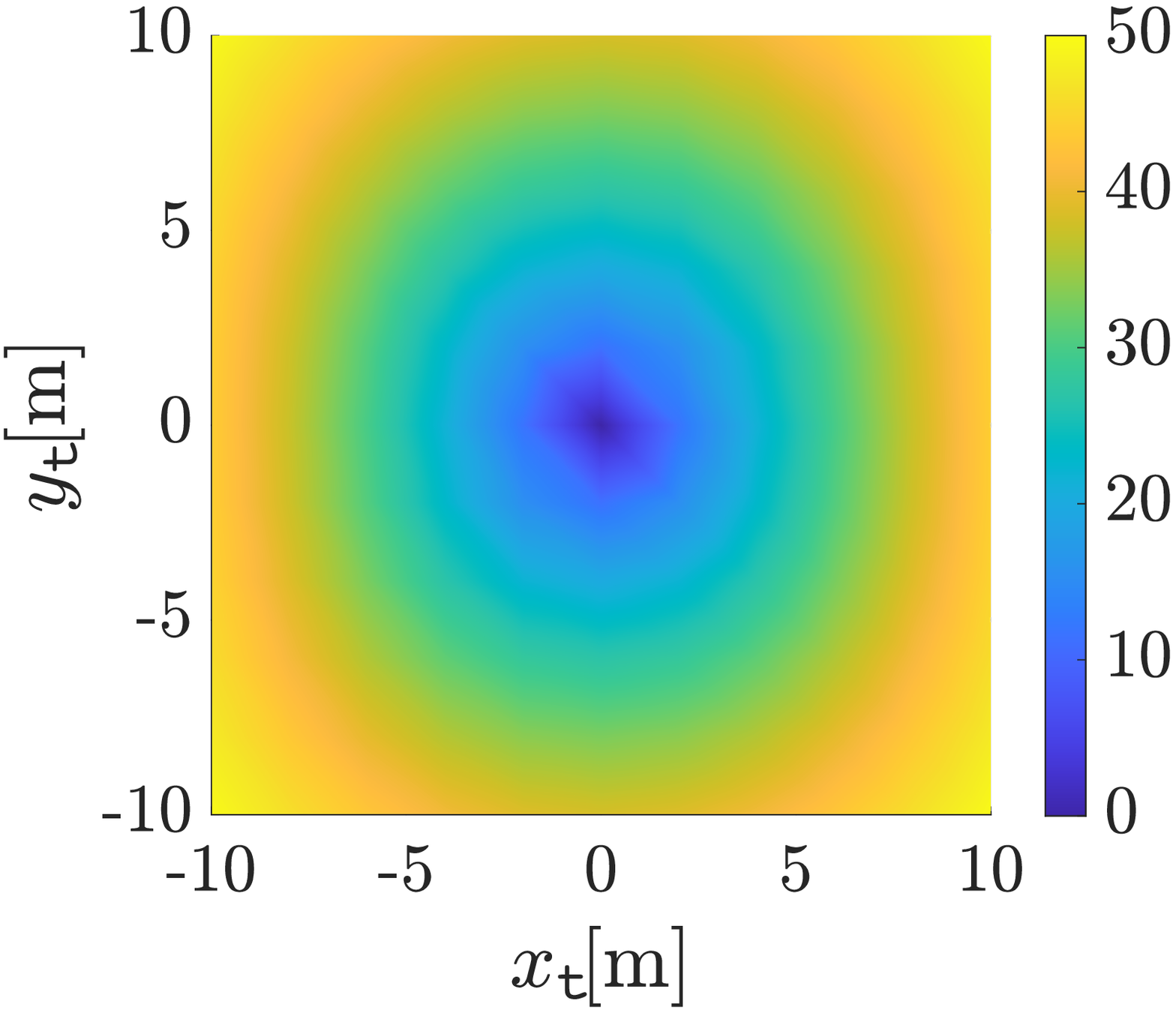}
		\label{fig:CRBzb}
	}
 \vfill
 \subfloat[{Normalized CRBs in 3D view.}]{
		\includegraphics[scale=0.45]{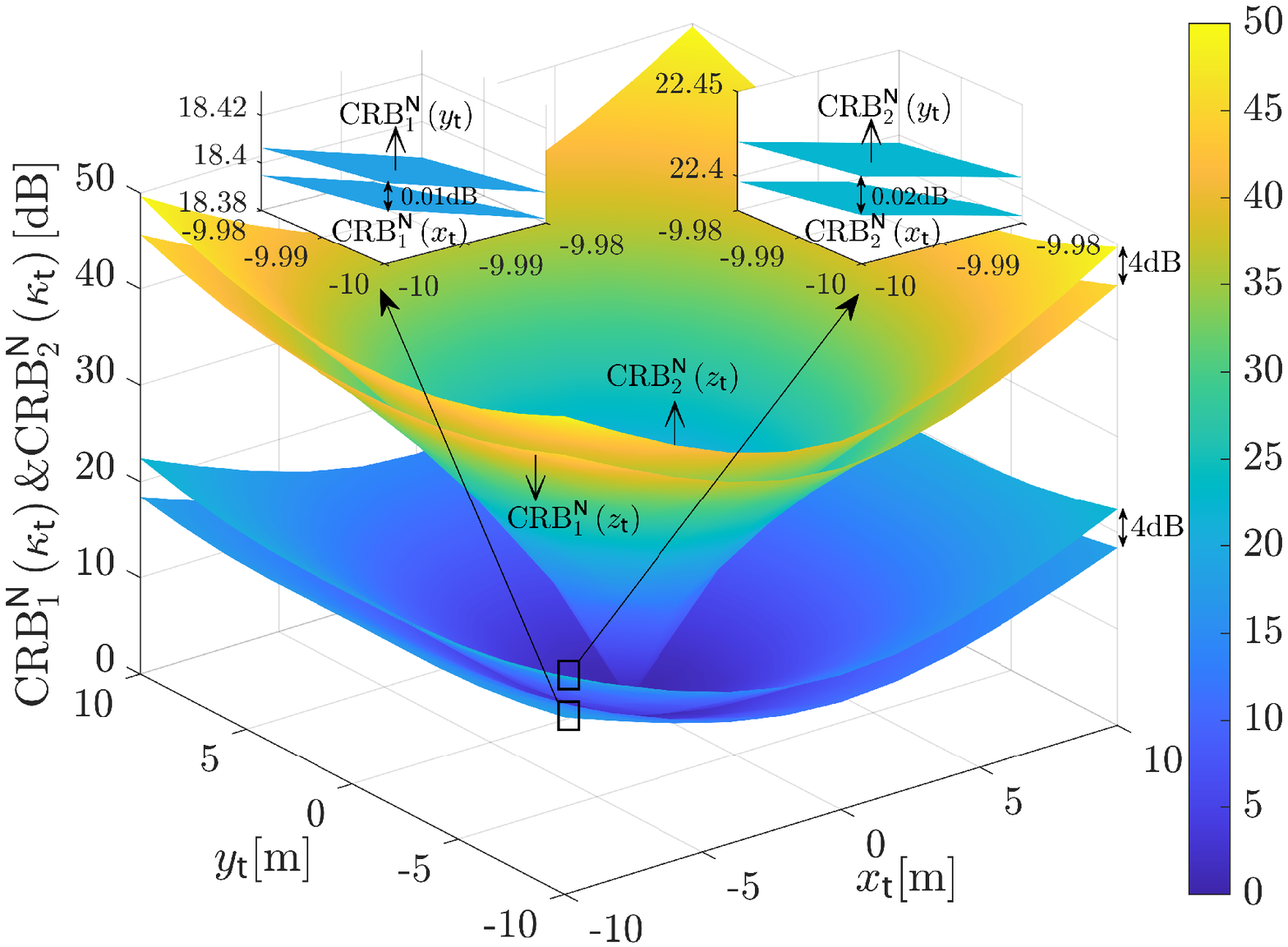}
		\label{fig:TVT1}
	}
 \caption{{Normalized CRBs, measured in $[\textrm{dB}]$, as a function of $x_{\tq}$ and $y_{\tq}$ for the terminal not on the CPL when using $\textit{VEF}$/$\textit{SEF}$, $z_{\tq} = 6 \textrm{m}$, $D_{\rq}=3\textrm{m}$, and $\lambda=0.01\textrm{m}$. We define that $\mathrm{CRB}_{1}^{\mathsf{N}}\left(\kappa_{\tq}\right)\triangleq10\log_{10}[\mathrm{CRB}_{1}\left(\kappa_{\tq}\right)/\mathrm{CRB}_{1}^{\mathsf{C}}\left(\kappa_{\tq}\right)]$ and $\mathrm{CRB}_{2}^{\mathsf{N}}\left(\kappa_{\tq}\right)\triangleq10\log_{10}[\mathrm{CRB}_{2}\left(\kappa_{\tq}\right)/\mathrm{CRB}_{2}^{\mathsf{C}}\left(\kappa_{\tq}\right)]$.}}
    \label{fig:simu1}
\end{figure}

\begin{figure}[!t]
	\centering
	\subfloat[$\mathrm{CRB}_{1}^{\mathsf{M}}\left(x_{\tq} \right)$ and $\mathrm{CRB}_{1}^{\mathsf{C}}\left(x_{\tq} \right)$ versus $D_{\rq}$.]{
	\includegraphics[scale=0.45]{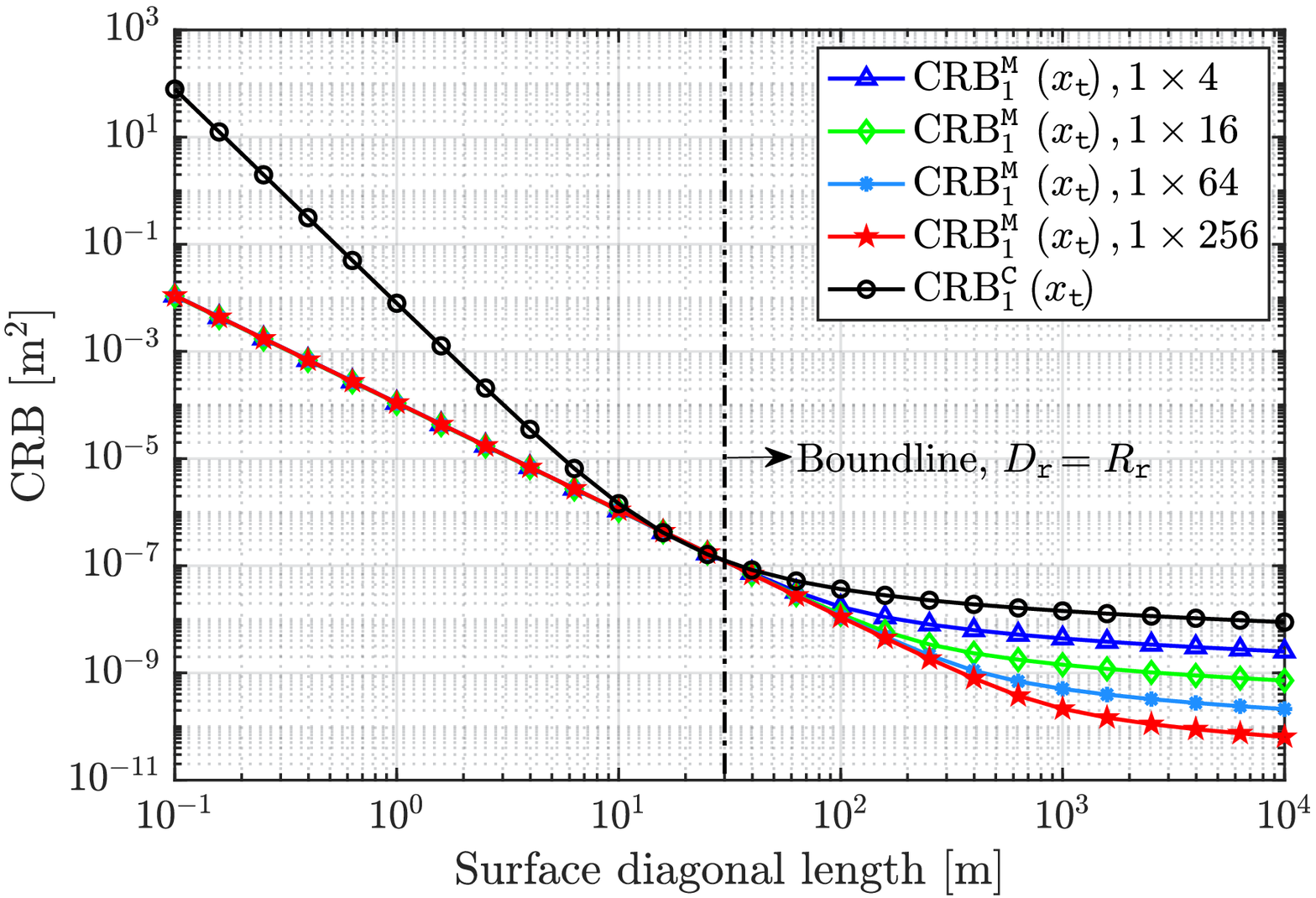}
		\label{fig:simo1}
	}
	\vfill
	\subfloat[$\mathrm{CRB}_{1}^{\mathsf{M}}\left(y_{\tq} \right)$ and $\mathrm{CRB}_{1}^{\mathsf{C}}\left(y_{\tq} \right)$ versus $D_{\rq}$.]{
		\includegraphics[scale=0.45]{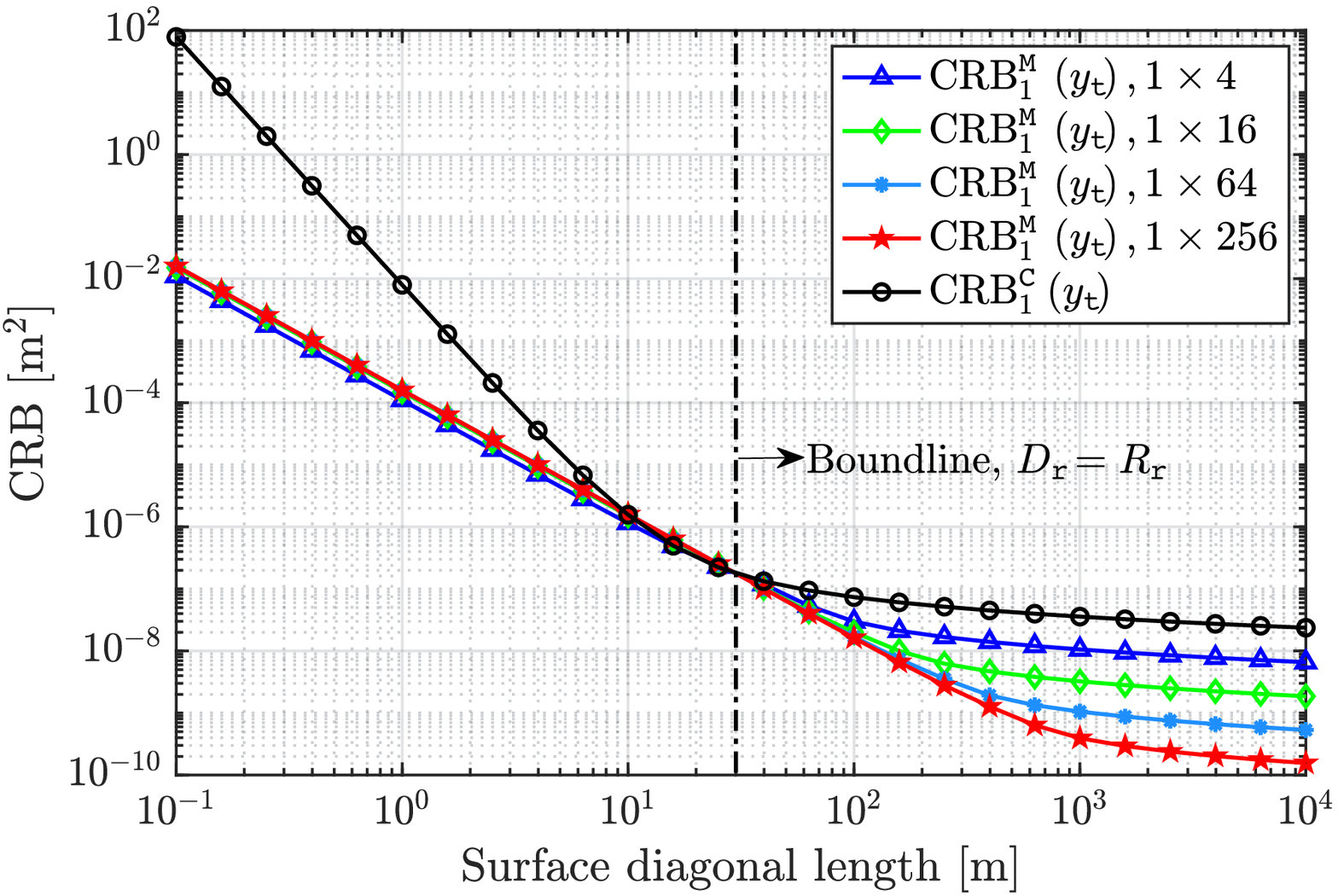}
		\label{fig:simo2}
	}
	\vfill
	\subfloat[$\mathrm{CRB}_{1}^{\mathsf{M}}\left(z_{\tq} \right)$ and $\mathrm{CRB}_{1}^{\mathsf{C}}\left(z_{\tq} \right)$ versus $D_{\rq}$.]{
		\includegraphics[scale=0.45]{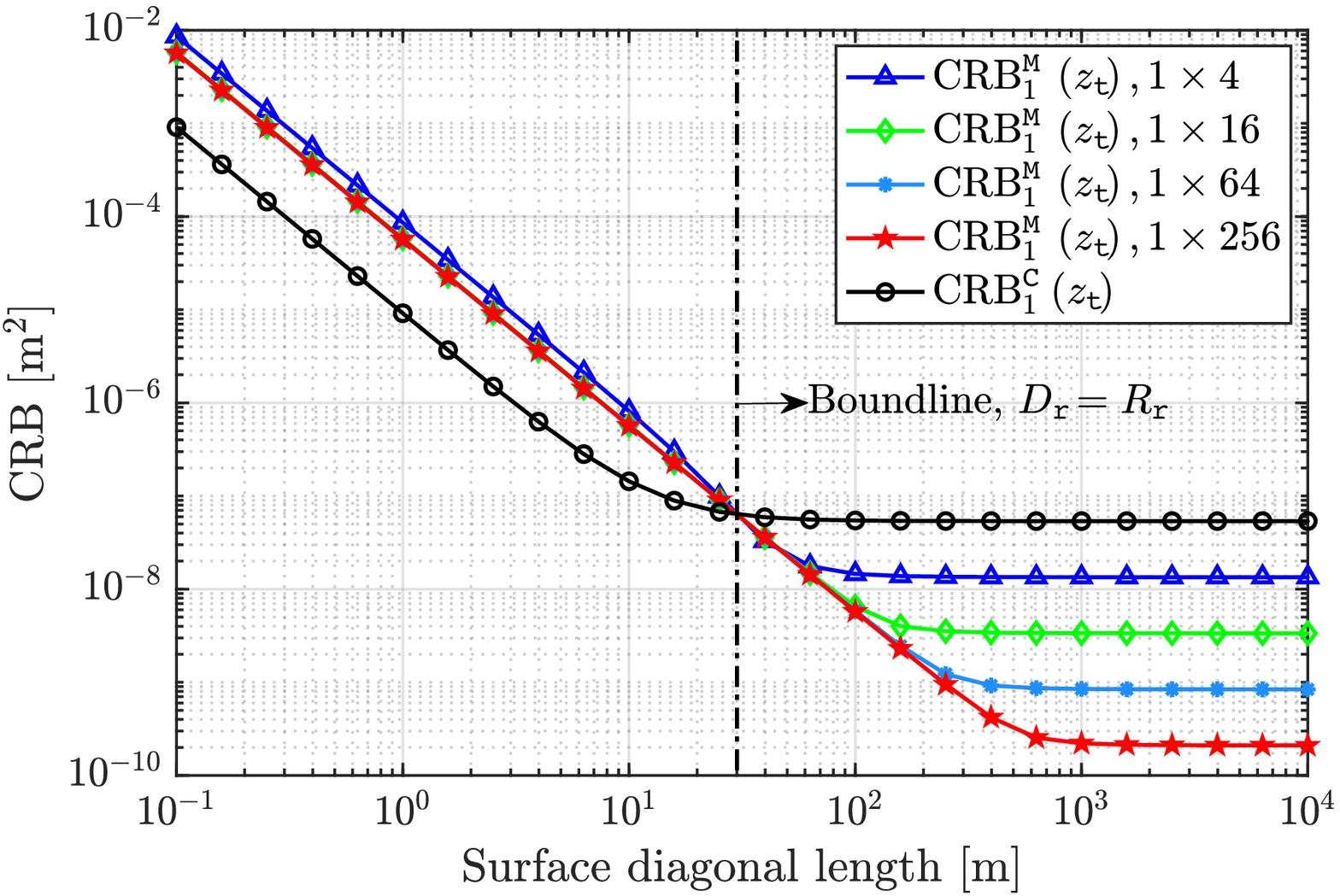}
		\label{fig:simo3}
	}
 \caption{CRBs with different numbers of small receiving antennas ($N_{\mathsf{s}}^{2}=1, 4, 16, 64, 256$, equivalently expressed as $\mathrm{CRB}_{1}^{\mathsf{M}}\left(\kappa_{\tq}\right)$, $1\times4, 1\times16,1\times64,1\times256$) with $R_{\rq}=30\textrm{m}$, $z_{\tq}=6\textrm{m}$, and $\lambda=0.001\textrm{m}$ when using \textit{VEF}.}
   \label{fig:simo}
\end{figure}
Fig. \ref{fig:simu1} demonstrates the  normalized CRBs for the terminal not on the CPL, versus $x_{\tq}$ and $y_{\tq}$ when $z_{\tq}=6\textrm{m}$, $D_{\rq} = 3\textrm{m}$, and using \textit{VEF} or \textit{SEF}. These normalized CRBs, measured in $[\textrm{dB}]$ and denoted as $\mathrm{CRB}_{1}^{\mathsf{N}}\left(\kappa_{\tq}\right)$ and $\mathrm{CRB}_{2}^{\mathsf{N}}\left(\kappa_{\tq}\right)$, are defined as the values of CRBs normalized by their minimum, which can be achieved when the terminal is on the CPL $(x_{\tq} = y_{\tq} = 0)$. To clearly illustrate the different behaviors of the CRBs when the target terminal moves away from the CPL, the color of the point $(x_{\tq}, y_{\tq})$ is used to measure the normalized CRB values corresponding to that point. In particular, the normalized CRB values are mapped to the \textit{color gamut}, in which \textit{warmer} colors represent higher values, and lower values are associated with \textit{cooler} colors. It shows that the CRB for estimating $z_{\tq}$ increases faster than those for $x_{\tq}$ and $y_{\tq}$ regardless of using \textit{VEF} or \textit{SEF}. In addition, the maximum normalized values of $\mathrm{CRB}_{1}\left(\kappa_{\tq}\right)$ (as shown in Fig. \ref{fig:CRBxa}, \ref{fig:CRBya}, and \ref{fig:CRBza}) and $\mathrm{CRB}_{2}\left(\kappa_{\tq}\right)$ (as shown in Fig. \ref{fig:CRBxb}, \ref{fig:CRByb}, and \ref{fig:CRBzb}) are $18.40\textrm{dB}$,
$18.41\textrm{dB}$, $45.68\textrm{dB}$, $22.41\textrm{dB}$, $22.43\textrm{dB}$, and $49.69\textrm{dB}$, respectively. {Further, to 
 distinguish the difference among  Fig. \ref{fig:CRBxa}, \ref{fig:CRBxb}, \ref{fig:CRBya}, \ref{fig:CRByb} in an obvious manner, Fig. \ref{fig:TVT1} demonstrates the normalized CRBs in \textit{three-dimensional} (3D) view. It shows that the CRBs utilizing \textit{SEF} have a more significant increase than using \textit{VEF}, and the difference is about $4\textrm{dB}$ for all dimensions.
 Additionally, as for utilizing the same electric field type, the normalized CRB for $y_{\mathsf{t}}$ is slightly larger than that for $x_{\mathsf{t}}$. In particular, the difference in the maximum increase is $0.01\textrm{dB}$ when using \textit{VEF}, and $0.02\textrm{dB}$ if using \textit{SEF}. }

\subsection{CRB for the SIMO Positioning System}\label{subsection:CRB_simo_numer}
Finally, we will evaluate the CRBs for the SIMO positioning system as
discussed in Sec. \ref{sec:simo}. We set $R_{\rq}=30\textrm{m}$, $z_{\tq}=6\textrm{m}$, and $\lambda=0.001\textrm{m}$. According to Proposition 4, we compare the CRBs for a terminal on the CPL with different numbers of small receiving antennas, i.e., $N_{\mathsf{s}}^{2}=1,4,16,64,256$. 

As shown in Fig. \ref{fig:simo}, when $D_{\rq}>R_{\rq}$, the SIMO positioning system renders lower CRBs than the SISO positioning system for all dimensions. More precisely, $\mathrm{CRB}_{1}^{\mathsf{M}}\left(\kappa_{\tq}\right)$ will be one-$N_{\mathsf{s}}^{2}$th of $\mathrm{CRB}_{1}^{\mathsf{C}}\left(\kappa_{\tq}\right)$ as $D_{\rq}$ increases infinitely, as in Corollary 13. Considering the space constraints, we are more interested in the range $D_{\rq}\leq R_{\rq}$, in which the total surface area covered by the small receiving antennas is smaller than $\mathcal{R}_{\mathsf{s}}$ (the large rectangular surface region). It shows that the CRBs for $x_{\tq}$ and $y_{\tq}$ are significantly improved when adopting the SIMO system in the above range of practical interest, although the estimation accuracy for $z_{\tq}$ becomes worse. For instance, the CRBs for $x_{\tq}$ and $y_{\tq}$ with $4$ small receiving antennas, each antenna has a surface diagonal length $0.025\textrm{m}$, can achieve the same CRBs by using a single receiving antenna equipped with 
$D_{\rq} \approx 0.9\textrm{m}$. In other words, the antenna surface area needed for estimating $X$- and $Y$-dimension by the SIMO positioning system is only $1.23\%$ of that by the SISO system when $D_{\rq}$ is smaller than $1\textrm{m}$. The CRB for estimating $z_{\tq}$ with $4$ small receiving antennas is around $10\textrm{dB}$ greater than $\mathrm{CRB}^{\mathsf{C}}_{1}\left(z_{\tq}\right)$ when $D_{\rq}$ is the same and less than $10\textrm{m}$. Moreover, we find that  $\mathrm{CRB}_{1}^{\mathsf{M}}\left(x_{\tq}\right)$ remains the same when the number of small receiving antennas changes, whereas $\mathrm{CRB}_{1}^{\mathsf{M}}\left(y_{\tq}\right)$ is slightly lower when $N_{\mathsf{s}}^{2}=4$ compared to $N_{\mathsf{s}}^{2}=16,64,256$, and  $\mathrm{CRB}_{1}^{\mathsf{M}}\left(z_{\tq}\right)$ is slightly larger when $N_{\mathsf{s}}^{2}$ equals $4$. In fact, to achieve synchronous cooperation and coupling calibration among the small receiving antennas, more stringent hardware equipment is required as the number of the small antennas rises. Therefore, in light of the performance of the near-field positioning system and the cost of hardware, the SIMO positioning system with $4$ small receiving antennas is an excellent option for estimating $x_{\tq}$ and $y_{\tq}$, whereas the SISO system is a better choice for estimating $z_{\tq}$, i.e., \textit{ranging}. It is worth noting that using \textit{SEF} in the SIMO system has the same rules as using \textit{VEF}. Using \textit{OSEF} in the SIMO system with $4$ small antennas still fails to estimate the three coordinates of the terminal, but when the number of small receiving antennas is large enough, using \textit{OSEF} can be approximated as using \textit{SEF}.

\section{Conclusions}\label{sec:con}
In this paper, we have developed a complete \textit{electromagnetic propagation model} (EPM) to characterize near-field signals intrinsically. A generic near-field positioning system considering three different observed electric field types and the universality of the terminal position has been proposed based on the EPM. The CRBs for the three-dimensional spatial coordinates of the terminal have been derived. Three electric field types (\textit{vector}, \textit{scalar}, and \textit{overall scalar electric field}) have been deeply investigated for different antenna paradigms with  three disparate observation capabilities. The CRB expressions are generic and shown to generalize the existing results in \cite{de2021cramer}, in which the terminal is restricted to be located on the CPL of the receiving  surface while only the \textit{vector electric
field} type is utilized. The correlation between estimation precision and observed electric field type has been discovered. Additionally, the generic CPL model has been expanded to account for systems with multiple receiving antennas, and their performance has been thoroughly discussed. Numerical results have indicated that centimeter-level accuracy can be achieved in the near-field of the receiving antenna of a practical size in the mmWave or sub-THz bands by using the \textit{vector} or \textit{scalar electric field}. The \textit{overall scalar electric field} observed by a conventional surface antenna could only be utilized for the basic ranging. Moreover, the multiple receiving antennas could enhance the estimation accuracy of dimensions parallel to the receiving antenna surface.
\begin{appendices}
\section{Proof of Proposition 2}\label{proof:VEF}
From \eqref{eq:E1}, \eqref{eq:dipole}, ${\mathbf{e}}^{\mathsf{v}}\left({\mathbf{p}}_{\mathsf{r}}\right)=-{G_{s}}(\rrt)I_{\mathsf{in}}l_{\tq}\sin{\theta}\hat{\bm{\theta}}$ is derived. Since $\hat{\bm{\theta}}=\cos{\theta}\cos{\phi}\hat{\mathbf{x}}-\sin{\theta}\hat{\mathbf{y}}-\cos{\theta}\sin{\phi}\hat{\mathbf{z}}$, we have
{\begin{small}\setlength\abovedisplayskip{3pt}
\setlength\belowdisplayskip{3pt}
    \begin{align}
&e_{x}^{\mathsf{v}}\left(\mathbf{p}_{\mathsf{r}} \right)={G_{s}}(\rrt)\left(-\sin{\theta}\cos{\theta}\cos{\phi}\right)I_{\mathsf{in}}l_{\tq},\label{eq:Ex_simply}\\
&e_{y}^{\mathsf{v}}\left(\mathbf{p}_{\mathsf{r}} \right)={G_{s}}(\rrt)\left(\sin^{2}{\theta}\right)
I_{\mathsf{in}}l_{\tq},\label{eq:Ey_simply}\\
&e_{z}^{\mathsf{v}}\left(\mathbf{p}_{\mathsf{r}} \right)={G_{s}}(\rrt)\left(\sin{\theta}
\cos{\theta}\sin{\phi}\right)I_{\mathsf{in}}l_{\tq}\label{eq:Ez_simply}.
\end{align}
\end{small}}The dependence of $e_{x}^{\mathsf{v}}\left(\mathbf{p}_{\mathsf{r}} \right)$, $e_{y}^{\mathsf{v}}\left(\mathbf{p}_{\mathsf{r}} \right)$, and $e_{z}^{\mathsf{v}}\left(\mathbf{p}_{\mathsf{r}} \right)$ on the position $\left(x_{\tq},y_{\tq},z_{\tq}\right)$ is hidden in $\left(\rrt,\theta,\phi \right)$. From $\cos \theta=\left(y_{\rq}-y_{\tq}\right)/{\rrt}$ and $\tan \phi={z_{\tq}}/\left({x_{\rq}-x_{\tq}}\right)$, it follows that
{\begin{small}\setlength\abovedisplayskip{3pt}
\setlength\belowdisplayskip{3pt}
\begin{align}
&\sin \theta \cos \theta \cos \phi={\left(x_{\rq}-x_{\tq}\right)\left(y_{\rq}-y_{\tq}\right)}/{r_{\mathsf{rt}}^{2}} ,\label{eq:sin1}\\
&\sin ^{2} \theta=1-{\left(y_{\rq}-y_{\tq}\right)^{2}}/{r_{\mathsf{rt}}^{2}},\label{eq:sin2} \\
&\sin \theta \cos \theta \sin \phi={z _{\tq}\left(y_{\rq}-y_{\tq}\right)}/{r_{\mathsf{rt}}^{2}}\label{eq:sin3}.
\end{align}\end{small}}Substituting ${G_{s}}(\rrt)=-{\imagunit \eta \mathrm{e}^{-\imagunit  k_{0}\rrt}}/\left(2 \lambda \rrt\right)$ and \eqref{eq:sin1} -- \eqref{eq:sin3} into \eqref{eq:Ex_simply} -- \eqref{eq:Ez_simply} yields \eqref{eq:exnear} -- \eqref{eq:eznear}.

\section{Some Complex Expressions}\label{experssion}
{In proof of Proposition 3, we should compute the following first-order derivatives to derive the elements of FIM $\mathbf{I}\left(\bm{\xi}\right)$.}
{
{\begin{small}\setlength\abovedisplayskip{3pt}
\setlength\belowdisplayskip{3pt}
\begin{subequations}
\begin{align}
&\frac{\partial  h_{x}^{\mathsf{v}}\left(\mathbf{p}_{\rq}\right)}{\partial x_{\tq}}=x_{\rq,\tq}^{2}y_{\rq,\tq}\left(\frac{3\imagunit}{r_{\mathsf{rt}}^{5}}-\frac{{\imagunit}}{{x_{\rq,\tq}^{2}r_{\mathsf{rt}}^{3}}}-\frac{k_{0}} {r_{\mathsf{rt}}^{4}}\right) \mathrm{e}^{-\imagunit k_{0} \rrt},\label{eq:hx_xt} \\
&\frac{\partial h_{x}^{\mathsf{v}}\left(\mathbf{p}_{\rq}\right)}{\partial y_{\tq}}=x_{\rq,\tq}y_{\rq,\tq}^{2}\left(\frac{3\imagunit}{r_{\mathsf{rt}}^{5}}-\frac{{\imagunit}}{{y_{\rq,\tq}^{2}r_{\mathsf{rt}}^{3}}}-\frac{k_{0}} {r_{\mathsf{rt}}^{4}}\right) \mathrm{e}^{-\imagunit k_{0} \rrt},  \\
&\frac{\partial h_{x}^{\mathsf{v}}\left(\mathbf{p}_{\rq}\right)}{\partial z_{\tq}}=x_{\rq,\tq}y_{\rq,\tq}z_{\tq}\left(-\frac{3\imagunit}{r_{\mathsf{rt}}^{5}}+\frac{k_{0}}{r_{\mathsf{rt}}^{4}}\right) \mathrm{e}^{-\imagunit k_{0} \rrt},\\
&\frac{\partial  h_{y}^{\mathsf{v}}\left(\mathbf{p}_{\rq}\right)}{\partial x_{\tq}}=x_{\rq,\tq}\left(\imagunit\frac{3y_{\rq,\tq}^{2}-r_{\mathsf{rt}}^{2}} {r_{\mathsf{rt}}^{5}}-k_{0}\frac{y_{\rq,\tq}^{2}-r_{\mathsf{rt}}^{2}} {r_{\mathsf{rt}}^{4}}\right) \mathrm{e}^{-\imagunit k_{0} \rrt}, \\
&\frac{\partial  h_{y}^{\mathsf{v}}\left(\mathbf{p}_{\rq}\right)}{\partial y_{\tq}}=y_{\rq,\tq}\left(\imagunit \frac{3y_{\rq,\tq}^{2}-3r_{\mathsf{rt}}^{2}} {r_{\mathsf{rt}}^{5}}-k_{0}\frac{y_{\rq,\tq}^{2}-r_{\mathsf{rt}}^{2}} {r_{\mathsf{rt}}^{4}}\right) \mathrm{e}^{-\imagunit k_{0} \rrt}, \\
&\frac{\partial h_{y}^{\mathsf{v}}\left(\mathbf{p}_{\rq}\right)}{\partial z_{\tq}}=z_{\tq}\left(\imagunit\frac{-3 y_{\rq,\tq}^{2}+r_{\mathsf{rt}}^{2}}{r_{\mathsf{rt}}^{5}}+k_{0}\frac{y_{\rq,\tq}^{2}-r_{\mathsf{rt}}^{2}}{r_{\mathsf{rt}}^{4}}\right) \mathrm{e}^{-\imagunit k_{0} \rrt}, \\
&\frac{\partial h_{z}^{\mathsf{v}}\left(\mathbf{p}_{\rq}\right)}{\partial x_{\tq}}=x_{\rq,\tq}y_{\rq,\tq}z_{\tq}\left(-\frac{3\imagunit}{r_{\mathsf{rt}}^{5}}+\frac{k_{0}}{r_{\mathsf{rt}}^{4}}\right) \mathrm{e}^{-\imagunit k_{0} \rrt}, \\
    &\frac{\partial h_{z}^{\mathsf{v}}\left(\mathbf{p}_{\rq}\right)}{\partial y_{\tq}}=y_{\rq,\tq}^{2}z_{\tq}\left(-\frac{3\imagunit}{r_{\mathsf{rt}}^{5}}+\frac{\imagunit}{y_{\rq,\tq}^{2}r_{\mathsf{rt}}^{3}}+\frac{k_{0}}{r_{\mathsf{rt}}^{4}}\right) \mathrm{e}^{-\imagunit k_{0} \rrt},\\
&\frac{\partial h_{z}^{\mathsf{v}}\left(\mathbf{p}_{\rq}\right)}{\partial z_{\tq}}=y_{\rq,\tq}z_{\tq}^{2}\left(\frac{3\imagunit}{r_{\mathsf{rt}}^{5}}-\frac{\imagunit}{z_{\tq}^{2}r_{\mathsf{rt}}^{3}}-\frac{k_{0}}{r_{\mathsf{rt}}^{4}}\right) \mathrm{e}^{-\imagunit k_{0} \rrt}, \label{eq:hz_zt}
\end{align}
\end{subequations}
\end{small}}where we have set $x_{\rq,\tq}\triangleq x_{\rq}-x_{\tq}$ and $y_{\rq,\tq} \triangleq y_{\rq}-y_{\tq}$.

The specific expressions of $\rho^{mn}_{11}$ and $\rho^{mn}_{12}$ are as follows.
{\begin{small}\setlength\abovedisplayskip{3pt}
\setlength\belowdisplayskip{3pt}
\begin{align}
&\rho_{11}^{11}=k_{0}^{2}\iint_{\mathcal{R}_{\rq}} \frac{x_{\rq,\tq}^{2}\left(x_{\rq,\tq}^{2}+z_{\tq}^{2}\right)}{r_{\mathsf{rt}}^{6}} d x_{\rq} d y_{\rq},\label{eq:rho_11^11}\\
&\rho_{12}^{11}=\iint_{\mathcal{R}_{\rq}} \frac{\left(x_{\rq,\tq}^{2}+y_{\rq,\tq}^{2}\right)r_{\mathsf{rt}}^{2}-3x_{\rq,\tq}^{2}y_{\rq,\tq}^{2}}{r_{\mathsf{rt}}^{8}} d x_{\rq} d y_{\rq}, \\
&\rho_{11}^{22}=k_{0}^{2}\iint_{\mathcal{R}_{\rq}}\frac{y_{\rq,\tq}^{2}\left(x_{\rq,\tq}^{2}+z_{\tq}^{2}\right)}{r_{\mathsf{rt}}^{6}} d x_{\rq} d y_{\rq},\\
&\rho_{12}^{22}=\iint_{\mathcal{R}_{\rq}}\frac{\left(x_{\rq,\tq}^{2}+z_{\tq}^{2}\right)\left(x_{\rq,\tq}^{2}+z_{\tq}^{2}+4y_{\rq,\tq}^{2}\right)}{r_{\mathsf{rt}}^{8}}d x_{\rq} d y_{\rq},\\
&\rho_{11}^{33}=k_{0}^{2}z_{\tq}^{2}\iint_{\mathcal{R}_{\rq}}\frac{x_{\rq,\tq}^{2}+z_{\tq}^{2}}{r_{\mathsf{rt}}^{6}} d x_{\rq} d y_{\rq},\\
&\rho_{12}^{33}=\iint_{\mathcal{R}_{\rq}}\frac{y_{\rq,\tq}^{2}\left(r_{\mathsf{rt}}^{2}-2z_{\tq}^{2}\right)+z_{\tq}^{2}\left(z_{\tq}^{2}+x_{\rq,\tq}^{2}\right)}{r_{\mathsf{rt}}^{8}}d x_{\rq} d y_{\rq},\label{eq:rho_12^33}\\
&\rho_{11}^{12}=k_{0}^{2}\iint_{\mathcal{R}_{\rq}} \frac{x_{\rq,\tq}y_{\rq,\tq}\left(x_{\rq,\tq}^{2}+z_{\tq}^{2}\right)}{r_{\mathsf{rt}}^{6}} d x_{\rq} d y_{\rq},\label{eq:rho_11^12}\\
&\rho_{12}^{12}=\iint_{\mathcal{R}_{\rq}} \frac{x_{\rq,\tq}y_{\rq,\tq}\left(x_{\rq,\tq}^{2}-2y_{\rq,\tq}^{2}+z_{\tq}^{2}\right)}{r_{\mathsf{rt}}^{8}} d x_{\rq} d y_{\rq}, \\
&\rho_{11}^{13}=k_{0}^{2}\iint_{\mathcal{R}_{\rq}} \frac{-x_{\rq,\tq}z_{\tq}\left(x_{\rq,\tq}^{2}+z_{\tq}^{2}\right)}{r_{\mathsf{rt}}^{6}} d x_{\rq} d y_{\rq},\\
&\rho_{12}^{13}=\iint_{\mathcal{R}_{\rq}} \frac{x_{\rq,\tq}z_{\tq}\left(2y_{\rq,\tq}^{2}-x_{\rq,\tq}^{2}-z_{\tq}^{2}\right)}{r_{\mathsf{rt}}^{8}} d x_{\rq} d y_{\rq}, \\
&\rho_{11}^{23}=k_{0}^{2}\iint_{\mathcal{R}_{\rq}} \frac{-y_{\rq,\tq}z_{\tq}\left(x_{\rq,\tq}^{2}+z_{\tq}^{2}\right)}{r_{\mathsf{rt}}^{6}} d x_{\rq} d y_{\rq},\\
&\rho_{12}^{23}=\iint_{\mathcal{R}_{\rq}} \frac{y_{\rq,\tq}z_{\tq}\left(2y_{\rq,\tq}^{2}-x_{\rq,\tq}^{2}-z_{\tq}^{2}\right)}{r_{\mathsf{rt}}^{8}} d x_{\rq} d y_{\rq}.
\label{eq:rho_12^23}
\end{align}
\end{small}}

Some first-order derivatives in proof of Corollary 5 are
{\begin{small}\setlength\abovedisplayskip{3pt}
\setlength\belowdisplayskip{3pt}
\begin{subequations}
\begin{align}
&\frac{\partial h^{\mathsf{s}}\left(\mathbf{p}_{\rq}\right)}{\partial x_{\tq}}=x_{\rq,\tq}\left({\imagunit k_{0}} r_{\mathsf{rt}}^{-\frac{7}{2}}+\frac{5}{2} r_{\mathsf{rt}}^{-\frac{9}{2}}-\frac{ r_{\mathsf{rt}}^{-\frac{5}{2}}}{f_{xz}}\right) f_{ez},\label{eq:h_xt}\\
&\frac{\partial h^{\mathsf{s}}\left(\mathbf{p}_{\rq}\right)}{\partial y_{\tq}}=y_{\rq,\tq}\left(\frac{5}{2} r_{\mathsf{rt}}^{-\frac{9}{2}}+{\imagunit k_{0}} r_{\mathsf{rt}}^{-\frac{7}{2}}\right) f_{ez},\\
&\frac{\partial h^{\mathsf{s}}\left(\mathbf{p}_{\rq}\right)}{\partial z_{\tq}}=\left(\frac{3 z_{\tq}^{2}+x_{\rq,\tq}^{2}}{2 z_{\tq}f_{xz}} r_{\mathsf{rt}}^{-\frac{5}{2}}-\imagunit k_{0} z_{\tq} r_{\mathsf{rt}}^{-\frac{7}{2}}-\frac{5}{2}z_{\tq} r_{\mathsf{rt}}^{-\frac{9}{2}}\right) f_{ez},\label{eq:h_zt}
\end{align}
\end{subequations}\end{small}}where $f_{xz}\triangleq x_{\rq,\tq}^{2}+z_{\tq}^{2}$ and $f_{ez}\triangleq \sqrt{z_{\tq}f_{xz}}\mathrm{e}^{-\imagunit k_{0} \rrt}$.

The specific expressions of $\rho^{mn}_{21}$ and $\rho^{mn}_{22}$ are as follows. 
{\begin{small}\setlength\abovedisplayskip{3pt}
\setlength\belowdisplayskip{3pt}
\begin{align}
&\rho_{21}^{11}=k_{0}^{2}z_{\tq}\iint_{\mathcal{R}_{\rq}} \frac{x_{\rq,
\tq}^{2}\left(x_{\rq,\tq}^{2}+z_{\tq}^{2}\right)}{r_{\mathsf{rt}}^{7}} d x_{\rq} d y_{\rq},\label{eq:rho_21^11}\\
&\rho_{22}^{11}=z_{\tq}\iint_{\mathcal{R}_{\rq}}\frac{x_{\rq,\tq}^{2}\left({25}f_{xz}/4-5r_{\mathsf{rt}}^{2}+f_{xz}^{-1}r_{\mathsf{rt}}^{4}\right)}{r_{\mathsf{rt}}^{9}}d x_{\rq} d y_{\rq},\\
&\rho_{21}^{22}=k_{0}^{2}z_{\tq}\iint_{\mathcal{R}_{\rq}} \frac{y_{\rq,\tq}^{2}\left(x_{\rq,\tq}^{2}+z_{\tq}^{2}\right)}{r_{\mathsf{rt}}^{7}} d x_{\rq} d y_{\rq},\\
&\rho_{22}^{22}=z_{\tq}\iint_{\mathcal{R}_{\rq}} \frac{{25}y_{\rq,\tq}^{2}\left(x_{\rq,\tq}^{2}+z_{\tq}^{2}\right)}{4r_{\mathsf{rt}}^{9}} d x_{\rq} d y_{\rq},\\
&\rho_{21}^{33}=k_{0}^{2}z_{\tq}^{3}\iint_{\mathcal{R}_{\rq}} \frac{x_{\rq,\tq}^{2}+z_{\tq}^{2}}{r_{\mathsf{rt}}^{7}} d x_{\rq} d y_{\rq},\\
&\rho_{22}^{33}=\iint_{\mathcal{R}_{\rq}}
\frac{\left[x_{\rq,\tq}^{2}\left(r_{\mathsf{rt}}^{2}-2z_{\tq}^{2}\right)+z_{\tq}^{2}\left(3y_{\rq,\tq}^{2}-2z_{\tq}^{2}\right)\right]^{2}}{4z_{\tq}\left(x_{\rq,\tq}^{2}+z_{\tq}^{2}\right)r_{\mathsf{rt}}^{9}}d x_{\rq} d y_{\rq},\label{eq:rho_22^33}\\
&\rho_{21}^{12}=k_{0}^{2}z_{\tq}\iint_{\mathcal{R}_{\rq}}\frac{x_{\rq,\tq}y_{\rq,\tq}\left(x_{\rq,\tq}^{2}+z_{\tq}^{2}\right)}{r_{\mathsf{rt}}^{7}} d x_{\rq} d y_{\rq}\label{eq:rho_21^12},\\
&\rho_{22}^{12}=z_{\tq}\iint_{\mathcal{R}_{\rq}}\frac{x_{\rq,\tq}y_{\rq,\tq}\left[25\left(x_{\rq,\tq}^{2}+z_{\tq}^{2}\right)/4-5r_{\mathsf{rt}}^{2}/2\right]}{r_{\mathsf{rt}}^{9}} d x_{\rq} d y_{\rq},\\
&\rho_{21}^{13}=k_{0}^{2}z_{\tq}^{2}\iint_{\mathcal{R}_{\rq}} \frac{-x_{\rq,\tq}\left(x_{\rq,\tq}^{2}+z_{\tq}^{2}\right)}{r_{\mathsf{rt}}^{7}} d x_{\rq} d y_{\rq},\\
&\rho_{22}^{13}=\iint_{\mathcal{R}_{\rq}} \frac{x_{\rq,\tq}\left(f_{5z}f_{xz}-f_{3z}-25z_{\tq}^{2}f_{xz}^{2}/2\right)}{2\left(x_{\rq,\tq}^{2}+z_{\tq}^{2}\right)r_{\mathsf{rt}}^{9}} d x_{\rq} d y_{\rq},\\
&\rho_{21}^{23}=k_{0}^{2}z_{\tq}^{2}\iint_{\mathcal{R}_{\rq}}\frac{-y_{\rq,\tq}\left(x_{\rq,\tq}^{2}+z_{\tq}^{2}\right)}{r_{\mathsf{rt}}^{7}} d x_{\rq} d y_{\rq},\\
&\rho_{22}^{23}=\iint_{\mathcal{R}_{\rq}} \frac{5y_{\rq,\tq}\left[f_{3z}/r_{\mathsf{rt}}^{2}-5z_{\tq}^{2}\left(x_{\rq,\tq}^{2}+z_{\tq}^{2}\right)\right]}{4r_{\mathsf{rt}}^{9}} d x_{\rq} d y_{\rq},\label{eq:rho_22^23}
\end{align}\end{small}}where $f_{3z}\triangleq \left(x_{\rq,\tq}^{2}+3z_{\tq}^{2}\right)r_{\mathsf{rt}}^{4}$ and $f_{5z}\triangleq 5\left(x_{\rq,\tq}^{2}+5z_{\tq}^{2}\right)r_{\mathsf{rt}}^{2}/2$.

In Corollary 6, the expressions of $\rho^{mn}_{3}$ are as follows.
{\begin{small}\setlength\abovedisplayskip{3pt}
\setlength\belowdisplayskip{3pt}
\begin{align}
&\rho_{3}^{11}=\frac{D_{\rq}^{2}}{2\alpha^{2}}\bigg|\sum_{i=1}^{\sqrt{\alpha}} \sum_{j=1}^{\sqrt{\alpha}}x_{i,\tq} g_{zx}\Big(g_{r}-\frac{z_{\tq}}{|g_{zx}|^{2}} r_{\mathsf{rt};i,j}^{-\frac{5}{2}}\Big)\bigg|^{2},\label{eq:rho_3^11}\\
&\rho_{3}^{22}=\frac{D_{\rq}^{2}}{2\alpha^{2}}\bigg|\sum_{i=1}^{\sqrt{\alpha}} \sum_{j=1}^{\sqrt{\alpha}}y_{j,\tq} g_{zx}g_{r}\bigg|^{2},\\
&\rho_{3}^{33}=\frac{D_{\rq}^{2}}{2\alpha^{2}}\bigg|\sum_{i=1}^{\sqrt{\alpha}} \sum_{j=1}^{\sqrt{\alpha}}g_{zx}\Big(\frac{3 z_{\tq}^{2}+x_{i,\tq}^{2}}{2 |g_{zx}|^{2}} r_{\mathsf{rt};i,j}^{-\frac{5}{2}}-z_{\tq}g_{r}\Big)\bigg|^{2},\\
&\rho_{3}^{12}=\frac{D_{\rq}^{2}}{2\alpha^{2}}\operatorname{Re}\bigg\{\Big(\sum_{i=1}^{\sqrt{\alpha}} \sum_{j=1}^{\sqrt{\alpha}}\mathscr{J}_{yg}\Big)\Big(\sum_{i=1}^{\sqrt{\alpha}} \sum_{j=1}^{\sqrt{\alpha}}x_{i,\tq} g_{zx}\mathscr{J}_{gzr}\Big)^{*}\bigg\},\\
&\rho_{3}^{13}=\frac{D_{\rq}^{2}}{2\alpha^{2}}\operatorname{Re}\bigg\{\Big(\sum_{i=1}^{\sqrt{\alpha}} \sum_{j=1}^{\sqrt{\alpha}}\mathscr{J}_{zxg}\Big)\Big(\sum_{i=1}^{\sqrt{\alpha}} \sum_{j=1}^{\sqrt{\alpha}}x_{i,\tq} g_{zx}\mathscr{J}_{gzr}\Big)^{*}\bigg\},\\
&\rho_{3}^{23}=\frac{D_{\rq}^{2}}{2\alpha^{2}}\operatorname{Re}\bigg\{\Big(\sum_{i=1}^{\sqrt{\alpha}} \sum_{j=1}^{\sqrt{\alpha}}\mathscr{J}_{zxg}\Big)\Big(\sum_{i=1}^{\sqrt{\alpha}} \sum_{j=1}^{\sqrt{\alpha}}y_{j,\tq} g_{zx}g_{r}\Big)^{*}\bigg\},\label{eq:rho_3^23}
\end{align}\end{small}}where $x_{i,\tq}\triangleq x_{i}-x_{\tq}$, $y_{j,\tq}\triangleq y_{j}-x_{\tq}$, $g_{r}\triangleq \frac{5}{2}r_{\mathsf{rt};i,j}^{-\frac{9}{2}}+\imagunit k_{0}r_{\mathsf{rt};i,j}^{-\frac{7}{2}}$, $g_{zx}\triangleq \sqrt{z_{\tq}\left(z_{\tq}^{2}+x_{i,\tq}^{2}\right)}\mathrm{e}^{-\imagunit k_{0} r_{\mathsf{rt};i,j}}$, $\mathscr{J}_{yg}\triangleq y_{j,\tq}g_{zx}g_{r}$, $\mathscr{J}_{gzr}\triangleq g_{r}-\frac{z_{\tq}}{|g_{zx}|^{2}} r_{\mathsf{rt};i,j}^{-\frac{5}{2}}$, and $\mathscr{J}_{zxg}\triangleq g_{zx}\left(\frac{3 z_{\tq}^{2}+x_{i,\tq}^{2}}{2 |g_{zx}|^{2}} r_{\mathsf{rt};i,j}^{-\frac{5}{2}}-z_{\tq}g_{r}\right)$.}

\section{The closed-form expressions}\label{ap:remark2}
{The double integral formulas \eqref{eq:12x}, \eqref{eq:12y}, \eqref{eq:11z} and \eqref{eq:12z} can be computed in the following closed-form expressions.
{\begin{small}\setlength\abovedisplayskip{3pt}
\setlength\belowdisplayskip{3pt}
\begin{align}
&\rho_{12x}=\frac{1}{\tau^{2}+8}\left[\frac{f_{\mathrm{tan}}}{2\sqrt{\tau^{2}+8}}-\frac{\tau^{2}\left(3\tau^{2}+16\right)}{\left(\tau^{2}+4\right)^{2}}\right],\label{eq:12x_close}\\
&\rho_{12y}=\frac{\left(9\tau^{4}+152\tau^{2}+544\right)}{2\left(\tau^{2}+8\right)^{5/2}\tau^{-1} f_{\mathrm{tan}}^{-1}}+\frac{\tau^{2}\left(3\tau^{4}+8\tau^{2}-32\right)}{\left(\tau^{2}+8\right)^{2}\left(\tau^{2}+4\right)^{2}},\label{eq:12y_close}\\
&\rho_{11z}=\frac{\tau}{\tau^{2}+8}\left[\frac{\left(3\tau^{2}+28\right)}{\sqrt{\tau^{2}+8}}f_{\mathrm{tan}}+\frac{2\tau}{\tau^{2}+4}\right],\label{eq:11z_close}\\
&\rho_{12z}=\frac{2\tau}{\left(\tau^{2}+8\right)^{2}}\left[\frac{\tau^{4}+16\tau^{2}+88}{\sqrt{\tau^{2}+8}f_{\mathrm{tan}}^{-1}}+\frac{16\tau\left(\tau^{2}+5\right)}{\left(\tau^{2}+4\right)^{2}}\right],\label{eq:12z_close}
\end{align}\end{small}}where $f_{\mathrm{tan}}\triangleq \arctan{\frac{\tau}{\sqrt{\tau^{2}+8}}}$.

To provide the closed-form upper and lower bounds of $\rho_{11x}$ and $\rho_{11y}$, we denote two circular domains $\mathcal{C}^{-}=\big \{(u,v):u^{2}+v^{2}\leq (\frac{\tau}{\sqrt{8}})^{2}\big\}$, $\mathcal{C}^{+}=\big \{(u,v):u^{2}+v^{2}\leq (\frac{\tau}{2})^{2}\big\}$ and two non-negative function $g_{11x}={u^{2}(u^{2}+1)}/{(u^{2}+v^{2}+1)^{3}}$, $g_{11y}={v^{2}(u^{2}+1)}/{(u^{2}+v^{2}+1)^{3}}$, then we have
{\begin{small}\setlength\abovedisplayskip{3pt}
\setlength\belowdisplayskip{3pt}
\begin{align}
&\iint_{\mathcal{C}^{-}}g_{11x}dudv<\rho_{11x}<\iint_{\mathcal{C}^{+}}g_{11x}dudv,\\
&\iint_{\mathcal{C}^{-}}g_{11y}dudv<\rho_{11y}<\iint_{\mathcal{C}^{+}}g_{11y}dudv.
\end{align}
\end{small}}Therefore, the closed-form upper and lower bounds of \eqref{eq:11x} and \eqref{eq:11y} can be derived as follows. 
{\begin{small}\setlength\abovedisplayskip{3pt}
\setlength\belowdisplayskip{3pt}
\begin{align}
&\iint_{\mathcal{C}^{+}}g_{11x}dudv=\frac{3\pi}{8}\ln{(1+\frac{\tau^{2}}{4})}-\frac{\pi\tau^{2}(5\tau^{2}+24)}{16(\tau^{2}+4)^{2}},\label{eq:11x+}\\
&\iint_{\mathcal{C}^{-}}g_{11x}dudv=\frac{3\pi}{8}\ln{(1+\frac{\tau^{2}}{8})}-\frac{\pi\tau^{2}(5\tau^{2}+48)}{16(\tau^{2}+8)^{2}},\label{eq:11x-}\\
&\iint_{\mathcal{C}^{+}}g_{11y}dudv=\frac{\pi}{8}\ln{(1+\frac{\tau^{2}}{4})}+\frac{\pi\tau^{2}(\tau^{2}-8)}{16(\tau^{2}+4)^{2}},\label{eq:11y+}\\
&\iint_{\mathcal{C}^{-}}g_{11y}dudv=\frac{\pi}{8}\ln{(1+\frac{\tau^{2}}{8})}+\frac{\pi\tau^{2}(\tau^{2}-16)}{16(\tau^{2}+8)^{2}}\label{eq:11y-}.
\end{align}\end{small}}Similarly, we denote $g_{2i\kappa}$, $i=1,2$ as the integrand functions of \eqref{eq:21x} -- \eqref{eq:22z}, then we have
{\begin{small}\setlength\abovedisplayskip{3pt}
\setlength\belowdisplayskip{3pt}
\begin{equation}
\rho_{2i\kappa}^{(l)}\triangleq \iint_{\mathcal{C}^{-}}g_{2i\kappa}dudv<\rho_{2i\kappa}<\iint_{\mathcal{C}^{+}}g_{2i\kappa}dudv \triangleq \rho_{2i\kappa}^{(u)}.
\end{equation}\end{small}}The closed-form upper and lower bounds of  $\rho_{21\kappa}$ and $\rho_{22\kappa}$ can be computed as follows.
{\begin{small}\setlength\abovedisplayskip{3pt}
\setlength\belowdisplayskip{3pt}
\begin{align}
&\rho_{21x}^{(u)}=\frac{8\pi}{15}-\frac{\pi\left(45\tau^{4}+320\tau^{2}+512\right)}{30\left(\tau^{2}+4\right)^{5/2}}\label{eq:21xu},\\
&\rho_{21x}^{(l)}=\frac{8\pi}{15}-\frac{\sqrt{2}\pi\left(45\tau^{4}+640\tau^{2}+2048\right)}{30\left(\tau^{2}+8\right)^{5/2}},\label{eq:21xl}\\
&\rho_{22x}^{(u)}=\frac{3\pi}{14}-\frac{\big[63\tau^{4}-112(\tau^{2}+4)^{3/2}+32(21\tau^{2}+40)\big]}{14\pi^{-1}\left(\tau^{2}+4\right)^{7/2}},\label{eq:22xu}\\
&\rho_{22x}^{(l)}=\frac{3\pi}{14}-\frac{\pi\big[63\sqrt{2}\tau^{4}-224\left(\tau^{2}+8\right)^{3/2}+64\sqrt{2}f_{\tau1}\big]}{7\left(\tau^{2}+8\right)^{7/2}},\\
&\rho_{21y}^{(u)}=\frac{4\pi}{15}-\frac{\pi\left(15\tau^{4}+160\tau^{2}+256\right)}{30\left(\tau^{2}+4\right)^{5/2}},\\
&\rho_{21y}^{(l)}=\frac{4\pi}{15}-\frac{\sqrt{2}\pi\left(15\tau^{4}+320\tau^{2}+1024\right)}{30\left(\tau^{2}+8\right)^{5/2}},\label{eq:21yl}\\
&\rho_{22y}^{(u)}=\frac{10\pi}{21}-\frac{5\pi\left(35\tau^{4}+448\tau^{2}+512\right)}{42\left(\tau^{2}+4\right)^{7/2}},\\
&\rho_{22y}^{(l)}=\frac{10\pi}{21}-\frac{5\sqrt{2}\pi\left(35\tau^{4}+896\tau^{2}+2048\right)}{21\left(\tau^{2}+8\right)^{7/2}}\label{eq:22yl},\\
&\rho_{21z}^{(u)}=\frac{8\pi}{15}-\frac{8\pi\left(5\tau^{2}+32\right)}{15\left(\tau^{2}+4\right)^{5/2}},\\
&\rho_{21z}^{(l)}=\frac{8\pi}{15}-\frac{16\sqrt{2}\pi\left(5\tau^{2}+64\right)}{15\left(\tau^{2}+8\right)^{5/2}},\label{eq:21zl}\\
&\rho_{22z}^{(u)}=\frac{13\pi}{42}-\frac{\big[7\tau^{4}(3\tau^{2}+32)+336(\tau^{2}+4)^{3/2}-f_{\tau 2}\big]}{42\pi^{-1}\left(\tau^{2}+4\right)^{7/2}}\label{eq:22zu},\\
&\rho_{22z}^{(l)}=\frac{13\pi}{42}-\frac{\pi\big[7\sqrt{2}\tau^{4}f_{\tau 3}+1344\left(\tau^{2}+8\right)^{3/2}-f_{\tau 4}\big]}{42\left(\tau^{2}+8\right)^{7/2}},\label{eq:22zl}
\end{align}
\end{small}}where \begin{small}$f_{\tau1}=21\tau^{2}+80$\end{small}, \begin{small}$f_{\tau2}=128\left(7\tau^{2}+8\right)$\end{small}, \begin{small}$f_{\tau3}=3\tau^{2}+64$\end{small}, and \begin{small}$f_{\tau4}=512\sqrt{2}\left(7\tau^{2}+16\right)$\end{small}.}

\section{Proof of Corollary 10}\label{ap:prop4}
{We need to prove that $k_{0}^{2}\rho_{11\kappa}\gg z_{\tq}^{-2}\rho_{12\kappa}$ and $k_{0}^{2}\rho_{21\kappa}\gg z_{\tq}^{-2}\rho_{22\kappa}$ for $z_{\tq}\gg \lambda$, then the approximation in Corollary 10 can be proved immediately.

When $z_{\tq}\gg \lambda$, we have $k_{0}^{2}\gg z_{\tq}^{-2}$. Observe that
\begin{equation}\small\setlength\abovedisplayskip{3pt}
\setlength\belowdisplayskip{3pt}
2\rho_{11x}>\iint_{\mathcal{R}_{\tau}}\frac{2u^{2}(u^{2}+1)}{(u^{2}+v^{2}+1)^{4}}dudv>\rho_{12x}>0,
\end{equation}
from which we  obtain $ k_{0}^{2}\rho_{11x}\gg z_{\tq}^{-2}\rho_{12x}$. Similarly, we have
\begin{equation}\small\setlength\abovedisplayskip{3pt}
\setlength\belowdisplayskip{3pt}
\frac{u^{2}+1}{(u^{2}+v^{2}+1)^{3}}\geq \frac{v^{4}+u^{2}v^{2}+1}{(u^{2}+v^{2}+1)^{4}}>0.
\end{equation}
Then, we have that $\rho_{11z}>\rho_{12z}$. Accordingly, we have that $k_{0}^{2}\rho_{11z}\gg z_{\tq}^{-2}\rho_{12z}$.
Observe that
\begin{equation}\small\setlength\abovedisplayskip{3pt}
\setlength\belowdisplayskip{3pt}
4\pi^{2}\rho_{21y}>\iint_{\mathcal{R}_{\tau}} \frac{4\pi^{2}v^{2}(u^{2}+1)}{(u^{2}+v^{2}+1)^{9/2}}dudv>\rho_{22y}>0.
\end{equation}
Consequently, $ k_{0}^{2}\rho_{21y}\gg z_{\tq}^{-2}\rho_{22y}$ can be proved.

The remaining inequalities are challenging to demonstrate analytically, so we provide numerical proofs. If we define the difference function $f_{dy1}(\tau)=\rho_{11y}-\rho_{12y}$, we can deduce from \eqref{eq:12y_close} and \eqref{eq:11y-} that the minimum value of the function $f_{dy1}(\tau)$ is greater than $-2.34$, which testifies that $k_{0}^{2}\rho_{11y}\gg z_{\tq}^{-2}\rho_{12y}$. Define the difference function $f_{dx2}(\tau)=\rho_{21x}-\rho_{22x}$, from \eqref{eq:21xl} and \eqref{eq:22xu}, we have that $f_{dx2}(\tau)>\rho_{21x}^{(l)}-\rho_{22x}^{(u)}$, and we derive that the minimum value of the function $f_{dx2}(\tau)$ is greater than $67\pi/210-1.23\approx-0.23$, therefore $k_{0}^{2}\rho_{21x}\gg z_{\tq}^{-2}\rho_{22x}$ can be proved.
Similarly, we define $f_{dz2}(\tau)=\rho_{21z}-\rho_{22z}$, then we have that $f_{dz2}(\tau)>\rho_{21z}^{(l)}-\rho_{22z}^{(u)}$ based on \eqref{eq:21zl} and \eqref{eq:22zu}. Next, we can deduce that the minimum value of the function $f_{dz2}(\tau)$ is greater than $47\pi/210-0.80\approx-0.10$, which verifies that $k_{0}^{2}\rho_{21z}\gg z_{\tq}^{-2}\rho_{22z}$.} 
\end{appendices}

\bibliographystyle{IEEEtran}
\bibliography{reference}

\begin{IEEEbiography}[{\includegraphics[width=1in,height=1.25in,clip,keepaspectratio]{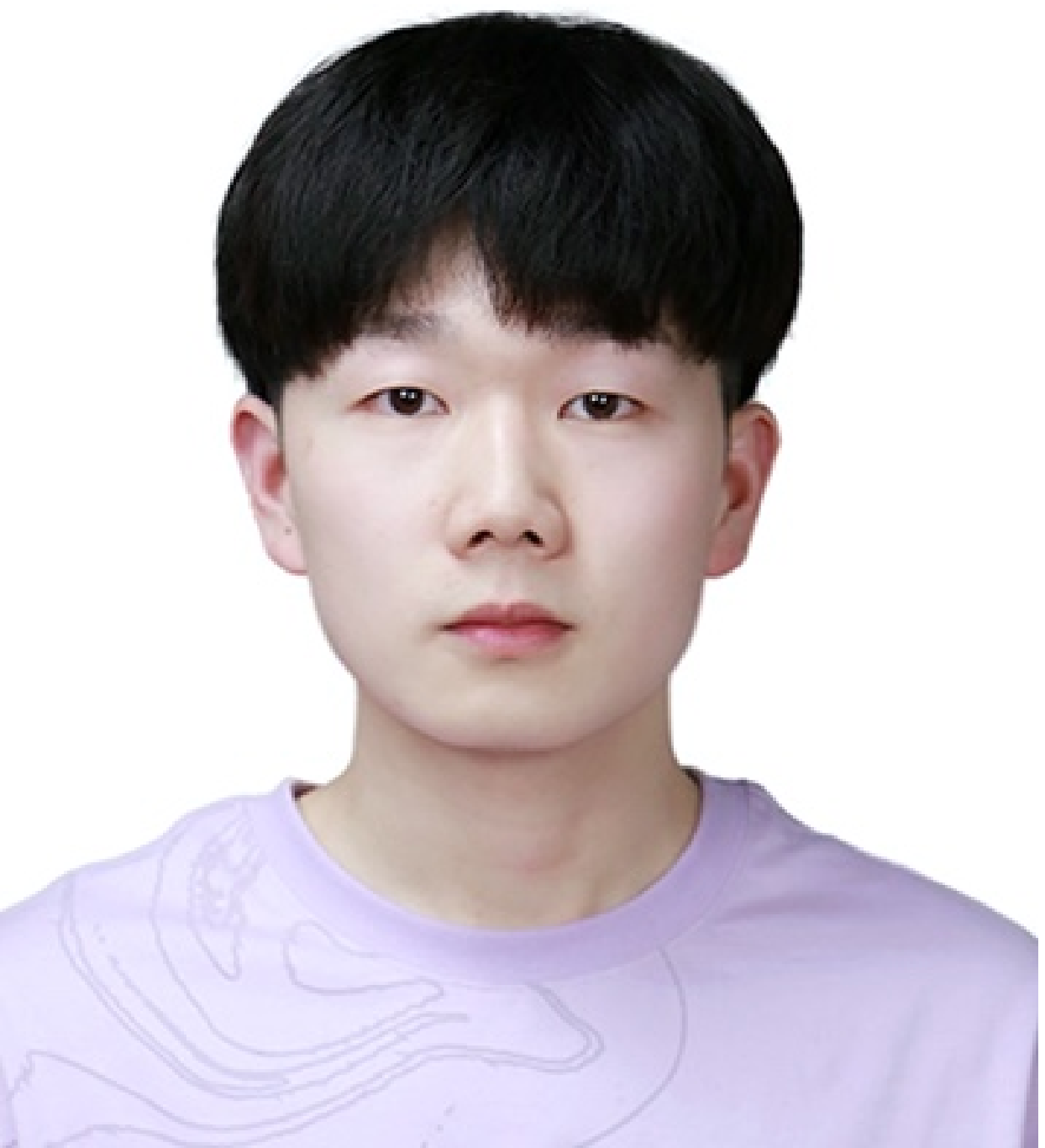}}]{Ang Chen}
    received the bachelor’s degree in communications engineering from the University of Electronic Science and Technology of China (UESTC), Chengdu, China, in 2021. He is currently pursuing the master’s degree with the Department of Electronic Engineering and Information Science, University of Science and Technology of China (USTC), Hefei, China. His research interests include near-field positioning and communication, statistical signal processing, and integrated sensing and communication.
\end{IEEEbiography}

\begin{IEEEbiography}[{\includegraphics[width=1in,height=1.25in,clip,keepaspectratio]{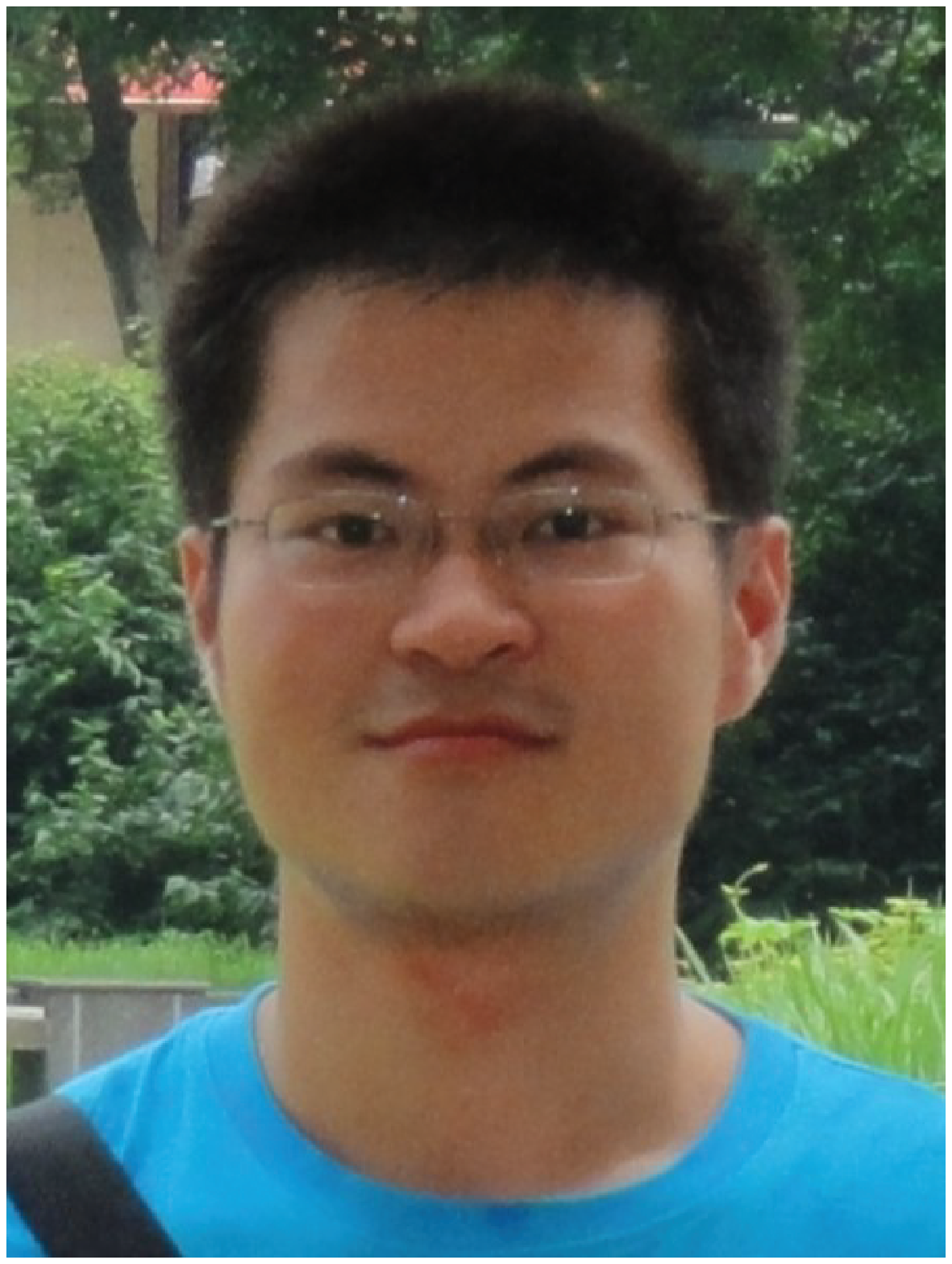}}]{Li Chen}
    (Senior Member, IEEE) received the B.E. in electrical and information engineering from Harbin Institute of Technology, Harbin, China, in 2009 and the Ph.D. degree in electrical engineering from the University of Science and Technology of China, Hefei, China, in 2014. He is currently an Associate Professor with the Department of Electronic Engineering and Information Science, University of Science and Technology of China. His research interests include integrated communication and computation, integrated sensing and communication and wireless IoT networks.
\end{IEEEbiography}

\begin{IEEEbiography}[{\includegraphics[width=1in,height=1.25in,clip,keepaspectratio]{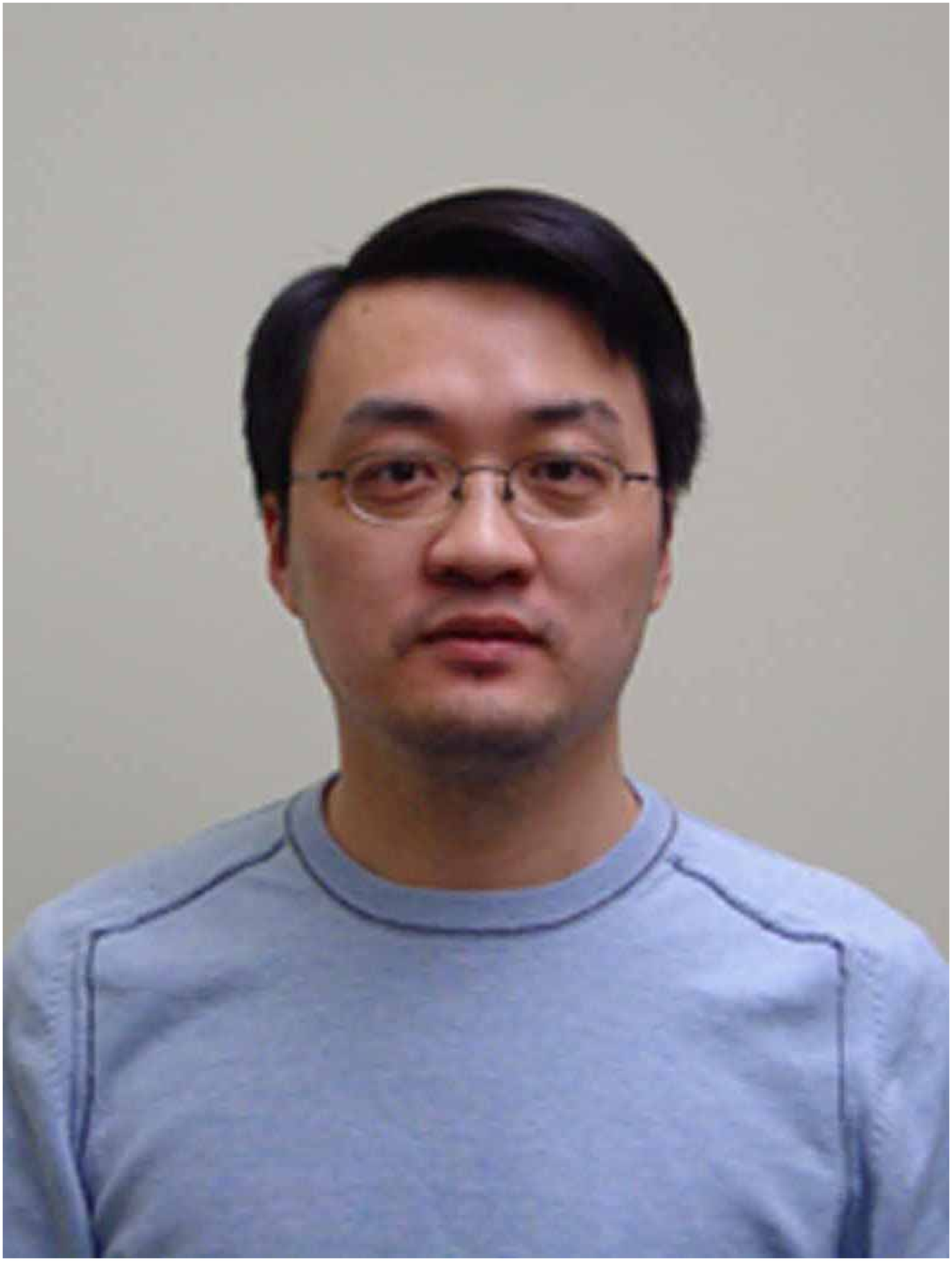}}]
{Yunfei Chen} (S'02-M'06-SM'10) received his B.E. and M.E. degrees in electronics engineering from Shanghai Jiaotong University, Shanghai, P.R.China, in 1998 and 2001, respectively. He received his Ph.D. degree from the University of Alberta in 2006. He is currently working as a Professor in the Department of Engineering at the University of Durham, UK. His research interests include wireless communications, cognitive radios, wireless relaying and energy harvesting.
\end{IEEEbiography}

\begin{IEEEbiography}[{\includegraphics[width=1in,height=1.25in,clip,keepaspectratio]{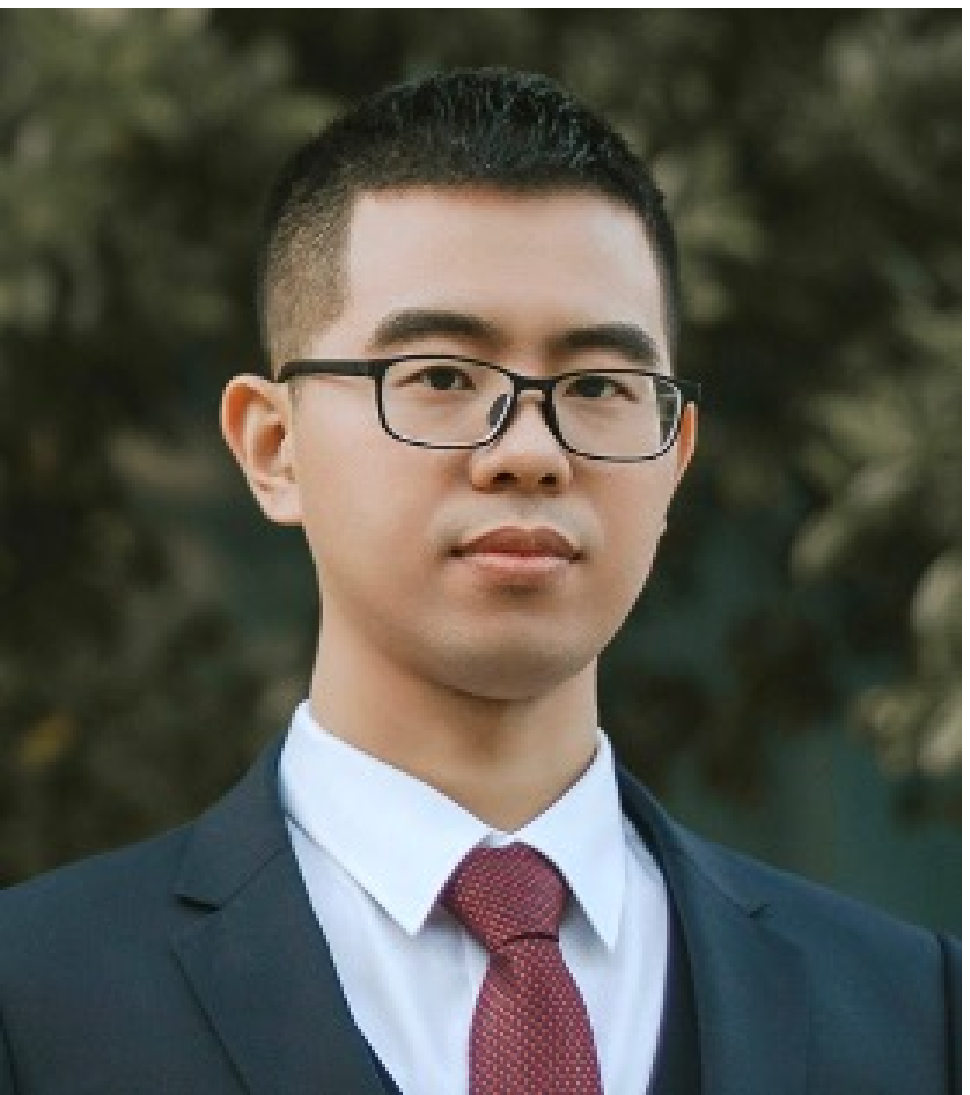}}]
{Changsheng You} (Member, IEEE) received his B.Eng. degree in 2014 from University of Science and Technology of China (USTC) and Ph.D. degree in 2018 from The University of Hong Kong (HKU). He is currently an Assistant Professor at Southern University of Science and Technology, and was a Research Fellow at National University of Singapore (NUS). His research interests include intelligent reflecting surface, UAV communications, edge learning, mobile-edge computing. Dr. You is an editor for IEEE Communications Letters (CL), IEEE IEEE Transactions on Green Communications and Networking (TGCN), and IEEE Open Journal of the Communications Society (OJ-COMS). He received the IEEE Communications Society Asia-Pacific Region Outstanding Paper Award in 2019, IEEE ComSoc Best Survey Paper Award in 2021, IEEE ComSoc Best Tutorial Paper Award in 2023. He is listed as the Highly Cited Chinese Researcher, Exemplary Reviewer of the IEEE Transactions on Communications (TCOM) and IEEE Transactions on Wireless Communications (TWC). 
\end{IEEEbiography}

\begin{IEEEbiography}[{\includegraphics[width=1in,height=1.25in,clip,keepaspectratio]{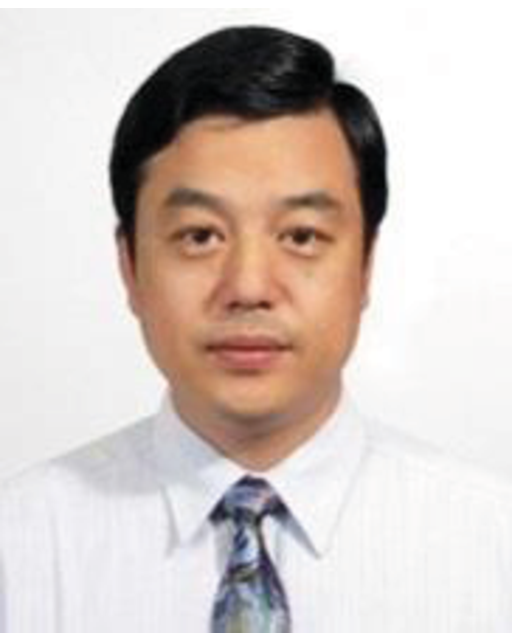}}]
{Guo Wei} received the B.S. degree in electronic engineering from the University of Science and Technology
of China (USTC), Hefei, China, in 1983 and
the M.S. and Ph.D. degrees in electronic engineering
from the Chinese Academy of Sciences, Beijing,
China, in 1986 and 1991, respectively.
He is currently a Professor with the School of Information
Science and Technology, USTC. His current
research interests include wireless and mobile
communications, wireless multimedia communications, and
wireless information networks.
\end{IEEEbiography}

\begin{IEEEbiography}[{\includegraphics[width=1in,height=1.25in,clip,keepaspectratio]{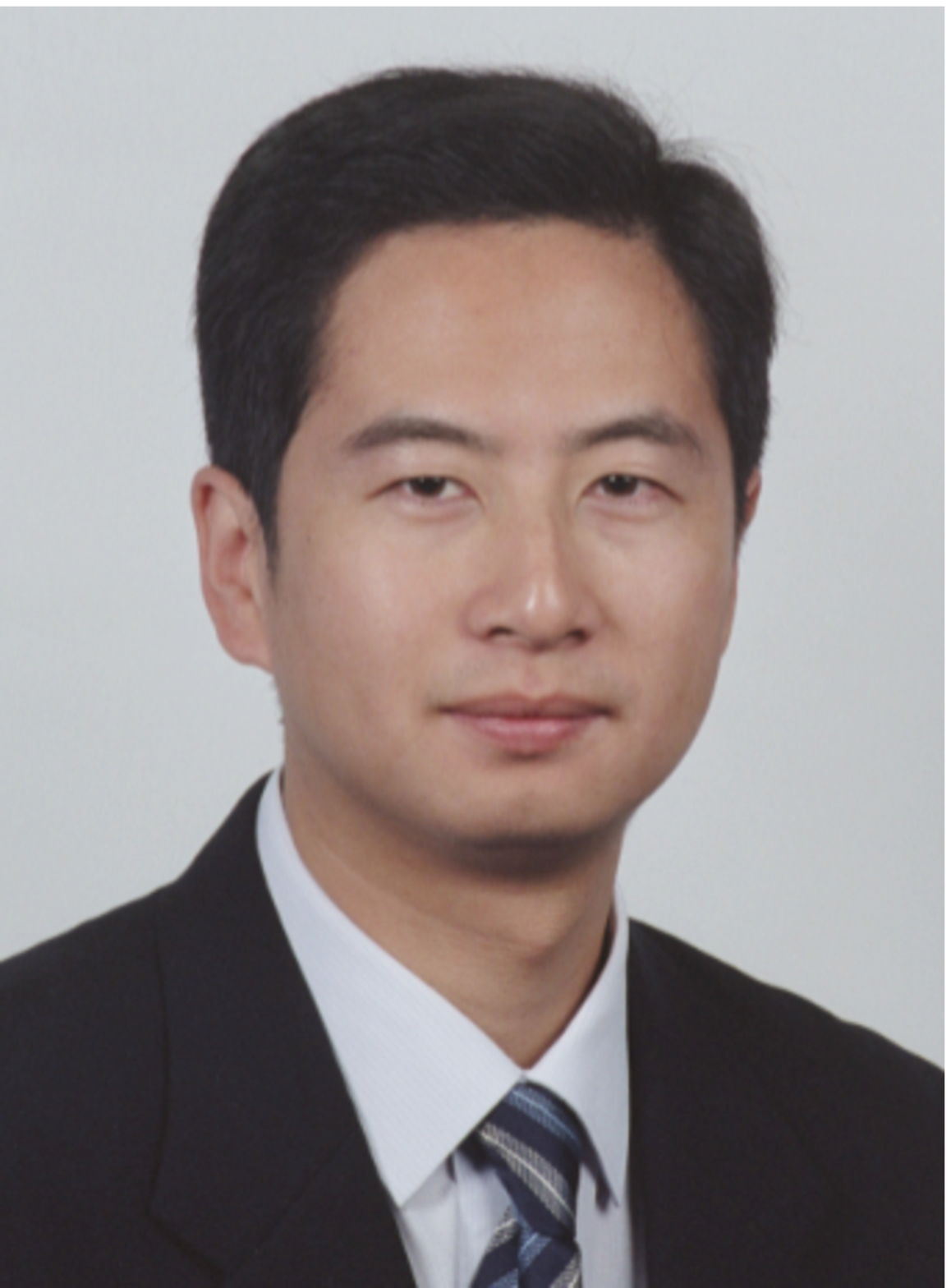}}]
{F. Richard Yu} (Fellow, IEEE)  received the PhD
degree in electrical engineering from the University
of British Columbia (UBC) in 2003. His research
interests include connected/autonomous vehicles, artificial intelligence, blockchain, and wireless systems. He has been named in the Clarivate Analytics
list of “Highly Cited Researchers” since 2019, and
received several Best Paper Awards from some first-tier conferences. He was a Board Member the
IEEE VTS and is the Editor-in-Chief for IEEE VTS
Mobile World newsletter. He is a Fellow of the
IEEE, Canadian Academy of Engineering (CAE), Engineering Institute of
Canada (EIC), and IET. He is a Distinguished Lecturer of IEEE in both VTS
and ComSoc.
\end{IEEEbiography}

\end{document}